\documentclass[12pt]{JHEP3}
\usepackage{amssymb}
\usepackage{amsmath}
\usepackage{graphicx}
\usepackage{epsfig}
\newcommand{\be}{\begin{equation}}
\newcommand{\ee}{\end{equation}}
\newcommand{\bea}{\begin{eqnarray}}
\newcommand{\eea}{\end{eqnarray}}
\newcommand{\ba}{\begin{eqnarray}}
\newcommand{\ea}{\end{eqnarray}}
\newcommand{\bln}{\begin{align}}
\newcommand{\eln}{\end{align}}
\newcommand{\bst}{\begin{split}}
\newcommand{\est}{\end{split}}
\newcommand{\bi}{\begin{itemize}}
\newcommand{\ei}{\end{itemize}}

\newcommand{\lt}{\left}
\newcommand{\rt}{\right}
\def\ov{\over}
\def\le{\left}
\def\ri{\right}
\def\ha{{1\over 2}}

\newcommand{\p}{\partial}
\newcommand{\lra}{\longrightarrow}

\def\lam{{\lambda}}
\def\Lam{{\Lambda}}
\def\al{{\alpha}}
\def\th{{\theta}}
\def\Om{{\Omega}}
\def \om {\omega}
\def \th{{\theta}}

\def \ga {\gamma}
\def \lam {\lambda}
\def\sig{{\sigma}}
\def\ep{{\epsilon}}
\newcommand{\eps}{\varepsilon}
\def\apr{{\alpha'}}

\newcommand{\OO}{\mathcal{O}}

\def\NN{{\cal N}}

\newcommand{\RR}{\mathbb{R}}

\def\det{{\rm det}}

\newcommand{\dbyd}[1]{\ensuremath{\left(\frac{\partial}{\partial #1}\right)}}

\relax

\title{A limiting velocity for quarkonium propagation in a strongly coupled
plasma via AdS/CFT}

\author{Qudsia J. Ejaz${}^{\,1}$, 
Thomas Faulkner${}^{\,1}$, Hong Liu${}^{\,1}$,
Krishna Rajagopal${}^{\,1}$ and Urs Achim Wiedemann${}^{\,2}$
\\
\vspace{0.1in}

${}^{\,1}$ Center for Theoretical Physics, MIT,
Cambridge, MA 02139, USA \vspace{0.1in}

${}^{\,2}$Department of Physics, CERN, Theory Division, CH-1211
Geneva 23 \vspace{0.1in}

E-mail addresses: {\tt ejazqj@mit.edu, thomasf@mit.edu, hong\_liu@mit.edu, krishna@ctp.mit.edu,
Urs.Wiedemann@cern.ch} }

\abstract{We study 
the dispersion relations of 
mesons in a particular hot strongly
coupled supersymmetric gauge theory plasma.  
We find that at large momentum $k$ the dispersion relations
become 
$\omega \simeq v_0 k + a + b/k+ \ldots$, where the limiting
velocity $v_0$ 
is the same for mesons with any quantum
numbers and depends only on the ratio of the temperature
to the quark mass $T/m_q$. 
We compute $a$ and $b$ in
terms of the meson quantum numbers and $T/m_q$. 
The limiting
meson velocity $v_0$ becomes much smaller than
the speed of light at temperatures below but close to $T_{\rm diss}$, 
the temperature above
which no meson bound states at rest in the plasma are found.
From our result for $v_0(T/m_q)$,
we find that the temperature above which no meson
bound states with velocity $v$ exist is 
$T_{\rm diss}(v) \simeq (1-v^2)^{1/4} T_{\rm diss}$, up to
few percent corrections.
We thus confirm by direct calculation of meson dispersion
relations a result inferred indirectly  
in previous work via analysis of
the screening length between a static quark and antiquark
in a moving plasma.  
Although we do not do our calculations in QCD,
we argue
that the qualitative features of the dispersion
relation we compute, including in particular the relation
between dissociation temperature and meson velocity, may apply
to bottomonium and charmonium mesons propagating in the strongly
coupled plasma of QCD.  We discuss
how our results can contribute to understanding 
quarkonium physics in heavy ion collisions.
}

\preprint{MIT-CTP-3912\\ CERN-PH-TH/2007-232\\ CAS-KITPC/ITP-024}

\numberwithin{equation}{section}
\begin{document}


\section{Introduction}
The radii of the tightly bound heavy quark-antiquark systems of the charmonium 
($J/\Psi$, $\Psi'$, $\chi_c$, ...) and bottomonium ($\Upsilon$, $\Upsilon'$, ...) families 
provide a unique set of decreasing length scales in strong interaction physics.
On general grounds, it is expected that the attraction between a heavy quark and an
anti-quark is sensitive to the medium in which the bound state is embedded, and that
this attraction weakens with increasing temperature. In the context of ultra-relativistic 
nucleus-nucleus collisions, the radii of 
some quarkonia states correspond to fractions of the natural length 
scale displayed by the medium produced in heavy ion collisions, namely fractions of its
inverse temperature $1/T$.   Such scale considerations support the 
idea that measurements of the medium-modification or dissociation of quarkonia
can characterize properties of the QCD matter produced in heavy ion collisions. 

Matsui and Satz were the first to highlight the role of quarkonium in the study of 
hot QCD matter~\cite{Matsui:1986dk}. They suggested that $J/\Psi$-suppression is a signature for the
formation of deconfined quark-gluon plasma (QGP). More precisely, they argued 
that in comparison to proton-proton or proton-nucleus collisions, the production 
of $J/\Psi$ mesons should be suppressed if quark-gluon plasma is formed in 
sufficiently energetic nucleus-nucleus collisions, since the screened interaction of a $c$ 
and a $\bar{c}$ in QGP would not bind them~\cite{Matsui:1986dk}. The theoretical 
basis for this argument has been clarified considerably within the last two 
decades~\cite{Satz:2005hx}. 
Model-independent calculations of the static potential between a heavy quark and anti-quark 
have been performed in lattice-regularized QCD, valid at strong 
coupling~\cite{Kuti:1980gh,Kaczmarek:2004gv,Petreczky:2004pz,Kaczmarek:2005ui,Maezawa:2006rd,Karsch:2006sf}. 
In lattice calculations without dynamical quarks, at temperature $T=0$ and large
separation $L$ this potential rises linearly with $L$,
consistent with confinement. At nonzero temperature, the potential
weakens and levels off at large distances; with increasing temperature,
the distance at which this screening occurs decreases. This behavior of
the static potential has been mapped out for hot QCD matter both 
without~\cite{Kaczmarek:2004gv} and with~\cite{Petreczky:2004pz,Kaczmarek:2005ui}
dynamical quarks. However, the physical interpretation of static potentials at finite
temperature rests on additional assumptions. For instance, even if a potential 
supports a bound state with several MeV binding energy, it remains unclear which physics
can be attributed to such a state in a heat bath of $\sim 200$ MeV temperature.
Such issues do not arise in a discussion of quarkonium mesons based directly
on their Minkowski space spectral functions or dispersion relations.
In recent years,  the spectral functions have been characterized by
lattice calculations of the Euclidean correlation functions to which they
are analytically related,
again in hot QCD matter both without~\cite{Asakawa:2000tr} and with~\cite{Morrin:2005zq}
dynamical quarks.  The use of these calculations of finitely many points
on a Euclidean correlator to constrain the Minkowski space spectral
function of interest via the Maximum Entropy Method
requires further inputs --- for example smoothness
assumptions or information on the analytic properties of the spectral
function~\cite{Asakawa:2000tr,Morrin:2005zq}.
At high enough temperatures that quark-gluon plasma becomes
weakly coupled, a complementary analytical approach based upon
resummed hard-thermal-loop perturbation theory becomes
available~\cite{Laine:2006ns}. These calculations have the
advantage that analytical continuation from Euclidean to Minkowski
space does not introduce additional uncertainties, but 
it remains unclear to what extent they can treat a strongly
coupled quark-gluon plasma.
In broad terms, all these calculations support the qualitative picture
behind the original suggestion of Matsui and 
Satz that color screening in the quark-gluon plasma is an efficient 
mechanism for quarkonium dissociation. In addition, these studies 
support the picture of a sequential dissociation 
pattern~\cite{Karsch:2005nk}, 
in which loosely bound, large,
quarkonia such as  the $\Psi'$ and $\chi_c$ 
cease to exist 
close to $T_c$, the temperature of the crossover
between hadronic matter and quark-gluon plasma,  
whereas more tightly bound, smaller,
states dissociate only at significantly higher temperatures. In particular,
$J/\Psi$ mesons continue to exist for a range of temperatures 
above the QCD phase transition and dissociate only above
a temperature that lies between $1.5\, T_c$ and $2.5\, T_c$~\cite{Karsch:2005nk}. 
The observation of bound-state-specific quarkonia
suppression patterns could thus provide 
detailed information about the temperature
attained in heavy ion collisions.

On the experimental side, there are by now data from the NA50 and NA60 experiments
at the CERN SPS and from the PHENIX experiment at RHIC demonstrating that the
production of $J/\Psi$ mesons is suppressed in ultra-relativistic nucleus-nucleus collisions 
compared to proton-proton or proton-nucleus collisions at the same center of mass 
energy~\cite{ExperimentalRefs}.
However, due to lack of statistics and resolution, an experimental characterization of
other charmonium states ($\Psi'$, $\chi_c$, ...) has not yet been 
possible at RHIC, and 
bottomonium states have not yet 
been characterized in any nucleus-nucleus collisions.
Moreover, the observed yield of $J/\Psi$ mesons is expected to receive 
significant decay contributions from $\Psi'$ and $\chi_c$, meaning that the
observed suppression of $J/\Psi$ mesons may originate only in
the suppression of the larger  $\Psi'$ and $\chi_c$ states~\cite{Karsch:2005nk},
or may indicate a suppression in the number of 
primary $J/\Psi$ mesons themselves in addition.
Thus, at present an experimental test of the 
sequential quarkonium suppression pattern is not in hand.
It is expected that the
LHC heavy ion program will furnish such a test, since two LHC 
experiments~\cite{Carminati:2004fp}
have demonstrated capabilities for
discriminating between the different states of the charmonium and bottomonium families.

From the existing data in ultra-relativistic heavy ion collisions and their phenomenological
interpretation, it has become clear that an unambiguous characterization of color screening 
effects in the quarkonium systems requires good experimental and theoretical control of 
several confounding factors. These include in particular control over the spatio-temporal 
evolution of the medium, control over the time scale and mechanism of quarkonium formation, 
as well as control over the effects of quarkonium propagation through the medium. We now
comment on these three sources of uncertainty in more detail: 

First, there is ample evidence
by now that the systems produced in ultra-relativistic heavy ion collisions display 
effects of position-momentum correlated motion (a.k.a. flow), which are as important as 
the effects of random thermal motion~\cite{RHICwhitepapers}. Moreover, the 
energy density achieved in these collisions drops rapidly with time
as the matter expands and falls apart after approximately 10 fm/c.
As a consequence, the
modeling of quarkonium formation in heavy ion collisions cannot be limited to a 
description of heavy quark bound states in a heat bath at constant temperature (which is
the information accessible in {\it ab initio} calculations in lattice-regularized QCD).
The effects of a rapid dynamical evolution during which the relevant
degrees of freedom in the medium change from partonic to hadronic must be
taken into account. 

Second, regarding the formation process, the conversion of a
heavy quark pair produced in a hard collision  into a bound
quarkonium state is not 
fully understood, even in the absence of a medium. There are different
production models, which all have known limitations and for which a systematic calculation
scheme remains to be fully established
(for a short review of these issues, see~\cite{Qiu:2006df}). The need for 
further clarification 
of the 
vacuum  case has even led to suggestions that nuclear matter could serve as a filter
to distinguish between different production mechanisms~\cite{Qiu:1998rz,Kharzeev:2005zr}.
However, it has also been pointed out that there may be a novel
quarkonium production mechanism operating only in 
ultra-relativistic heavy ion collisions at RHIC and at the LHC~\cite{Thews:2000rj}:
charm quarks may be so abundant in these collisions that 
$c$ and $\bar{c}$ quarks produced separately in different primary 
hard scattering
interactions may find each other and combine,
contributing
significantly to charmonium production at soft and intermediate 
transverse momentum.  To a lesser extent, this mechanism 
may also contribute to the production of Upsilon mesons. Identifying and characterizing such
a novel formation process is of considerable interest, since recombination is likely to be 
quadratically sensitive to the 
phase space density of charm and thus to properties of the
produced matter. On the other hand, if realized in nature recombination also implies that
quarkonium spectra at soft and intermediate transverse momenta 
are determined
predominantly
during the late hadronization stage and cannot be viewed as probes
which test color screening in the quark gluon plasma. This would indicate that the
high transverse momentum regime (say above 5-8 GeV) of quarkonium spectra,
which should not be significantly affected by recombination, is better suited for
tests of the fundamental color screening effects predicted by QCD.
However, the sensitivity of 
high transverse momentum spectra to properties of the medium remains to be established. 
In particular, quarkonium formation or dissociation proceeds on a time scale
comparable to the size of the bound state in its rest frame, meaning that 
quark-antiquark pairs with very high transverse velocity may escape the
finite-sized droplet of hot matter produced in a heavy ion collision
before they have time to form a meson, meaning 
in turn that screening 
effects cease to play a role in quarkonium production above some 
very high transverse momentum~\cite{Karsch:1987zw}.
At lower transverse momenta, where screening
does play a role, one must nevertheless understand for how
long quarkonium is exposed to the medium and how readily it dissociates if moving relative to 
that medium. For quarkonium at high transverse momentum, the time of exposure to the medium
depends on the geometry of the collision region, which determines
the in-medium path length, and it depends on the propagation velocity. The results contained
in this paper give novel input to modeling this process by demonstrating 
that the real part of the
finite temperature quarkonium dispersion relation can differ significantly from the vacuum
one, and can imply a limiting quarkonium propagation velocity which 
is much smaller than $c$, the
velocity of light in vacuum.  Our results indicate that at temperatures 
close to but below that at which a given quarkonium state dissociates,
these mesons move through a strongly
coupled quark-gluon plasma at a velocity that is much smaller than $c$ even if
they have arbitrarily high transverse momentum.  Certainly this
means that the formation time arguments of~\cite{Karsch:1987zw}
will need rethinking before they can be applied quantitatively. 

Third, we turn to the question of how the relative motion of quarkonium with respect
to the local rest frame of the medium affects quarkonium production. As discussed above, 
the standard vacuum relation between the momentum of a quarkonium state 
and its velocity can be altered in the presence of a medium and this effect may be  
phenomenologically relevant.  In addition, it is 
expected that a finite relative velocity
between the medium and the bound state enhances the 
probability of dissociation~\cite{Chu:1988wh}. 
In a recent strong coupling calculation of hot ${\cal N}=4$ supersymmetric QCD,
three of us have
have shown~\cite{Liu:2006nn,Liu:2006he} 
that the sceening length $L_s$ for a heavy quark-antiquark 
pair decreases with increasing velocity as $L_s(v,T) \sim L_s(0,T)/\sqrt{\gamma}$,
with $\gamma=1/\sqrt{1-v^2}$ the Lorentz boost factor.  This suggests 
that a quarkonium state that is bound at $v=0$ at a given temperature could
dissociate above some transverse momentum due to the increased screening,
providing a significant 
additional source of quarkonium suppression at finite transverse momentum. 
The present work
started from the motivation to establish how this velocity scaling manifests itself 
in a description of mesons at finite temperature, rather than via drawing inferences
from a calculation of the screening length that characterizes the quark-antiquark
potential.  This motivation is analogous to that behind
going from lattice QCD calculations
of the static potential in QCD to calculations of the Minkowski space
meson spectral function.  We shall do our calculation in a different strongly
coupled gauge theory plasma, in which we are able to do this investigation
for mesons with nonzero velocity.
We shall see that the critical velocity
for the dissociation of quarkonium inferred from the velocity scaling of the screening length
also appears as a limiting 
velocity for high-momentum quarkonium propagation in the hot 
non-abelian plasma.   

Finally, the characterization of color screening also depends on the
experimental and theoretical ability to separate its effects on quarkonium
production from effects arising during
the late time hadronic phase of the heavy ion
collision.
In particular, it has been noted early on that significant charmonium
suppression may also
occur in confined hadronic matter~\cite{Brodsky:1988xz}. However, it has been argued on
the basis of model estimates for the hadronic $J/\Psi$ dissociation cross
section~\cite{Kharzeev:1994pz}
that dissociation in a hadronic heat bath is much less efficient than in a
partonic one.
The operational procedure for separating such hadronic phase effects is to
measure
them separately in proton-nucleus collisions~\cite{Kharzeev:1995id}, and to establish
then to what
extent the number of $J/\Psi$ mesons produced in
nucleus-nucleus collisions drops below the yield
extrapolated from
proton-nucleus collisions \cite{ExperimentalRefs,Adler:2005ph}.

The above discussion highlights the extent to which an understanding of quarkonium
production in heavy ion collisions relies on theoretical modelling as the bridge between
experimental observations and the underlying properties of hot QCD matter. This task
involves multiple steps. It is of obvious interest to validate or constrain by first
principle calculations as many steps as possible, even in a simplified theoretical
setting. The present work is one of a number of recent 
developments~\cite{RecentDevelopments}
that explore to what extent
techniques from string theory, in particular the AdS/CFT correspondence, can 
contribute to understanding processes in hot QCD by specifying how these processes
manifest themselves in a large class of hot strongly coupled
non-abelian gauge theories. Although
it is not known how to extend the AdS/CFT correspondence to QCD, there are 
several motivations for turning to this technique. First, there are a growing number of
explicit examples which indicate that a large class of
thermal non-abelian field theories with gravity duals share commonalities 
such that
their properties in the thermal sector are either universal
at strong coupling, i.e. independent of the 
microscopic dynamics 
encoded in the particular quantum field theory under study,
or their properties are related to each other by simple scaling laws e.g. depending
on the number of elementary degrees of freedom. This supports
the working hypothesis that by learning something about a large class of 
strongly coupled thermal
non-abelian quantum field theories, one can gain guidance towards understanding
the thermal sector of QCD. 
Second, the AdS/CFT 
correspondence allows for a technically rather simple formulation of problems involving
real-time dynamics. This is very difficult in finite temperature lattice-regularized 
calculations, which exploit the imaginary time formalism. In particular, this is the reason
why so far lattice QCD calculations treat only static quark-antiquark pairs in the plasma,
and why the only nonperturbative calculation of the velocity dependence of quarkonium
dissociation exploits the AdS/CFT correspondence. 
Third,  data from experiments at RHIC 
pertaining to many aspects of the matter produced in heavy ion collisions
indicate that this matter is strongly coupled.
Since the AdS/CFT correspondence provides a
mapping of difficult nonperturbative calculations 
in a quantum field theory with strong coupling onto
relatively simple, semi-classical calculations in a gravity dual, it 
constitutes a novel --- and often
the only --- technique for addressing dynamical questions about
hot strongly coupled non-abelian matter, questions that are being
raised directly by experimental results on QCD matter coming from
RHIC. 

We have focussed in this Section on the larger
context for our results. In Section 2,  which is an
introduction in a more narrow sense, we review the past
results which serve as an immediate motivation for our
work, in particular the screening length that characterizes
the potential between a static quark and antiquark in a moving
plasma wind.  
Adding fundamental quarks with finite mass $m_q$,
and hence mesons, into ${\cal N}=4$ SYM theory
requires adding a D7-brane in the dual gravity theory, as
we review in Section 3.  The fluctuations of the D7-brane
are the mesons, as we review for the case of zero temperature in Section 3.
In Section 4 we set up the analysis of the mesons at nonzero
temperature, casting the action for the D7-brane fluctuations
in a particularly geometric form, written entirely in terms
of curvature invariants. Parts of the derivation are explained
in more detail in Appendix A.  With all the groundwork in place,
in Section 5 we derive the meson dispersion relations. In addition
to obtaining them numerically without taking any limits
as has been done previously~\cite{Mateos:2007vn}, we
are able to calculate them analytically in three limits:
first, upon taking the low temperature limit at fixed $k$;
second, upon taking the low temperature limit at fixed $kT$;
and third, using insights from the first
two calculations, at large $k$ for any temperature.
At large $k$ we find
\be
\omega = v_0 k + a + \frac{b}{k} + \ldots
\ee
where $v_0$ is independent of meson quantum numbers, depending
only on $T/m_q$. $v_0$ turns out to be given by the local speed
of light at the ``tip of the D7-brane'', namely the place in
the higher dimensional gravity dual theory where the D7-brane
comes closest to the black hole~\cite{Mateos:2007vn}.
We compute $a$ and $b$ in terms of meson
quantum numbers and $T/m_q$. Our result for the limiting
velocity $v_0$ for mesons at a given temperature $T$ can
be inverted, obtaining $T_{\rm diss}(v)$, the temperature
above which no mesons with velocity $v$ are found.  We find
that up to few percent corrections, our result can
be summarized by 
\be
T_{\rm diss}(v)=(1-v^2)^{1/4}T_{\rm diss}\ ,
\ee
where $T_{\rm diss}$ is the temperature at which zero-velocity
mesons dissociate, obtained in previous work and introduced in
Section 3.  As we discuss in Section 2, our results 
obtained by direct calculation of meson dispersion
relations confirm inferences reached (in two different ways)
from the analysis of the screened potential between
a static quark and antiquark in a hot plasma wind.
In Section 6, we close with a discussion of potential
implications of these dispersion relations for quarkonia
in QCD as well as a look at open questions.   The dispersion
relations that we calculate in this paper
describe how mesons propagate and so affect a class of observables,
but determining whether quarkonium meson
formation from a precursor quark-antiquark pair is 
suppressed by screening is a more dynamical question that can
at present be addressed only by combining our calculation
and the more heuristic results of \cite{Liu:2006he}.

\section{From screening in a hot wind to moving mesons}

In the present work, we shall
use the AdS/CFT correspondence to study the propagation of
mesonic excitations moving through
a strongly coupled hot quark-gluon plasma. In this Section, however,
we introduce what we have learned from the simpler calculation
of the potential between a test quark-antiquark pair moving through
such a medium.  This will allow us to pose the questions
that we shall address in the present paper.

The simplest example of the AdS/CFT correspondence is provided by the duality between
${\cal N}=4$ super Yang-Mills (SYM) theory and classical gravity in 
$AdS_5\times S_5$~\cite{AdS/CFT}. ${\cal N}=4$
super Yang-Mills (SYM) theory is a conformally invariant theory
with two parameters: the rank of the gauge group $N_c$ and the 't
Hooft coupling $\lambda = g_{\rm YM}^2 N_c$. In the large $N_c$
and large $\lam$ limit, gauge theory problems can be solved using
classical gravity in $AdS_5 \times S_5$ geometry. We shall work
in this limit throughout this paper.

In $\NN=4$ SYM theory at zero temperature, the static potential
between a heavy external quark and antiquark separated by
a distance $L$ is given in the large $N_c$ and large
$\lam$ limit by~\cite{Rey:1998ik,Maldacena:1998im}
 \be \label{zepo}
 V (L) = -\frac{4\pi^2}{\Gamma(\frac{1}{4})^4}\frac{\sqrt{\lam}}{L}\, ,
 \ee
where the $1/L$ behavior is required
by conformal invariance.  
This potential is obtained by computing
the action of an extremal string world sheet, bounded at 
$r\rightarrow\infty$ ($r$ being the fifth dimension of $AdS_5$)
by the world lines of the quark and antiquark and ``hanging down''
from these world lines toward smaller $r$.
At nonzero temperature, the potential
becomes~\cite{Rey:1998bq}
 \bea \label{Awe}
 V(L,T) \approx & \sqrt{\lam} f (L) \qquad & L < L_c \nonumber\\
         \approx & \lam^0 g(L) \qquad & L > L_c\ .
 \eea
In (\ref{Awe}), at $L_c = 0.24/T$  there is a change
of dominance between different saddle points and
the slope of the potential changes discontinuously. 
When $L<L_c$, the potential is determined as at 
zero temperature by the area
of a string world sheet bounded by the worldlines
of the quark and antiquark, but now the world sheet hangs
down into a different five-dimensional spacetime: 
introducing nonzero temperature
in the gauge theory is dual to introducing a black hole
horizon in the five-dimensional spacetime. 
When $L \ll L_c $,
$f(L)$ reduces to its zero temperature behavior (\ref{zepo}). 
When $L \gg L_c$, $g(L)$ has the behavior~\cite{Bak:2007fk}
  \be
  g(L) \propto c_1 - c_2 e^{- m_{\rm gap} L}\ ,
 \ee
with $c_1$, $c_2$ and $m_{\rm gap}$ constants all
of which are proportional to $T$.  This large-$L$ potential 
arises from two disjoint strings, each separately
extending downward from the quark or antiquark all the
way to the black hole horizon, 
exchanging supergravity
modes the lightest
of which has a mass given by $m_{\rm gap} = 2.34\, \pi T$. 
(There are somewhat lighter modes with nonzero R-charge, but these
are not relevant here~\cite{Amado:2007pv}.)
It is
physically intuitive to interpret $L_c$ as the screening length
$L_s$ of the plasma since at $L_c$ the qualitative behavior of the
potential changes.  Similar criteria are used in the
definition of screening length in QCD~\cite{Karsch:2006sf}, although
in QCD there is no sharply defined length scale
at which screening sets in.   Lattice calculations of the 
static potential between a heavy quark
and antiquark in QCD indicate a screening length $L_s \sim 0.5/T$ in hot
QCD with two flavors of light quarks~\cite{Kaczmarek:2005ui} and
$L_s\sim 0.7/T$ in hot QCD with no dynamical
quarks~\cite{Kaczmarek:2004gv}.   The fact that there {\it is} a sharply
defined $L_c$ in (\ref{Awe}) is an artifact of the limit
in which we are working.\footnote{The theoretical advantage of using
$1/m_{\rm gap}$ to define a screening length 
as advocated in~\cite{Bak:2007fk} is
that it can be precisely defined in  ${\cal N}=4$ SYM theory
at finite $\lambda$ and $N_c$, as well as in QCD,
as it characterizes the behavior
of the static potential in the $L\rightarrow\infty$ limit.  The disadvantage
of this proposal from a phenomenological point of view is that
quarkonia are not sensitive to the potential at distances much
larger than their size.  
For questions relevant to the
stability of bound states,  therefore, the 
length scale determined
by the static potential that is phenomenologically most important
is that at which the potential flattens.  Although this length
is not defined sharply in QCD, it is apparent in lattice 
calculations and can be defined operationally 
for practical purposes~\cite{Kaczmarek:2004gv,Kaczmarek:2005ui}.
This $L_s$
seems most analogous to $L_c$ in (\ref{Awe}), and we shall therefore
continue to
refer to $L_s\equiv L_c$
as the screening length, as in the original literature~\cite{Rey:1998bq}.
Note that $L_c$ is larger than $1/m_{\rm gap}$ by a purely numerical
factor $\simeq 1.8$.}

In~\cite{Liu:2006nn,Liu:2006he}, three of us studied the velocity
scaling of the screening length $L_s$
 in $\NN=4$ super-Yang-Mills theory and found 
that\footnote{In~\cite{Liu:2006nn,Liu:2006he}
 $L_s$ was defined using a slightly different 
quantity than $L_c$ in (\ref{Awe}), such that 
$L_s=0.28/T$ for a quark-antiquark at rest.  For
technical reasons, this other definition was more
easily generalizable to nonzero velocity.  
}
\be
L_s(v,\theta,T) = \frac{f(v,\theta)}{\pi T}\left(1-v^2\right)^{1/4}\, ,
\label{lmaxwithf}
\ee
where $\th$ is the angle between the orientation of the
quark-antiquark dipole and the velocity of the moving thermal
medium in the rest frame of the dipole.  $f (v, \th)$ is only weakly
dependent on both of its arguments. That is, it is close to
constant. So, to a good approximation we can write
\be
 L_{s} (v,T) \approx L_s (0,T) (1-v^2)^{1/4} \propto \frac{1}{T}(1-v^2)^{1/4}\ .
 \label{lmax}
 \ee
This result, also obtained in~\cite{Peeters:2006iu} 
and further
explored in~\cite{Avramis:2006em,Caceres:2006ta,Natsuume:2007vc}, 
has proved robust
in the sense that it applies in various strongly
coupled plasmas other than 
${\cal N}=4$ SYM~\cite{Avramis:2006em,Caceres:2006ta,Natsuume:2007vc}.
The velocity dependence of the screening
length (\ref{lmax}) suggests that in
 a theory containing dynamical heavy quarks and meson
bound states (which $\NN=4$ SYM does not) 
the dissociation 
temperature $T_{\rm diss}(v)$, defined as the temperature above which mesons with
a given velocity do not exist,
should scale with velocity as~\cite{Liu:2006nn}
 \be \label{rro}
 T_{\rm diss} (v) \sim  T_{\rm diss} (v=0) (1-v^2)^{1/4} \ ,
   \ee
since $T_{\rm diss}(v)$ should be the temperature at which
the screening length $L_s(v)$ is comparable to the size of
the meson bound state.
The scaling (\ref{rro}) then
indicates that slower mesons can exist up to
higher temperatures than faster ones.
In this paper, we shall replace the inference that
takes us from the calculated result (\ref{lmax}) to the conclusion (\ref{rro})
by a calculation of the properties of mesons themselves, specifically
their dispersion relations.
We shall reproduce (\ref{rro}) in this more nuanced setting, finding
few percent corrections to the basic scaling result inferred previously. 

The results (\ref{lmax}) and (\ref{rro}) have a simple physical
interpretation which suggests that they could be applicable to a wide
class of theories regardless of specific details.
First, note that since $L_s (0) \sim {1 \ov T}$,  both (\ref{lmax})
and (\ref{rro}) can be interpreted as if in their rest frame the
quark-antiquark dipole experiences a higher effective temperature 
$T\sqrt{\ga}$.  Although this is not literally the case in
a weakly coupled theory, in which the dipole will see
a redshifted momentum distribution
of quasiparticles coming at it from some directions and
a blueshifted distribution from others~\cite{Chu:1988wh}, we
give an argument below for how
this interpretation can nevertheless
be sensible. The result (\ref{lmax})
can then be seen as validating the relevance of this interpretation
in a strongly coupled plasma.  
The argument is based on the idea that quarkonium propagation and
dissociation
are mainly sensitive to the local energy density of the medium. Now, in the
rest frame of the dipole, 
the energy density (which we shall denote by $\rho$) is
blue shifted by a factor $\sim\ga^2$ and since 
$\rho \propto T^4$ in a conformal theory, the result (\ref{lmax}) is
as if quarks feel a higher effective temperature given
by $T\sqrt{\ga}$.\footnote{Applying a Lorentz boost  to $\rho$
yields $\ga^2 (1+\frac{1}{3}v^2)\rho$.  Including the
$(1+\frac{1}{3}v^2)$ factor
makes this argument reproduce the 
result (\ref{lmaxwithf}), including
the weak velocity dependence in the
function $f$, more quantitatively than merely tracking
the powers of $\gamma$.}
Lattice calculations indicate that the quark-gluon
plasma in QCD is nearly conformal over a range of
temperatures $1.5 T_c < T \lesssim 5 T_c$,
with an energy
density $\rho \approx b T^4$ where $b$ is a constant about $80\%$
of the free theory 
value~\cite{Karsch:2000ps}. So it does not seem far-fetched to
imagine that (\ref{lmax}) could apply to QCD. 
We should also
note that AdS/CFT calculations in other strongly coupled gauge
theories with a gravity description are consistent with the
interpretation above~\cite{Caceres:2006ta} and that
for near conformal theories the deviation from
conformal theory behavior appears to be small~\cite{Caceres:2006ta}.  
If a velocity scaling like (\ref{lmax}) and (\ref{rro}) holds for
QCD, it can potentially have important implications for quarkonium
suppression in heavy ion collisions, as we have discussed in
Section 1 and will return to in Section~6.

While the argument leading from (\ref{lmax}) to (\ref{rro}) is
plausible, it is more satisfying to have a set-up within which one can
study mesons directly. 
Direct study of meson bound states will also 
yield more insights than the
study of the screening length from
the potential. It is the purpose of this paper to examine this issue
in a specific model with dynamical flavors. 

Before beginning
our analysis, let us first note a curious feature regarding the quark
potential observed in~\cite{Liu:2006nn,Liu:2006he}. There one introduces a
probe brane near the boundary of the AdS$_5$ black hole geometry
with open strings ending on it corresponding to fundamental ``test quarks'' of mass
$m_q \gg \sqrt{\lam} T$. It was found that for any given quark
mass $m_q$, there exists a maximal velocity $v_c$ given by
 \be \label{oep}
 v_c^2 = 1 - {\lam^2 T^4 \ov 16 m_q^4}\ ,
 \ee
beyond which there is no ${\cal O}(\sqrt{\lambda})$ potential
between the pair for any value of their separation larger than
their Compton wavelength, i.e. for any distance at which a potential
can be defined.
This result can be interpreted as saying that
for any given $T$ and $m_q$,
it is impossible to obtain bound states beyond (\ref{oep}), i.e.
as indicating that
there is a velocity bound (a ``speed limit'') for the mesons. One can also turn
(\ref{oep}) around and infer that for any 
large $m_q$ and $v$ close to 1,
the dissociation temperature is given by
 \be \label{PWw}
 T_{\rm diss} = {2 m_q \ov \sqrt{\lam}} (1-v^2)^{1 \ov 4}\ ,
 \ee
which is consistent with (\ref{rro}). Note that the above argument
is at best heuristic since  $\NN=4$ SYM itself does not contain
dynamical quarks and thus genuine mesons do not exist. 
In the present paper, however, we shall see by deriving
them from meson dispersion relations that
(\ref{oep}) and (\ref{PWw}) are precisely correct in
the limit of large quark mass once we introduce fundamentals, 
and hence mesons, into the theory.  We shall also find that the
more dynamical, albeit heuristic, interpretation of (\ref{oep}) as
a velocity beyond which a quark and antiquark do not feel a potential
that can bind them remains of value.

\section{D3/D7-brane construction of mesons}

In this Section we review the gravity dual description of 
strongly coupled $\NN=4$ SYM theory with gauge group $SU(N)$
coupled to $N_f \ll N$ $\NN=2$ hypermultiplets in the fundamental
representation of $SU(N)$, introduced in~\cite{Karch:2002sh} and
studied in~\cite{Kruczenski:2003be,Babington:2003vm,Kruczenski:2003uq,Hong:2003jm,Kirsch:2004km,Kirsch:2005uy,Myers:2006qr,Mateos:2006nu,Hoyos:2006gb,Mateos:2007vn,Myers:2007we,Peeters:2007ti,Erdmenger:2007cm}.
We will first describe the theory at
zero temperature and then turn to nonzero temperature.
We will work
in the limit $N \to \infty$, $\lam = g_{YM}^2 N \to \infty$ and
$N_f$ finite (in fact $N_f =1$). In the deconfined 
strongly coupled plasma
that this theory describes, heavy
quark mesons exist below a dissociation temperature 
that, for mesons at rest, is given by 
$T_{\rm diss} = 2.166 \,m_q/\sqrt{\lam}$~\cite{Babington:2003vm,Mateos:2006nu,Hoyos:2006gb,Mateos:2007vn}.
In Section 5 we shall calculate the dispersion relations
for these mesons, namely the meson spectrum at nonzero
momentum $k$ and in so doing determine
$T_{diss} (v)$ directly, rather than by inference
as described in Section 2.

\subsection{Zero temperature}

Consider a stack of $N$ coincident D3-branes and $N_f$ coincident
D7-branes in 9+1-dimensional
Minkowski space, which we represent by the array
 \be
\begin{array}{rccccccccccl}
\mbox{D3:}\,\,\, & 0 & 1 & 2 & 3 & \cdot & \cdot & \cdot 
& \cdot & \cdot & \cdot & \, \\
\mbox{D7:}\,\,\, & 0 & 1 & 2 & 3 & 4 & 5 & 6 & 7 & \cdot & \cdot & \,
\label{array}
 \end{array}
 \ee
which denotes in which of the 9+1 dimensions the D3- and D7-branes
are extended, and in which they occupy only points.
The D3-branes sit at the origin of the 89-plane,
with $L$ denoting the distance between the D3- and the
D7-branes in the 89-plane. Without loss of generality (due to
rotational symmetry in 89-plane), we can take the D7-branes to be at
$x_8 = L, \; x_9 =0$. This is a stable configuration and preserves
one quarter of the total number of supersymmetries, meaning that it describes
an ${\cal N}=2$ supersymmetric gauge theory as
we now sketch~\cite{Karch:2002sh}.

The open string sector of the system contains 3-3 strings, 
both of whose ends lie on one of the $N$
D3-branes, 7-7 strings ending on $N_f$ D7-branes, 
and 3-7 and 7-3 strings
stretching between D3- and D7-branes. In the low energy limit
 \be \label{mals}
 \apr \to 0, 
  \qquad {L \ov 2 \pi \apr} = {\rm finite}, 
 \ee
all the stringy modes decouple except for: (i) the lightest modes of
the 3-3 strings, which give rise to an $SU(N)$ $\NN=4$ SYM theory
in 3+1-dimensional Minkowski space; (ii) the lightest modes of  the 
3-7 and 7-3 strings, which give rise to
$N_f$ hypermultiplets in the ${\cal N}=2$
gauge theory transforming under the fundamental
representation of $SU(N)$. The whole theory thus has $\NN=2$
supersymmetry. We will call $N_f$ hypermultiplets quarks below
even though they contain both fermions and bosons. The mass of the
quarks is given by
 \be \label{qma}
 m_q = {L \ov 2\pi \apr} \ ,
 \ee
where $1/(2\pi\apr)$ is the tension of the strings.

In the limit
 \be
N \to \infty, \qquad N_f = {\rm finite}, \qquad \lam = g_{YM}^2 N
\gg 1\ ,
 \ee
the above gauge theory has a gravity
description~\cite{Karch:2002sh} in terms of D7-branes in the
near-horizon geometry of the D3-branes, which is $AdS_{5} \times
S_{5}$ with a metric
 \begin{eqnarray}
ds^2  &= &  {r^2 \ov R^2} \le(-  dt^2 + dx_1^2 + dx_2^2 + dx_3^2
\ri)+ \frac{R^2 }{ r^2} dr^2 + R^2 d \Om_5^2 \nonumber 
 \\
 & = & {r^2 \ov R^2} \le(-  dt^2 + dx_1^2 + dx_2^2 + dx_3^2
\ri)+ \frac{R^2 }{ r^2} \sum_{i=4}^9 dx_i^2\ ,
 \label{OneF}
 \end{eqnarray}
where $r^2 = \sum_{i=4}^9 x_i^2$ and $d\Omega_5^2$ is the metric
on a 5-sphere. $R$ is the curvature radius of
AdS and is related to the Yang-Mills theory 't~Hooft coupling by
 \be\label{aprlamrelation}
 {R^2 \ov \apr} = \sqrt{\lam} \ .
 \ee
The string coupling constant $g_s$ is related to
the gauge theory parameters by 
\be\label{gstring}
4\pi g_s=g^2_{YM}=\frac{\lambda}{N}\ ,
\ee
where $g^2_{YM}$ is defined according to standard field
theory conventions and is twice as 
large as the Yang-Mills gauge coupling
defined according to standard 
string theory conventions. 
In this zero temperature setting, 
the embedding of the D7-branes in the $AdS_{5} \times S_{5}$ geometry
(\ref{OneF}) can be read directly from (\ref{array}). The
D7-brane worldvolumes  fill the $(t, x_i)$ 
coordinates, with $i = 1, \ldots , 7$, 
and are located at the point 
$x_8 = L, \; x_9 =0$ in the 89-plane. Since $N_f$ remains
finite in the large $N$ limit, the gravitational back-reaction of
the D7-branes on the spacetime of the D3-branes (\ref{OneF}) 
may be neglected.

The
dictionary between 
the gauge theory and its dual gravity description 
can thus be summarized as
follows. On the gauge theory side we have two sectors: excitations
involving adjoint degrees of freedom only and excitations
involving the fundamentals. The first type of excitations
correspond to closed
strings in $AdS_5 \times S_5$ as in the standard AdS/CFT story.
The second type is described by open strings ending on the
D7-branes\footnote{We will not consider baryons in this paper.}.
In particular, the low-lying (in a sense that we shall define later)
meson spectrum of the gauge theory can
be described by fluctuations of $x_{8,9}$ and 
gauge fields on D7-branes. We shall focus on
the fluctuations of $x_{8,9}$ on the D7-brane (equivalently,
the fluctuations of the position of the D7-brane in the $(x_8,x_9)$
plane) which describe scalar mesons.  There are also gauge fields
localized within the D7-brane, and their fluctuations describe
vector mesons.  The description of the vector mesons is
expected to be similar to that of the scalar mesons.
We shall limit our presentation entirely to the scalar mesons.
We shall take $N_f=1$,  
meaning that
the gauge theory is specified
by the parameters $N$, $\lambda$ and $m_q$ which
are related to their counterparts in the dual
gravity theory by (\ref{aprlamrelation}), (\ref{gstring})
and (\ref{qma}).  We see that the $N\rightarrow \infty$ 
limit corresponds to $g_s\rightarrow 0$, making the string 
theory weakly coupled.  
Considering the theory with the parameter $\lambda$
taken to $\infty$ corresponds to taking the string tension
to infinity.  These limits justify the use of the classical
gravity approximation in which we
consider strings moving in a background spacetime.

For later generalization to finite temperature, it is convenient
to describe the D7-brane in a coordinate system
which makes the symmetries of its embedding more manifest. We split
the $\RR^6$ factor in the last term of (\ref{OneF}) into $\RR^4
\times \RR^2$ (i.e. parts longitudinal and transverse to the
D7-brane) and express them in terms of polar coordinates
respectively. More explicitly,
 \bea
 & r^2 = \rho^2 + y^2, \qquad \rho^2 = x_4^2+ x_5^2 + x_6^2 + x_7^2, \qquad
  y^2 = x_8^2 + x_9^2, \nonumber \\
  & x_8 = y \cos \phi , \qquad x_9 = y \sin \phi \ .
  \eea
The metric (\ref{OneF}) then becomes
  \begin{eqnarray}
ds^2  =  {\rho^2 + y^2 \ov R^2} \le(-  dt^2 + d \vec x^2 \ri)+
\frac{R^2 }{\rho^2 + y^2}  \le(d \rho^2 + \rho^2 d \Om_3^2 + dy^2
+ y^2 d \phi^2 \ri) \ .
 \label{twoF}
 \end{eqnarray}
The D7-brane now covers $(t,\vec x)=(t,x_1,x_2,x_3,\rho,\Om_3)$ and
sits at $y=L$ and $\phi=0$. Note that in the radial direction the
D7-brane extends from $\rho = 0$, at which the size
of the three-sphere $\Om_3$ becomes zero, to $\rho=\infty$.  The
point $\rho=0$ corresponds to $r=L$.

We now briefly describe how to find the low-lying meson
spectrum described by the fluctuations of $x_{8,9}$.
The action of the D7-brane is given by the Dirac-Born-Infeld 
action
  \be
 S_{D7} = -\mu_7\int d^8 \xi \, \sqrt{-\det \tilde {h}_{ij}}\ ,
\label{daction}
 \ee
where the $\xi^i$ (with $i=0,1,\ldots, 7$) denote 
the worldvolume coordinates
of the D7 brane and ${\tilde h}_{ij}$ is the induced metric
in the worldvolume
 \be \label{eie}
  {\tilde h}_{ij} = G_{\mu \nu} (X) {\p X^\mu
  \ov \p \xi^i} {\p X^\nu \ov \p\xi^j}
   \ .
  \ee
The value of the 
D7-brane tension, $\mu_7=(2\pi)^{-6}g_s^{-1}\alpha'^{-4}$,
will play no role in our considerations.
The spacetime metric $G_{\mu \nu}$ is given by (\ref{twoF}) and
$X^\mu (\xi)$ describe the embedding of the D7-brane, where $\mu$
runs through all spacetime coordinates. The action (\ref{daction})
is invariant under the coordinate transformations $\xi \to \xi'
(\xi)$. We can use this freedom to set $\xi^i = (t,\vec x,
\rho, \Om_3)$, and the embedding described below equation
(\ref{twoF}) then corresponds to the following solution to the
equations of motion of (\ref{daction}):
 \be \label{d7em}
 y (\xi) = L, \quad \phi (\xi) = 0 \qquad {\rm or} \qquad
 x_8 (\xi) = L, \quad x_9 (\xi) =0 \ .
 \ee
To find the meson spectrum corresponding to the fluctuations of
the brane position, we let
  \be
x_8=L+2\pi\apr \psi_1 (\xi) \ , \qquad x_9=0+2\pi\apr \psi_2 (\xi)
, \label{embed}
 \ee
and expand the action 
(\ref{daction}) to quadratic order in $\psi_{1,2}$, obtaining
  \be
 S_{D7} \simeq \mu_7 \int d^8 \xi \,
 \rho^3 \left(-1- \ha (2\pi \apr R)^2 \frac{h^{ij}}{\rho^2 +L^2}
 (\p_i \psi_1 \p_j
\psi_1 + \p_i \psi_2 \p_j \psi_2)\right)\ . \label{lag}
 \ee
In (\ref{lag}), $h_{ij}$ denotes the induced metric on the
D7-brane for the embedding (\ref{d7em})
in the absence of any fluctuations, i.e.
 \be  \label{3F}
 ds^2 = h_{ij} d \xi^i d \xi^j =  {\rho^2 + L^2 \ov R^2} \le(-  dt^2 + d\vec x^2 \ri)
 + \frac{R^2 }{\rho^2 + L^2}  \le(d \rho^2 + \rho^2 d
\Om_3^2 \ri) \ .
 \ee
Note that when $L =0$, the above metric reduces to $AdS_5 \times
S^3$, reflecting the
fact that in the massless quark limit the Yang-Mills theory
is conformally invariant in the large $N/N_f$ limit.

The equation of motion following from (\ref{lag}) 
is 
\be \label{spec}
 {R^4 \ov (\rho^2 + L^2)^2} \p_\al \p^\al \psi + {1 \ov \rho^3}\,
 \frac{\p}{\p\rho} \left(\rho^3 \frac{\p}{\p\rho}\psi\right) + {1 \ov \rho^2} \nabla^2 \psi
 =0\ ,
 \ee
where $\psi$ denotes either $\psi_1$ or $\psi_2$,
where $\al = 0\ldots 3$, and where $\nabla^2$ denotes the Laplacian
operator on the unit $S^3$. 
Eq. (\ref{spec}) can be solved exactly and
normalizable solutions have a discrete spectrum. It was found
in~\cite{Kruczenski:2003be} that the four dimensional mass
spectrum is given by
 \be \label{mesp}
 m_{nl} = {4 \pi m_q \ov \sqrt{\lam}} \sqrt{(n+l+1) (n+l+2)}, \qquad
 n=0,1,\ldots\  , \quad l =0,1,\ldots\  ,
 \ee
 with degeneracy $(\ell+1)^2$,
where $l$ is the angular momentum on $S^3$. 
The $(\ell + 1)^2$ degeneracy is understood in
the field theory as arising because the scalar mesons
are in the $(\ell/2,\ell/2)$ representation of
a global $SU(2)\times SU(2)$ symmetry corresponding
to rotations in the $S^3$ in the dual gravity theory~\cite{Kruczenski:2003be}.

The mass scale
appearing in (\ref{mesp}) can also be deduced without
calculation via a scaling
argument. Letting
  \begin{align}
 t \rightarrow \frac{R^2}{L} t, \qquad \vec  x  \rightarrow
\frac{R^2}{L} \vec x ,
 \qquad  \rho  \rightarrow L\rho ,
 \label{eq:rcme}
 \end{align}
 the metric (\ref{3F}) can be solely expressed in terms of
 dimensionless quantities:
  \be  \label{31F}
 {ds^2 \ov R^2} =  {(\rho^2 + 1)} \le(-  dt^2 + d\vec x^2 \ri)
 + \frac{1 }{\rho^2 + 1}  \le(d \rho^2 + \rho^2 d
\Om_3^2 \ri) \ .
 \ee
Thus, the mass scale for the 
mesonic fluctuations must be  
 \be \label{msma}
 M \equiv {L \ov R^2} = {2 \pi m_q \ov \sqrt{\lam}} \ ,
 \ee
as is indeed apparent in the explicit result (\ref{mesp}).
We see that 
the mesons are very tightly
bound in the large $\lam$ limit with a mass $M$ that is parametrically
smaller than the rest mass of a separated quark and antiquark, $2m_q$.
This means that the binding energy is $\approx - 2 m_q$. From this
fact and the Coulomb potential~(\ref{zepo}), one can also
estimate that the size of the bound states is parametrically 
of order $\sim {1 \ov M} \sim {\sqrt{\lam} \ov m_q}$.

Finally, we can now explain the sense in which our analysis is limited 
to low-lying mesons.  We are only analyzing those scalar mesons
whose mass is of order $M$.  There are other, stringy, excitations
in the theory with meson quantum numbers whose masses
are of  order $L/(R\sqrt{\alpha'})\sim M \lambda^{1/4}\sim m_q/\lambda^{1/4}$
and of order $L/\alpha'\sim M\lambda^{1/2} \sim m_q$~\cite{Kruczenski:2003be}.
They are parametrically heavier than the mesons we analyze, and can
be neglected in the large $\lambda$ limit even though those
with masses $\sim m_q/\lambda^{1/4}$ are also tightly bound, since
their masses are also parametrically small compared to $m_q$.
In Section 5, we shall see again in a different way that our
analysis of the dispersion relations for the mesons with
masses $\sim m_q/\sqrt{\lambda}$ that we focus on is controlled
by the smallness of $1/\lambda^{1/4}$.


\subsection{Nonzero Temperature}

We now put the Yang-Mills theory at nonzero temperature, in which case
the $AdS_5$  part of the metric (\ref{OneF}) is replaced by the metric of an AdS
Schwarzschild black hole
 \begin{eqnarray}
ds^2 & = &  - f(r) dt^2 + \frac{r^2 }{ R^2} d \vec x^2 + \frac{1 }{
f(r)} dr^2 + R^2 d \Om_5^2\, ,
\label{3.3}\\
f(r) &\equiv& \frac{r^2 }{ R^2} \left(1 - \frac{r_0^4 }{ r^4}
\right)\, . \label{3.4}
 \end{eqnarray}
The temperature of the gauge theory is equal to the Hawking
temperature of the black hole, which is
\begin{equation}
 T = \frac{r_0 }{ \pi R^2}\, .
   \label{3.6}
 \end{equation}
This is the one addition  at nonzero temperature to the dictionary 
that relates the parameters of the (now hot) gauge theory to those
of its dual gravity description.

At nonzero temperature, the embedding of the D7-brane is
modified because the D7-brane now feels a gravitational
attraction due to the presence of the black hole. To
find the embedding, it is convenient to use coordinates which are
analogous to those in (\ref{twoF}). For this purpose, we introduce a
new radial coordinate $u$ defined  by
 \be
 {dr^2 \ov f(r)} = {R^2 du^2 \ov u^2}, \quad {\rm i.e.} \quad  u^2  = \ha \left(r^2 + \sqrt{r^4 - r_0^4}\right),
 \ee
in terms of which (\ref{3.3}) can then be written as
  \bea \label{pejs}
  ds^2 & = &   - f dt^2 + {r^2 \ov R^2}d \vec x^2 + {R^2 \ov u^2} ( du^2 + u^2
d \Omega_5^2) \\
 & = & - f dt^2 + {r^2 \ov R^2} d \vec x^2  + {R^2 \ov
u^2} \le(d \rho^2 + \rho^2 d \Om_3^2 + dy^2 + y^2 d \phi^2 \ri)  \
. \label{uew}
 \eea
As in (\ref{twoF}), we have split the last term of (\ref{pejs}) in
terms of  polar coordinates on $\RR^4 \times \RR^2$, with
 \be\label{uvsrho}
  u^2 = y^2 + \rho^2 \ .
  \ee
In (\ref{pejs}) and (\ref{uew}), $f$ and $r$ should now be
considered as functions of $u$,
 \be
 r^2 = u^2 + {r_0^4 \ov 4 u^2}, \qquad  f(u) =
\frac{( u^4 - r_0^4/4)^2}{ u^2 R^2 ( u^4 + r_0^4/4)} \ .
 \ee
In terms of $u$, the horizon is now at $u_0 = {r_0 \ov \sqrt{2}}$.

The D7-brane again covers $\xi^i = (t, \vec{x}, \rho,\Omega_3)$
and its embedding $y (\xi), \phi (\xi)$ in the $(y,\phi)$ plane will
again be determined by extremizing the Dirac-Born-Infeld action
(\ref{daction}). Because of the rotational symmetry in the $\phi$ direction,
we can choose $\phi (\xi) =0$. Because of the translational symmetry in the
$(t , \vec x)$ directions and the rotational symmetry in $S^3$, $y$
can depend on $\rho$ only. Thus, the embedding is fully specified by a
single function $y (\rho)$. The induced metric on the D7-brane
worldvolume can be written in terms of this function as
\begin{eqnarray}
h_{ij} d\xi^i d\xi^j &=& - f(u) dt^2 + {r^2 \ov R^2} d\vec{x}^2 +
\frac{R^2}{u^2} \left( (1+y'(\rho)^2) d\rho^2 +  \rho^2
d\Omega_3^2 \right)\ , \label{eq:yr}
\end{eqnarray}
where $u$ in (\ref{uvsrho}) and hence $f(u)$ are functions of $\rho$ and $y(\rho)$. 
Substituting (\ref{eq:yr}) into (\ref{daction}), one finds
 \begin{equation}
S_{D7} \propto \int d\rho \frac{\rho^3}{u (\rho) ^8} \left( 16
\left(\frac{u(\rho)}{r_0}\right)^8 - 1 \right) \sqrt{ 1+
y'(\rho)^2 }\ ,
\end{equation}
which leads to the equation of motion
 \begin{equation}
\label{eq:yrom} \frac{ y''}{1+ y'^2} + \frac{3 y'}{\rho} + \frac{8
r_0^8}{u^2} \frac{(\rho y' - y)}{16 u^8 - r_0^8} = 0
\end{equation}
for $y (\rho)$, where $u^2(\rho) = \rho^2 + y^2(\rho)$.

To solve (\ref{eq:yrom}) one imposes the boundary condition that
$y \to L$ as $\rho \to \infty$, and that the induced metric
(\ref{eq:yr}) is non-singular everywhere. $L$ determines the bare
quark mass as in (\ref{qma}).
 It is convenient to introduce a parameter
 \be\label{epsinftydefn}
 \ep_\infty \equiv {u_0^2 \ov  L^2} = {r_0^2 \ov 2 L^2} = { \lam T^2
 \ov 8 m_q^2} = {\pi^2 T^2 \ov 2 M^2}\ ,
 \ee
where we have used (\ref{3.6}) and
(\ref{msma}). Because $\NN=4$ SYM is scale invariant
before introducing the massive fundamentals, meaning that
all dimensionful quantities must be proportional to appropriate
powers of $T$, when we introduce the fundamentals the only
way in which the quark mass $m_q$ can enter is through the 
dimensionless ratio $m_q/T$.  Scale invariance alone does
not require that this ratio be accompanied by a $\sqrt{\lambda}$,
but it is easy to see that, after rescaling to dimensionless
variables as in (\ref{eq:rcme}), the only combination of
parameters that enters (\ref{eq:yrom})
is $\ep_\infty$. The small
$\ep_\infty$ regime can equally well be 
thought of as a low temperature
regime or a heavy quark regime.  
In the remainder of this section, 
we shall imagine $m_q$ as fixed and describe the physics
as a function of varying $T$, i.e. varying horizon radius $r_0$.

\FIGURE[t]{
\centerline{\hbox{\psfig{figure=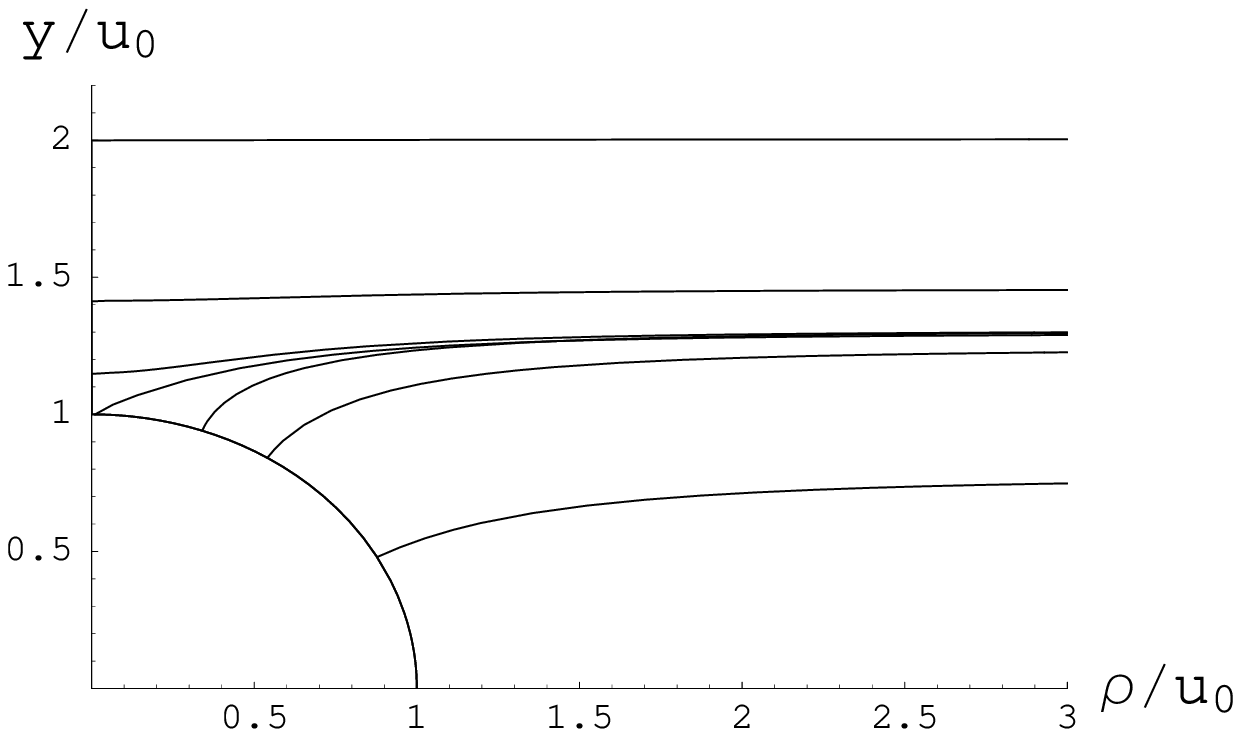,width=12cm,height=8cm}}}
\caption{Some possible D7-brane embeddings $y(\rho)$. 
The quark mass to temperature ratio is determined by $y(\infty)=L$. 
Specifically,
$\sqrt{8} m_q/(T\sqrt{\lambda})= y(\infty)/u_0\equiv 1/\sqrt{\ep_\infty}$.
The top three curves are Minkowski embeddings, with
$y(\rho)$ extending from $\rho=0$ to $\rho=\infty$.  
The bottom three curves
are black hole embeddings, in which the D7-brane begins at the black
hole horizon at $y^2+\rho^2=u_0$.  
The middle curve is the critical embedding.
The seven curves, ordered from top to bottom as they
occur in the left part of the figure,  are drawn for temperatures 
specified by $\ep_\infty=$ 0.249, 0.471, 0.5865, 
0.5948, 0.5863, 0.647 and 1.656. 
Note that the $\ep_\infty=0.5863$ black hole embedding
crosses both the $\ep_\infty=0.5948$ critical embedding
and the $\ep_\infty=0.5865$ Minkowski embedding.
 \label{fig:1}
 }}

The equation of motion (\ref{eq:yrom}) that specifies the D7-brane embedding
can be solved numerically. 
Upon so doing, one finds that
there are three types of solutions with different 
topology~\cite{Babington:2003vm,Mateos:2006nu,Mateos:2007vn}:
\begin{itemize}
\item Minkowski embeddings: The D7-brane extends all the way to
$\rho=0$ with $y (0) > u_0 = {r_0 \ov \sqrt{2}}$ (see e.g. the
upper three curves in 
Fig.~\ref{fig:1}). 
In order for the solution to be regular one needs
$y' (0) =0$. This gives rise to a one-parameter family of
solutions parameterized by $y(0)$.  The topology of the
brane is $\mathbb{R}^{1,7}$.
\item Critical embedding: The D7-brane just touches the horizon,
i.e. $y(0) = u_0$ (see e.g. the middle curve in Fig.~\ref{fig:1}). 
The worldvolume metric is singular at the point
where the D7-brane touches the horizon.
\item Black hole embeddings:  The D7-brane ends on the horizon 
$u_0= r_0/\sqrt{2}$ at some $\rho > 0$ (see e.g. the lower three
curves in Fig.~\ref{fig:1}). The topology of the D7-brane is then
$\mathbb{R}^{1,4} \times S^3$.
\end{itemize}

It turns out~\cite{Hoyos:2006gb,Mateos:2007vn}
that Minkowski embeddings that begin at $\rho=0$ with $y$ 
close to $r_0/\sqrt{2}$, 
just above the critical embedding, 
can cross the critical embedding, ending up at $\rho\rightarrow\infty$
with $y(\infty)$ just below that for the critical embedding.  Similarly,
embeddings that begin just below the critical embedding 
can end up just above it. Furthermore, those embeddings that
begin even closer to the critical embedding can cross it more than once.
This means that there is a range of values around the critical 
$\epsilon_\infty^c=0.5948$ for which there are three or more embeddings for
each value of $\epsilon_\infty$.  At low temperatures (precisely,
for $\ep_\infty < 0.5834$) this does not occur: there is only a single
Minkowski embedding for each value of $\ep_\infty$.  
At high temperatures (precisely, for $\ep_\infty > 0.5955$)
there is only a single black hole embedding per  value of $\ep_\infty$.
In the intermediate range of temperatures $0.5834 < \ep_\infty < 0.5955$,
one needs to compare the free energy of 
each of the three or more different D7-brane embeddings
that have the same value of $\ep_\infty$ to determine which is favored.
One finds that there is a first order phase transition 
at a temperature $T_c$ at which $\ep_\infty = 0.5863$,
where the favored embedding jumps discontinuously from a Minkowski
embedding to a black hole embedding as a function of increasing 
temperature~\cite{Hoyos:2006gb,Mateos:2007vn}.\footnote{The critical 
embedding occurs at an $\ep_\infty=0.5948$ which is greater than
the $\ep_\infty$ at which the first order phase transition occurs, meaning
that at $\ep_\infty=0.5948$ there is a black hole embedding that has 
a lower free energy than the critical embedding.}

As we shall study in detail in Section 4, fluctuations about a Minkowski
embedding describe a discrete meson spectrum with a 
mass gap of order $O(M)$. In contrast, fluctuations about a black hole
embedding yield a continuous spectrum~\cite{Hoyos:2006gb,Mateos:2007vn}.
A natural interpretation of the first
order transition is that $T_c=T_{\rm diss}$, the temperature above which the
mesons dissociate~\cite{Hoyos:2006gb,Mateos:2007vn}.  It is
interesting, and quite unlike what is expected in QCD, that all the
mesons described by the zero temperature spectrum (\ref{mesp})
dissociate at the same temperature.  This is presumably related
to the fact that the mesons are so tightly bound, again unlike in QCD.
We shall therefore focus on the velocity-dependence of
the meson spectrum at nonzero temperature --- in other words, 
the meson dispersion relations first studied
in~\cite{Mateos:2007vn}.  As we have explained in Section 1, the
velocity-dependence is currently inaccessible to lattice QCD calculations.
Hence, even qualitative results are sorely needed.  Furthermore, inferences drawn
from a previous calculation of
the potential between a moving quark-antiquark pair lead to
a velocity-scaling (\ref{rro}) of $T_{\rm diss}$ that has a simple
physical interpretation, which suggests
that it could be applicable in varied
theories~\cite{Liu:2006nn}.  We shall see this velocity dependence
emerge from the meson dispersion relations in Section 5.

\begin{figure}[t]
\begin{center}
$\begin{array}{cc} \psfig{figure=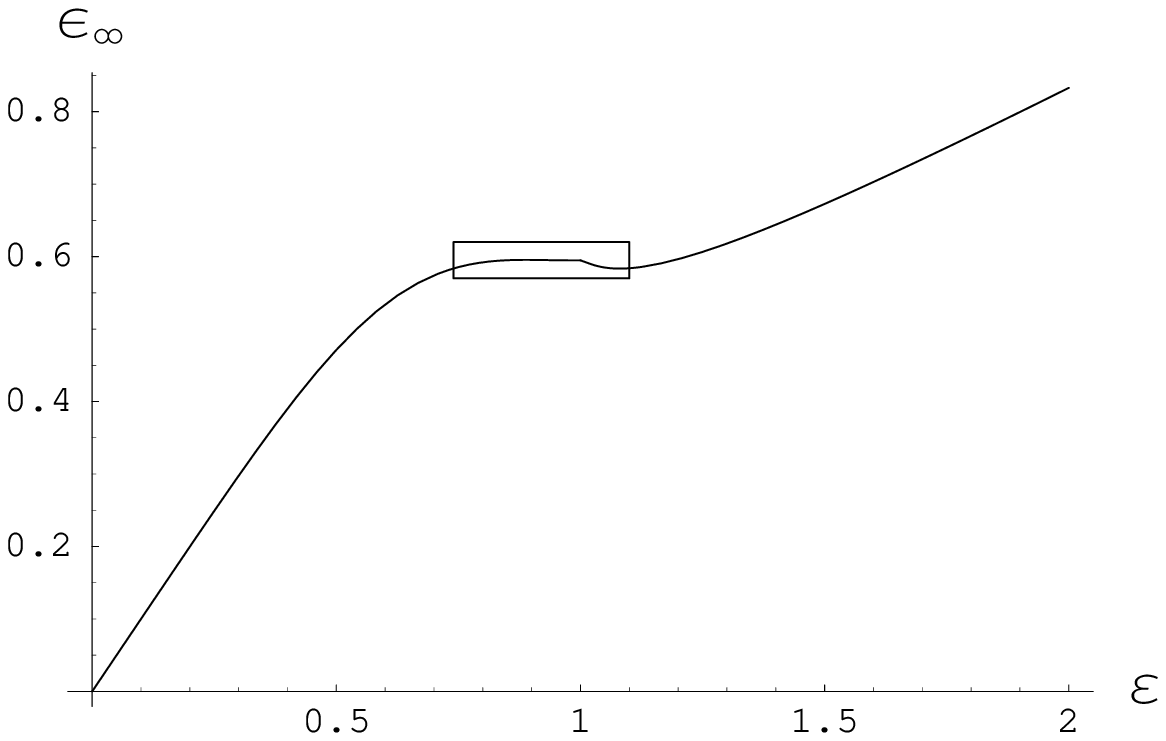,width=9cm,height=6cm}  &
\psfig{figure=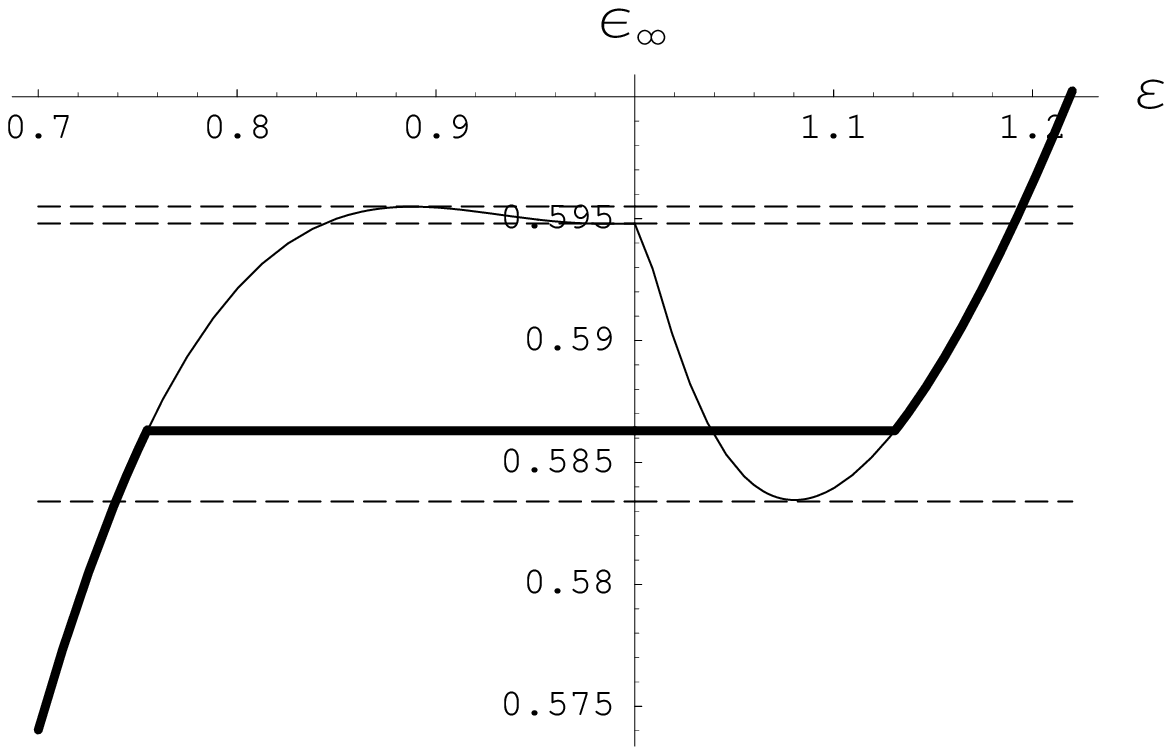,width=9cm,height=6cm}
\end{array}$
\caption{
$\ep_\infty$ (determined by the embedding $y$ at infinity) versus $\eps$ (determined
either by $y(0)$, for Minkowski embeddings with $\eps<1$, or by 
where the embedding intersects the horizon, for $\eps>1$).  The right panel
zooms in on the vicinity of the critical embedding at $\eps=1$. The stable 
embeddings and the first order phase transition are indicated by the thick
curve; the metastable embeddings are indicated by the thin curves.
\label{fig:epsepsinf}}
\end{center}
\end{figure}

It is interesting to return to the qualitative estimate of $T_{\rm diss}$ 
obtained from the static quark-antiquark potential in Section 2,
and see how it compares to the $T_{\rm diss}=T_c$ obtained from
the analysis of the mesons themselves.  Equating the size of a 
meson with binding energy $2 m_q$, determined by the zero-temperature
potential (\ref{zepo}), with the screening length $L_s=L_c=0.24/T$, 
determined by the potential (\ref{Awe}) at nonzero temperature, yields
the estimate that $T_{\rm diss}$ should be 
$\sim 2.1 m_q/\sqrt{\lambda}$. This is in 
surprisingly good agreement with
$T_{\rm diss} = \sqrt{8 \ep_\infty }\,m_q/\sqrt{\lambda} 
= 2.166 \,m_q/\sqrt{\lambda}$
for $\ep_\infty=0.5863$.

In subsequent sections, we shall derive the dispersion relations
for mesons at $T<T_{\rm diss}$.  We close this section
by introducing some new notation that simplifies the analysis
of the Minkowski embedding of the D7-brane, whose fluctuations we shall
be treating.    We first introduce
parameters
 \be\label{epsilondefn}
 L_0\equiv y (0)\ , \qquad \varepsilon \equiv {u_0^2 \ov L_0^2} = {r_0^2 \ov 2 L_0^2}
 \ .
 \ee
For Minkowski embeddings, $\varepsilon$ takes value in the range $[0,1]$,
with $\eps =0$ corresponding to zero temperature, and $\varepsilon=1$ to
the critical embedding.  Although $\ep_\infty$ that we introduced
earlier has the advantage of being directly related to the fundamental
parameters of the theory according to (\ref{epsinftydefn}), the new
parameter has the advantage that there is only one embedding
for each value of $\varepsilon$. And,
$\varepsilon$ will turn out to be convenient for analyzing the equations
of motion (\ref{eq:yrom}) and the fluctuations on D7-branes.  When
$\ep_\infty \ll 1$, $\varepsilon \approx \ep_\infty$.  A full analytic
relation between $\varepsilon$ and $\ep_\infty$ is not known, but given
an $\varepsilon$ one can readily find  
the corresponding $\ep_\infty$ numerically.
For example, at $T=T_c$, $\varepsilon = 0.756$ 
and  $\ep_\infty = 0.586$ while for
the critical embedding, $\varepsilon =1$ and $\ep_\infty =\ep_\infty^c= 0.5948$.
We depict the relation between $\ep_\infty$ and $\eps$ in
Fig.~\ref{fig:epsepsinf}.  In order to make this figure,
for the black hole embeddings 
we have defined $\eps=1/\sin^2\theta$ 
where $\theta$ is the angle in the $(y,\rho)$ plane of Fig.~\ref{fig:1}
at the point at which the black hole embedding $y(\rho)$ intersects the
black hole horizon $y^2+\rho^2=u_0^2$. That is,
$1 < \eps< \infty$ parametrizes black hole embeddings which
begin at different points along the black hole horizon.  
The seven embeddings in Fig.~\ref{fig:1} have 
$\eps=0.25$, 0.5, 0.756, 1.00, 1.13, 1.41 and 4.35, from
top to bottom as they are ordered on the left, i.e. at the
tip of the D7-brane at $y=0$ for the Minkowski embeddings
and at the horizon for the black hole embeddings.

Finally, it will also prove convenient to introduce
dimensionless coordinates by a rescaling according to
 \begin{align}
 t \lra \frac{R^2}{L_0} t, \qquad x_i  \lra \frac{R^2}{L_0} x_i ,
 \qquad
 \rho  \lra L_0 \rho , \qquad y  \lra L_0 y,
 \label{eq:rce}
 \end{align}
after which the spacetime metric becomes
 \begin{eqnarray}
{ds^2  \ov R^2} &=& G_{\mu\nu} dx^\mu dx^\nu 
= - f(u) dt^2 + {r (u)^2} d\vec{x}^2 +
\frac{1}{u^2} \left(  d\rho^2 +  \rho^2 d\Omega_3^2 + dy^2 + y^2 d
\phi^2 \right) \label{eyr}
\end{eqnarray}
and the induced metric becomes
 \begin{eqnarray}
 {ds^2_{D7} \ov R^2} = h_{ij} d\xi^i d\xi^j = - f(u) dt^2 + {r^2 } d\vec{x}^2 +
\frac{1}{u^2} \left( (1+y'(\rho)^2) d\rho^2 +  \rho^2 d\Omega_3^2
\right) \label{eq:Nyr}
\end{eqnarray}
with
 \be \label{vqu}
u^2 = y^2  + \rho^2, \qquad f(u)
=\frac{(u^4-\eps^2)^2}{u^2(u^4+\eps^2)}, \qquad r^2 (u) = u^2 +
{\eps^2 \ov  u^2} \ ,
 \ee
where both $G_{\mu\nu}$ and $h_{ij}$ are now dimensionless.
The equation of motion for $y (\rho)$ becomes
\begin{align}
\label{eq:em} \frac{y''}{1+y'^2}+3\frac{y'}{\rho}+ \frac{8}{u^2}
\lt(\frac{\rho y'-y}{\eps^{-4}u^8-1}\rt)=0,
\end{align}
with the boundary conditions
\begin{align}
y(0)=1, \qquad  y'(0)=0 \ .
\end{align}
This form of the equations of motion 
that determine the embedding $y(\rho)$
will be useful in subsequent sections.

\section{Meson Fluctuations at Nonzero Temperature}

In this section we derive linearized equations of motion that describe the
small fluctuations of the D7-brane position. A version of these
equations have been derived and solved numerically by various
authors (see
e.g.~\cite{Babington:2003vm,Kruczenski:2003uq,Hoyos:2006gb,Mateos:2007vn}).
Here we will rederive the equations in a different form by
choosing the worldvolume fields parameterizing the fluctuations in
a more geometric way. The new approach gives a nice geometric
interpretation for the embedding and small fluctuations. It also
simplifies the equations dramatically, which will enable us to
extract analytic information for the meson dispersion relations in
the next section.
We present the main ideas
and results in this Section but we leave technical details to Appendix A.
In that Appendix, we shall also present a general discussion of the fluctuations
of a brane embedded in any curved spacetime.

The action for small perturbations of the D7-brane location can be
obtained by inserting
 \begin{equation}
X^\mu(\xi) = X_0^\mu(\xi^i) + \delta X^{\mu}(\xi^i)
\label{eq:pertq}
\end{equation}
into the D-brane action~(\ref{daction}) and (\ref{eie}), where
$X_0^\mu (\xi)$ denotes the background solution that
describes the embedding in the absence of fluctuations, and $\delta X^\mu$
describes
small fluctuations transverse to the brane. For the D7-brane under
consideration, in the coordinates used in (\ref{eyr}) the general expression
(\ref{eq:pertq}) becomes
 \be \label{nss}
 y(\xi) = y_0(\rho) + \delta y (\xi), \qquad \phi(\xi) = \delta \phi(\xi)
 \ee
with $y_0 (\rho)$ the embedding solution obtained by solving
(\ref{eq:em}). The choice of the worldvolume fields $\delta y,
\delta \phi$  is clearly far from unique. Any two
 independent functions of $\delta y$ and $\delta \phi$ will also do.
(This freedom corresponds to the freedom to choose different
coordinates for the 10-dimensional space within which the D7-brane
is embedded.)
In fact, it is awkward to
use $\delta y $ and $\delta \phi$ as worldvolume fields since they
are differences in coordinates and thus do not transform nicely
under coordinate changes. Using them obscures the geometric
interpretation of the equations. Below we will adopt a coordinate
system 
which makes the geometric
interpretation manifest. Since our discussion is rather general,
not specific to the particular system under consideration, we will
describe it initially using general language.

Consider a point $X_0(\xi)$ on the brane. The tangent space at
$X_0$ perpendicular to the D7-brane
is a two-dimensional subspace $V_0$ spanned by unit
vectors $n_1^\mu, n_2^\mu$ which are orthogonal to the branes,
i.e.
 \begin{eqnarray}
\label{eq:units}
n_1^\mu &\propto& \dbyd{y}^\mu - y'_0(\rho) \dbyd{\rho}^\mu \\
n_2^\mu &\propto& \dbyd{\phi}^\mu \ .
 \label{n2}
\end{eqnarray}
Any vector $\eta^\mu$ in $V_0$ can be written as
 \be
 \eta^\mu  = \chi_1 n_1^\mu + \chi_2 n_2^\mu \ .
 \ee
We can then establish a map from $(\chi_1, \chi_2)$ to small
perturbations $\delta X^\mu$ in (\ref{eq:pertq}) by 
shooting out
geodesics of unit affine parameter from $X_0$ with tangent
$\eta^\mu$. Such a map should be one-to-one for $\chi_1, \chi_2$
sufficiently small. Clearly $\chi_1$ and $\chi_2$ behave like
scalars under coordinate changes and we will use them as the
worldvolume fields parameterizing small fluctuations 
of the position of the brane. By solving
the geodesic equation, $\delta X^\mu$  can be expressed in terms of
$\chi_{1,2}$ as
 \begin{equation}
\label{eq:eta} \delta X^{\mu} = \eta^\mu -  \frac{1}{2}
\Gamma^\mu_{\alpha\beta} \eta^\alpha \eta^\beta + \ldots \ ,
\qquad
\end{equation}
where $\Gamma^\mu_{\alpha\beta}$ are the Christoffel symbols of the 
10-dimensional metric. 
Note that the choice of $\chi_{1,2}$ is not unique. There is in fact
an $SO(2)$ ``gauge'' symmetry under which $\chi_{1,2}$ transform
as a vector, since one can make
different choices of basis vectors $n_1, n_2$ that
are related by a local 
$SO(2)$ transformation.

We now insert (\ref{eq:eta}) and (\ref{eq:pertq}) into the
Dirac-Born-Infeld action (\ref{daction}) and, after some algebra
discussed further in Appendix A, we find that the equations of
motion
satisfied by $X_0$ (i.e. which determine the embedding
in the absence of fluctuations)
can be written as
 \be \label{baeom}
 K_s =0, 
 \ee
and the quadratic action for small fluctuations $\chi_{1,2}$ 
about $X_0$ takes the form
 \begin{align}
S_{D7} =   \mu_7 R^8 \int d^8\xi \sqrt{-\det h_{ij}} & \left(  -
\frac{1}{2} D_i \chi_s D^i \chi_s
 - \frac{1}{2} \chi_s \chi_t \left( - K_{s ij} K_t^{ij}  +
R_{sijt} h^{ij}   \right) \right)\, ,
 \label{bss}
\end{align}
where $s,t =1,2$ and where we have defined the
following quantities:
 \bea
 && h_{ij} = G_{\mu \nu} \p_i X_0^\mu \p_j X_0^\nu\ , \qquad
  R_{sijt} = n_s^\al n_t^\beta \p_i X_0^\mu \p_j X_0^\nu
 R_{\al \mu \nu \beta}\ ,
 \\
 && K_{s ij} = \p_i X_0^\mu \p_j X_0^\nu  \nabla_\mu n_{s \nu}\ ,
 \qquad K_s = K_{s ij} h^{ij} \ ,\\
 && D_i \chi_s = \p_i \chi_s + U_{ist} \chi_t\ , \qquad  U_{ist}
  =  n_{s \nu} \p_i X_0^\mu \nabla_\mu n_t^\nu\ .
 \eea
Note that $h_{ij}$ is the induced metric on the brane and $i,j$
are raised by $h^{ij}$. 
$R_{\al \mu\nu\beta}$ is the Riemann tensor for the 10-dimensional
spacetime. 
$K_{sij}$ is the extrinsic curvature 
of the brane along
the direction $n_s^\mu$. 
$U_{ist}$ (which is antisymmetric in $s,t$) is an
$SO(2)$ connection for the $SO(2)$ gauge symmetry and $D_i$ is the
corresponding covariant derivative.
We see that the embedding equations of motion (\ref{baeom}) have a
very simple geometric interpretation as 
requiring that the trace of the
extrinsic curvature in each orthogonal direction has to vanish,
which is what we expect since this is equivalent to 
the statement that the volume of the D7-brane is extremal.

The symmetries of the D7-brane embedding that we are
analyzing allow us to further simplify the action (\ref{bss}).
Because $n_2^\mu$ in (\ref{n2}) is 
proportional to a Killing vector and is hypersurface orthogonal, 
$U_{i12}$ and $K_{2ij}$ vanish identically. (See Appendix A for
a proof, and for the definition of hypersurface orthogonal.) 
With $K_2=0$ satisfied as an identity,
the remaining equation of motion specifying
the embedding, namely $K_1=0$, is then
equivalent to the 
equation of motion for $y$ that we derived 
in Section 3, namely Eq.~(\ref{eq:em}). 
After some further algebra (see Appendix A)
we find that the action (\ref{bss}) for small fluctuations 
reduces to
 \begin{equation}
\label{eq:finalaction} S_{D7} =   \mu_7 R^8 \int
d^8\xi \sqrt{-\det h_{ij}} \left( 
 - \frac{1}{2}
(\partial \chi_1)^2 - \frac{1}{2} (\partial \chi_2)^2 -
\frac{1}{2}
m_1^2 \chi_1^2 - \frac{1}{2} m_2^2 \chi_2^2 
 \right)
\end{equation}
with
\begin{eqnarray} \label{effM}
m_1^2 &=&  R_{11} + R_{2112} + 2 R_{22} +  \left.^{(8)}R\right. - R\ ,\nonumber\\
m_2^2 &=& - R_{22} - R_{2112}\ ,
\end{eqnarray}
where we have defined
\begin{eqnarray}\label{R2112defn}
R_{2112} &=& n^\mu_2 n_1^\nu n_1^\sig n_2^\tau R_{\mu\nu\sig\tau}\ ,\nonumber\\
R_{11} &=& n_1^\nu n_1^\sig R_{\nu\sig}\ ,\nonumber\\
R_{22} &=& n_2^\nu n_2^\sig R_{\nu\sig}\ ,
\end{eqnarray}
and where $R$ is the Ricci scalar for the 10-dimensional spacetime
while $^{(8)}R$ is the Ricci scalar for
the induced metric $h_{ij}$ on the D7 brane. The background metric
$h_{ij}$ is given by (\ref{eyr}). The ``masses'' $m_1^2$ and
$m_2^2$ are nontrivial functions of $\rho$. Since the worldvolume
metric is regular for Minkowski embeddings, they are well defined
for $\rho \in [0, \infty)$.

Our result in the
form (\ref{bss}) is very general, applicable to the embedding of
any codimension-two branes in any spacetime geometry. 
For example, we can
apply it to the embedding of D7-branes at zero temperature given by
(\ref{d7em}) and learn that the meson fluctuations at zero temperature
are described by (\ref{eq:finalaction}) with
 \begin{equation} \label{epp}
m_1^2 = m_2^2 = - \frac{ 3 \rho^2 + 4}{1 + \rho^2}
\end{equation}
and with $h_{ij}$ in~(\ref{eq:finalaction}) given by (\ref{3F}). 
It is
also straightforward to check that equations of motion derived
from~(\ref{eq:finalaction}) with (\ref{epp}) and $h_{ij}$ given by
(\ref{3F}) are equivalent to (\ref{spec}). At zero temperature,
(\ref{lag}) and (\ref{3F}) are 
already simple enough and the formalism we have described here
does not gain us further advantage.
However, at nonzero temperature the
equations of motion
obtained from~(\ref{eq:finalaction}) yield both technical
and conceptual simplification.   In Section 5 we shall
use the formalism that we have developed to obtain the dispersion relations
at large momentum analytically.

\begin{figure}[t]
\begin{center}
$\begin{array}{cc} \psfig{figure=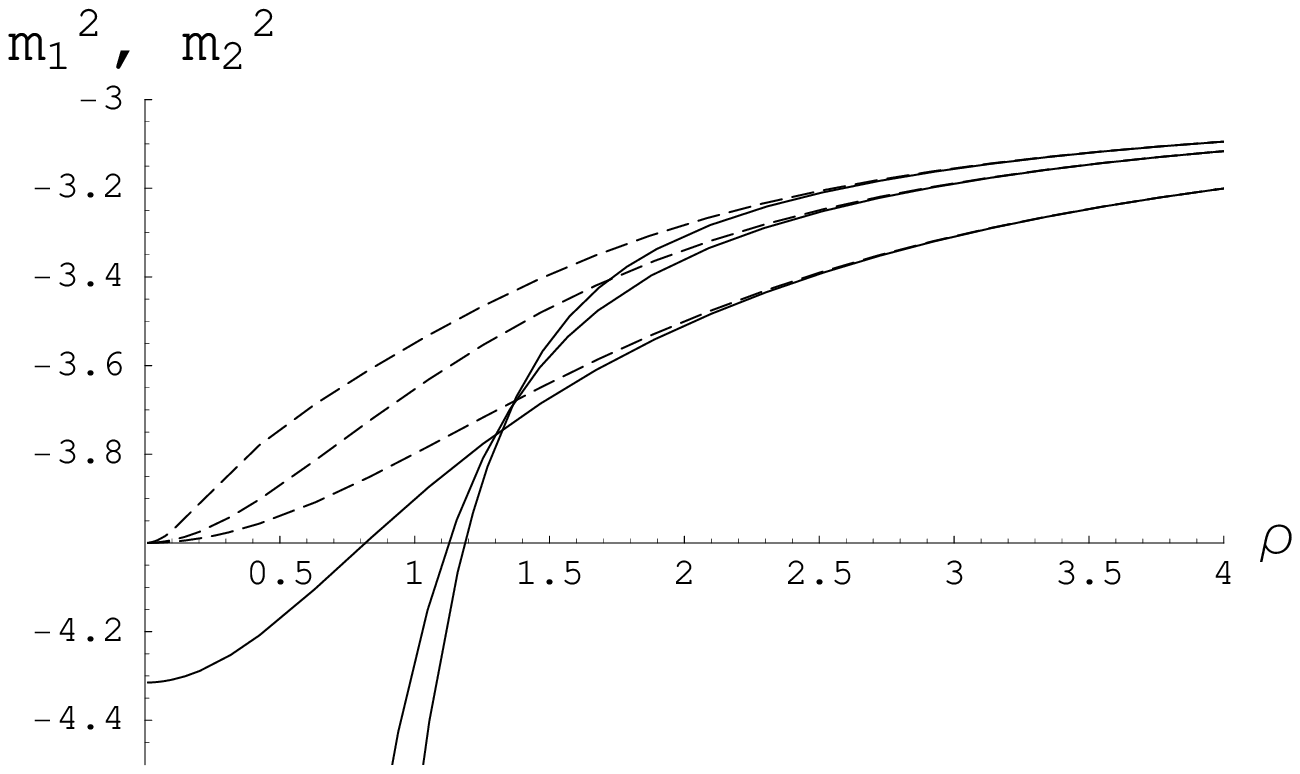,width=9cm,height=6cm}  &
\psfig{figure=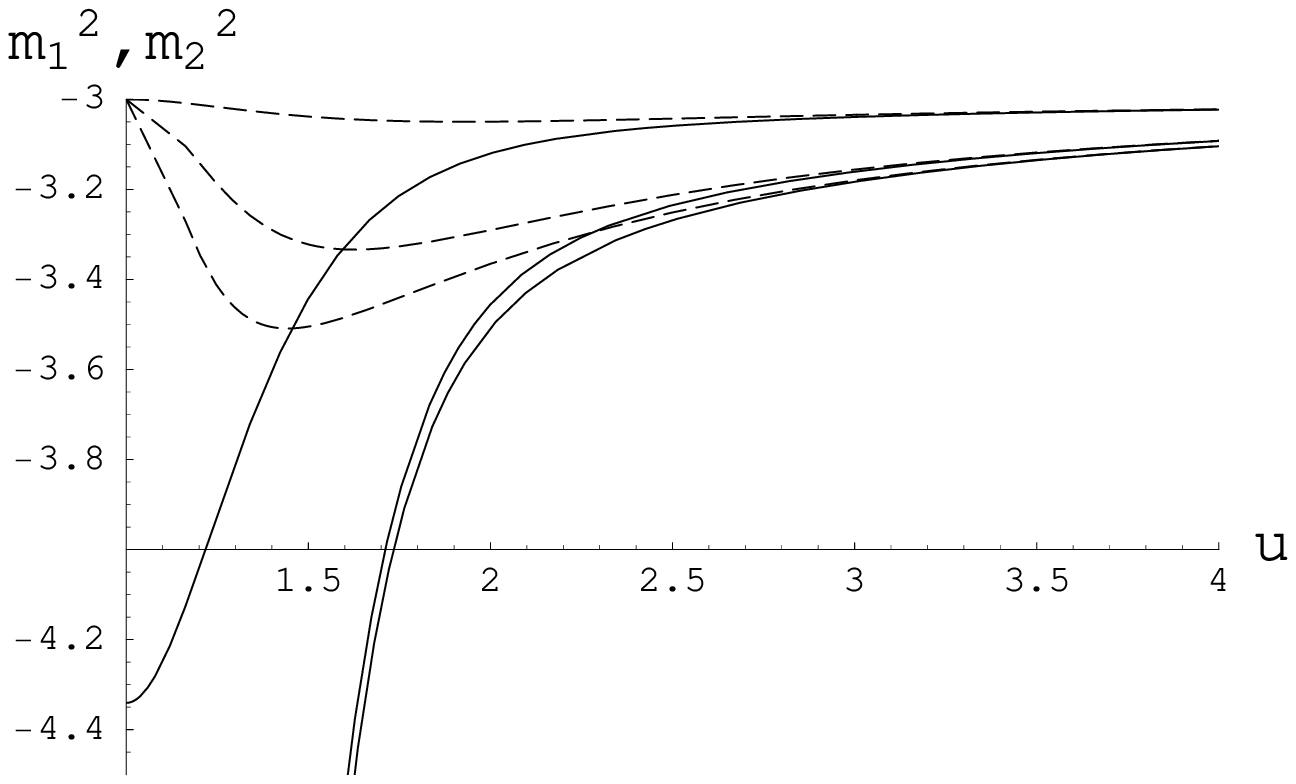,width=9cm,height=6cm}
\end{array}$
\caption{The squared ``masses''  of the two orthonormal geometric
modes of the D7-brane fluctuations for
Minkowski embeddings (left panel) and black
hole embeddings (right panel). In each figure, $m_1^2$
($m_2^2$) is plotted as a solid (dashed) line for three
values of $\ep_\infty$.  The Minkowski embeddings
have $\ep_\infty=0.587$, 0.471 and 0.249 (top to bottom) and
the black hole embeddings have $\ep_\infty=1.656$, 0.647 and 0.586 (again
top to bottom, this time with temperature increasing from top to bottom.)
The Minkowski embedding is plotted as
a function of $\rho$ and the black hole embedding as a function of $u$ with
the horizon on the left at $u=1$.  
\label{fig:masses}}
\end{center}
\end{figure}

Before turning to the dispersion relations, we plot the ``masses''
$m_1^2$ and $m_2^2$ for various D7-brane embeddings at nonzero
temperature in Fig.~\ref{fig:masses}.  
Using a numerical solution for $y(\rho)$, it is straightforward to
evaluate (\ref{effM}), obtaining the masses in the figure. 
For the black hole embeddings, the D7-brane begins at the
black hole horizon at $u=1$
rather than at $\rho=0$, see Fig.~\ref{fig:1},
making it more convenient to plot the masses as a function
of $u$ rather than $\rho$.
We can infer several important features from the masses
plotted in Fig.~\ref{fig:masses}.
As $\rho \to \infty$,  both
$m_1^2$ and $m_2^2$ approach $-3$ for
all the embeddings. This implies that 
$\chi_1$ and $\chi_2$ 
couple to boundary operators of dimension
$3$, as shown in~\cite{Myers:2007we}
by explicit construction of the operators
in the boundary theory which map onto
$\chi_1$ and $\chi_2$.
As $\eps \to 1$ from below for the Minkowski embeddings
(from above for the black hole embeddings),
the behavior of $m^2_1$
at the tip of the D7-brane at $\rho=0$ 
(at $u=1$) becomes singular, diverging
to minus infinity. 
This is a reflection of
the curvature divergence at the tip of the critical embedding
at $\rho=0$ ($u=1$).

We have referred to $m_1^2$ and $m_2^2$ as ``masses'' in quotes
because the equations of motion obtained by straightforward
variation of the
action (\ref{eq:finalaction}) 
in which they arise yields
\begin{equation}
\label{eq:kg}  {1 \ov \sqrt{-h}} \p_i (\sqrt{-h} h^{ij} \p_j
\chi_s) - m_s^2 \chi_s = 0 , \quad s=1,2 \ 
\end{equation}
with $h\equiv {\rm det}h_{ij}$, which is a Klein-Gordon 
equation in a curved spacetime with spatially varying ``masses".
If we could cast the equations of motion in such a way
that they take the form of a Schr\"odinger equation with
some potential, this would make it possible to infer
qualitative implications for the nature of the meson
spectrum immediately via physical intuition, which
is not possible to do by inspection of the
curves in Fig.~\ref{fig:masses}.
To achieve this, we recast the equations of
motion as follows. We introduce a
``tortoise coordinate'' $z$ defined by
 \be \label{tor}
 dz^2 = {1 \ov u^2 f(u)} \left(1 + y_0'(\rho)^2\right) d \rho^2\ ,
 \ee
in terms of which the induced metric on the brane 
takes the simple form
 \begin{equation}
\label{eq:mtort} \frac{ds^2_{D7}}{R^2} = f ( - dt^2 + dz^2) +
r^2(u) d\vec{x}^2 + \frac{\rho^2}{u^2} d\Omega_3^2\ .
\end{equation}
(We choose the additive constant in the definition of $z$ so
that $z=0$ at $\rho =0$.)
Then, we seek solutions to the equations of motion (\ref{eq:kg})
that separate according to the ansatz
\begin{equation}\label{wavefunctionansatz}
\chi_s =  {\psi_s (z)  \ov Z} \, e^{-i \om t + i \vec k\cdot \vec x}
\, Y_{\ell m\tilde{m}}(\Omega_3)
\end{equation}
with
 \be
  \label{eq:frs}
Z \equiv 
\le({\sqrt{-h} \ov f}\ri)^\ha = \le({r \rho \ov u}\ri)^{3 \ov 2}\ .
\end{equation}
Such a solution is the wave function for a scalar meson 
of type $s=1$ or $s=2$ with energy 
$\omega$ and wave vector $\vec k$ (note the plane wave
form for the dependence on (3+1)-dimensional Minkowski space
coordinates) and with quantum numbers $\ell$, $m$ and $\tilde m$
specifying the angular momentum spherical harmonic
on the ``internal'' three-sphere.
(Rotation symmetry of the three-sphere guarantees that
the quantum numbers $m$ and $\tilde m$ will not appear in any
equations.)
The $\psi_s(z)$ that we must solve for are the wave functions
of the meson states in the fifth dimension.  

\begin{figure}[t]
\begin{center}
$\begin{array}{cc} \psfig{figure=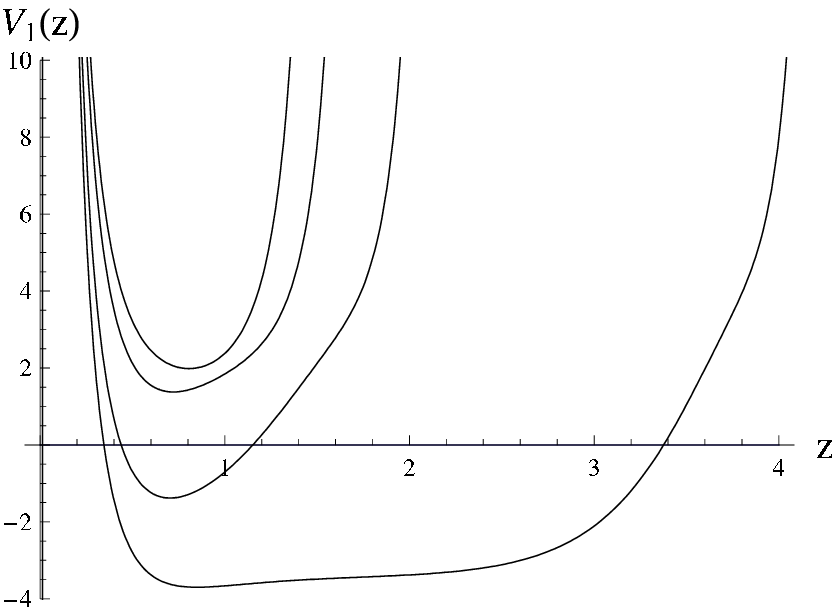,width=9cm,height=6cm}
&
\psfig{figure=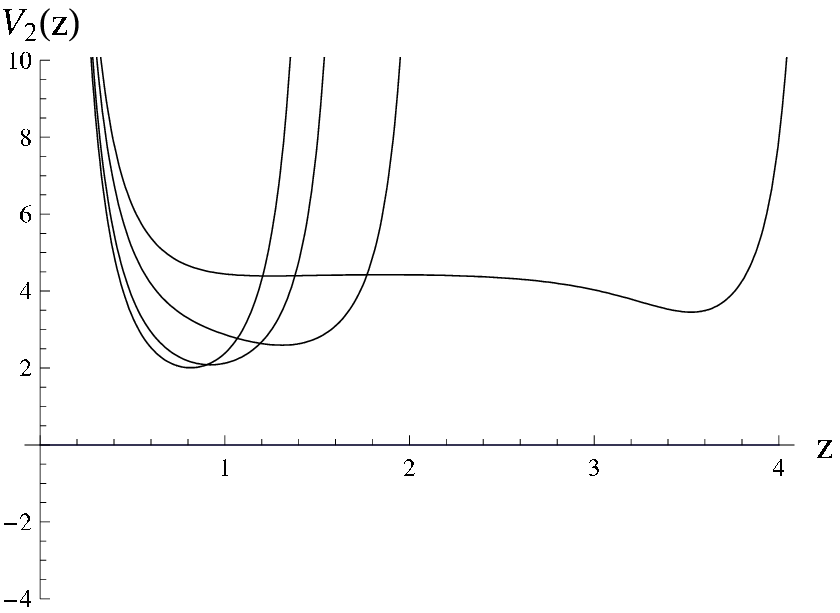,width=9cm,height=6cm} \\
\end{array}$
\caption{Potentials $V_s(z)$ for Minkowski
embeddings at various temperatures, 
all with $k=\ell=0$.  
The left (right) panel is for $s=1$ ($s=2$).
In each panel, the potentials are drawn for
$\ep_\infty=0.249$, 0.471, 0.586 and 0.5948, with the potential
widening as the critical embedding is approached,
i.e. as $\ep_\infty$ is increased. The $\ep_\infty=0.586$ potential
is that for the Minkowski embedding at the first order transition; the
widest potential shown describes the fluctuations of a metastable
Minkowski embedding very close to the critical embedding. The potential
becomes infinitely wide as the critical embedding is approached, but
it does so only logarithmically in $\ep_\infty^c-\ep_\infty$.
Note that the tip of the D7-brane is at $z=0$, on the left side
of the figure, whereas $\rho=\infty$ has been mapped to
a finite value of the tortoise coordinate $z=z_{\rm max}$, corresponding
to the ``wall'' on the right side of each of the potentials in the figure.
}
\end{center}
 \label{fig3}
\end{figure}

\begin{figure}[t]
\begin{center}
$\begin{array}{cc} 
\psfig{figure=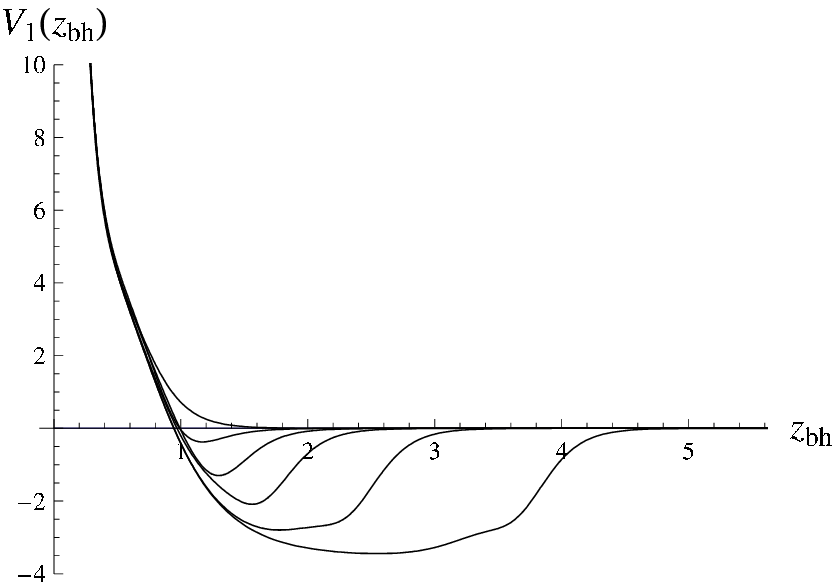,width=9cm,height=6cm}  
&
\psfig{figure=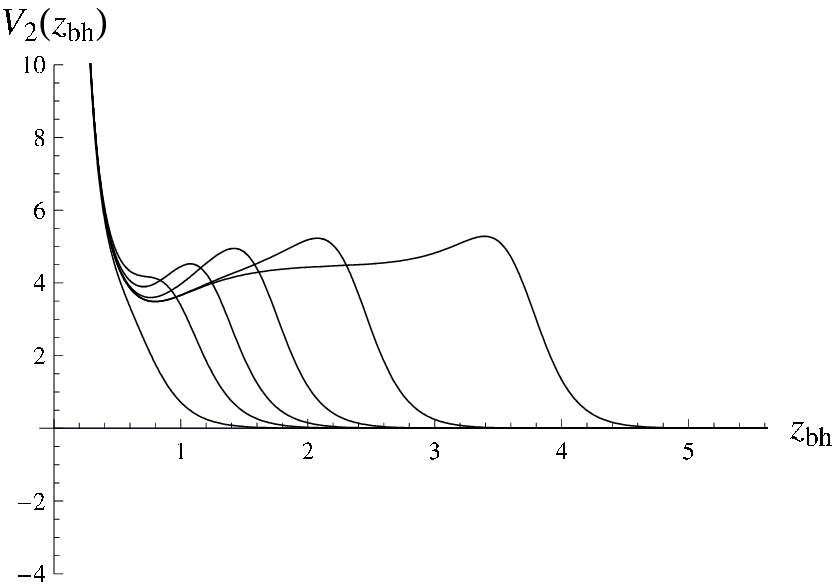,width=9cm,height=6cm} \\
\end{array}$
\caption{Potentials $V_s(z_{bh})$ for black hole
embeddings at various temperatures, 
all with $k=\ell=0$.  
The left (right) panel is for $s=1$ ($s=2$).
In each panel, the potentials are drawn for
$\ep_\infty=3584.$, 0.647, 0.586, 0.586,  0.5940 and 0.5948,
from narrower to wider,
with the potential
widening as the critical embedding is approached
from the right along the curve in Fig.~{\protect\ref{fig:epsepsinf}}.
Note that $z_{bh}$ is defined such that the horizon is
at $z_{bh}=\infty$, and $\rho=\infty$ is at $z_{bh}=0$.
The narrower (wider) of the two potentials with $\ep_\infty=0.586$ is
that for the stable (unstable) black hole embedding: at this $\ep_\infty$,
there is a first order transition (see Fig.~{\protect\ref{fig:epsepsinf}})
between the stable Minkowski embedding (whose
potential is found in Fig.~{\protect\ref{fig3}}) and the stable black hole
embedding.  The potentials at $\ep_\infty=0.5940$ and 0.5948
describe fluctuations of metastable black hole embeddings, with the
latter being a black hole embedding very close to the critical embedding.
\label{fig3BH} }
\end{center}
\end{figure}

The reasons for
the introduction of the tortoise coordinate $z$ 
and the ansatz (\ref{wavefunctionansatz}) for the form of the solution
become apparent when we discover that the equations of motion 
(\ref{eq:kg}) now take the Schr\"odinger form
\begin{equation} \label{eow}
- \frac{\p^2}{\p z^2 } \psi_s + V_s(k,z) \psi_s = \omega^2 \psi_s\ ,
\end{equation}
with potentials for each value of $k=|{\vec k}|$ and for each of
the two scalar mesons labelled by $s=1,2$ given by
\begin{equation} \label{pot}
V_s(k,z) = {Z'' \ov Z} + f m_s^2 + {f k^2 \ov r^2}
 +\frac{l(l+2)  f u^2}{\rho^2}\ .
\end{equation}
Here, the prime denotes differentiation with respect to $z$.
Recall that $u^2 = \rho^2 + y_0^2 (\rho)$ and it should be
understood that $\rho$, $u$, and $y_0$ 
are all functions of the tortoise
coordinate $z$. In Figs.~\ref{fig3} and \ref{fig3BH}, we provide
plots of $V_s(z)$ with $k=\ell=0$ for $s=1,2$ and for
Minkowski (Fig.~\ref{fig3})) and black hole (Fig.~\ref{fig3BH})
embeddings at various temperatures.
With the tortoise coordinate $z$ defined as we have described,
in a Minkowski embedding 
$z$ extends from $z=0$, which corresponds to the tip of the D7-brane, to
\be
z=z_{\rm max}\equiv\int_0^\infty \frac{d\rho}{u}\sqrt{\frac{1+y_0'(\rho)^2}{f(u)}}\ ,
\label{tortoiseintegral}
\ee
which corresponds to $\rho=\infty$. Here,
$u(\rho)$ and $f(u)$ are given in (\ref{vqu}).  This defines
the width of the potentials for the Minkowski embeddings shown 
in Fig.~\ref{fig3}, which get wider and wider as the critical embedding
is approached.  

If we used the same tortoise coordinate for the black
hole embeddings, the lower limit of the integral (\ref{tortoiseintegral})
is then the $\rho$ at which $y(\rho)$ intersects the horizon
and $f(u)$ vanishes, making the integral divergent.  This means
that $\rho=\infty$ is mapped onto $z=\infty$ for black hole embeddings.
It is more convenient to define $z_{bh}$ by first
choosing the integration constant such that $\rho=\infty$ corresponds
to $z_{bh}=0$, and then multiplying by -1.  This is the tortoise
coordinate that we have used in Fig.~\ref{fig3BH}

The qualitative implications for the meson spectrum
can be inferred immediately from Figs.~\ref{fig3} and \ref{fig3BH}, since
we have intuition for solutions of the Schr\"odinger equation.
We can see immediately that the Minkowski embeddings all
have a discrete spectrum of meson excitations, while the
fluctuations of the 
black hole embeddings all have continuous spectra.
This justifies the identification of the first order 
phase transition from Minkowski to black hole embeddings
that we described in Section 3 as the transition at which
mesons dissociate.

Other phenomena that are discussed quantitatively
in~\cite{Mateos:2007vn,Hoyos:2006gb,Myers:2007we} can be inferred qualitatively
directly from the potentials in Figs.~\ref{fig3} and \ref{fig3BH}.
For example we see from the left panel in Fig.~\ref{fig3BH} that,
in addition to the continuous spectrum characteristic
of all black hole embeddings,  those embeddings that are close to the critical
embedding will have discrete bound states 
for the $\psi_1$ fluctuations.  These bound states will always have
negative mass-squared, 
representing an instability.  This instability arises only in
a regime of temperatures at which the black hole embeddings
already have a higher free energy than the stable Minkowski
embedding, that is, at temperatures below the first order
transition~\cite{Mateos:2007vn}.  They therefore represent an
instability of the branch of the spectrum that was already
metastable.   Similarly, the left panel of Fig.~\ref{fig3}
shows that Minkowski embeddings close to the critical
embedding also have negative mass-squared bound states;
again, this instability only occurs for embeddings that were
already only metastable~\cite{Mateos:2007vn}.
We see from the right panel
of Fig.~\ref{fig3BH} that resonances may also occur in the $\psi_2$ channel
for the black hole embedding. They are interpreted as quasi-normal modes;
close to the transition these resonances become more well defined
and may be interpreted as quasi-particle meson 
excitations~\cite{Hoyos:2006gb,Myers:2007we}.


\section{Dispersion relations}

We have now laid the groundwork needed to evaluate the
dispersion relations for the $\psi_1$ and $\psi_2$ scalar mesons, 
corresponding in the gravity dual
to fluctuations of the position of the D7 brane.  These 
fluctuations are governed by (\ref{eow}), which are Schr\"odinger
equations with the potentials  $V_1(k,z)$ or $V_2(k,z)$
given by (\ref{pot}) and (\ref{effM}) 
and depicted in Fig.~\ref{fig3}.  The eigenvalues of
these Schr\"odinger equations are $\omega^2$  for the mesons.
So, it is now a straightforward numerical task to find the 
square root of the eigenvalues
of the Schr\"odinger equation with, say, potential $V_1(k,z)$, at a 
sequence of values of $k$.  At $k=0$, this will reproduce the
results that we reviewed in Section 3.2.  As we increase $k$,  we
map out the dispersion relation $\omega$ of each of the $\psi_1$ 
mesons.
In Fig.~\ref{fig:NumericalResults} in Section~\ref{sect:NumericalResults}
below, we show the dispersion relations for the ground state $\psi_1$ meson
at several values of the temperature.   Such dispersion relations
have also been obtained numerically in~\cite{Mateos:2007vn}.
In order to more fully understand the dispersion equations, and their
implications, we shall focus first on analytic results.
The potentials are complicated enough that we
do not have analytic solutions for the general case.
We shall show, however, that in the low temperature and/or
the large-$k$ limit, the equations simplify sufficiently that
we can find the dispersion relations analytically.
It is the large-$k$ limit that is of interest to us, but
it is very helpful to begin first at low temperatures,
before then analyzing the dispersion relations in the 
large-$k$ limit at any temperature
below the dissociation temperature.

Readers who are only interested in the final results can proceed 
directly to Section~\ref{sect:DispRelSummary}, where we summarize 
and discuss our
central results for the dispersion relations.

\subsection{Low temperature}

At low temperature, $\eps \ll 1$, the D7-branes are far from
the horizon of the black hole.   In this regime, 
we can expand various quantities that
occur in the potentials (\ref{pot}) as power
series in $\eps^2$.    We shall then be able to
determine the dispersion relations analytically 
to order $\eps^2$ in 
two limits: (i) $\eps\rightarrow 0$ at fixed $k$, meaning
in particular that $\eps k \rightarrow 0$; and (ii) $\eps\rightarrow 0$
at fixed, large, $\eps k$, meaning that $k\rightarrow \infty$ as  $\eps\rightarrow 0$.

We begin by seeing how  the equation (\ref{eq:em}) that determines the embedding 
$y(\rho)$ in the absence of fluctuations simplifies at small $\eps$.
Expanding $y (\rho)$  as a power series in $\eps$, one immediately
finds that $y(\rho)$ is modified only at order $\eps^4$, i.e.
 \be \label{Eje}
 y (\rho) = 1 + \OO(\eps^4)\ ,
 \ee
 which in turn implies that
\be 
 \ep_\infty = \eps \le(1 + \OO(\eps^4) \ri)\ .
 \ee
Thus, if we work only to order $\eps^2$, we can
treat the embedding as being $y(\rho)=1$,  as at zero temperature,
and can neglect the difference between $\eps$ and $\ep_\infty$ (which is to
say the difference between $y(0)$
and $y(\infty)$). 
From (\ref{vqu}), then,
 \begin{align}
 u^2 = 1 + \rho^2 + \OO(\eps^4) , \qquad
 f(u)\approx u^2 \le(1- {3\eps^2\ov u^4}+\OO(\eps^4) \ri) \ .
 \end{align}
By expanding the curvature invariants in (\ref{effM}) to order 
$\eps^2$, we find that
\be
m_1^2=m_2^2 = - \frac{4 + 3 \rho^2}{1+\rho^2} + \OO(\eps^4)\ ,
\ee
meaning that to order $\eps^2$
the mass terms occurring in (\ref{eq:em}) are as
in (\ref{epp}) at zero temperature. 
Next, we expand the tortoise coordinate (\ref{tor}), finding
 \begin{align}
 z &=\tan^{-1}{\rho} + \eps^2 g(\rho)+\OO(\eps^4),
   &\text{with}\;\;
    g(\rho)&={3\ov 16}\lt( 3\tan^{-1}{\rho}
           + {\rho(5+3\rho^2)\ov(1+\rho^2)^2}\rt)\ .
           \label{eq:tse1}
 \end{align}
We can then invert (\ref{eq:tse1}) to obtain $\rho$ in terms of $z$:
\begin{align} \label{ejsp}
\rho&=\tan z - \eps^2 {g(\tan z)\ov\cos^2 z}+\ldots \ .
\end{align}
Using these equations, we find that the potential (\ref{pot})
is given to order $O(\eps^2)$ by
\begin{align} \label{nepo}
V(z)= k^2 +  V^0(z)  - 4 \eps^2 k^2 \cos^4 z  +  \eps^2
h(z)+\OO(\eps^4, \eps^4 k^2),
\end{align}
where
\begin{align} \label{sbn}
V^0(z)&\equiv  {4 \al_\ell \ov\sin^2 2z}-1, \quad {\rm with} \quad \al_\ell \equiv {3 \ov 4} +
\ell(\ell+2)
\end{align}
is the potential at zero temperature, and 
\be
h(z)=\frac{ 3 \al_\ell \left( \sin^2(2z) + 6 z \cot(2z) - 3 \right)}
{2 \sin^2 (2z)} + \frac{9}{4} \sin^2 (2z)\ .
\ee
We shall not use the explicit form of $h(z)$ in the following.

\subsubsection{Low temperature at fixed $k$}

At zero temperature ($\eps=0$), solving the Schr\"odinger equation
(\ref{eow}) with potential $V^0 (z)$ yields the eigenvalues (and
hence the dispersion relations)
 \be
 \om^2 - k^2 = m_{n\ell}^2, \qquad n=1,2, \ldots\ , \quad l=0,1, \ldots\ ,
 \ee
with $m_{n\ell}$ given  by (\ref{mesp}) (after restoring its dimensions).   
If we work in the limit $\eps\rightarrow 0$ with $k$ fixed, then both
the $\OO(\eps^2)$ and the $\OO(\eps^2 k^2)$ terms 
that describe the effects of nonzero but small temperature in the
potential (\ref{sbn}) can be treated using quantum mechanical
perturbation theory.  
To first order in $\eps^2$, the
dispersion relation becomes
 \be \label{doss}
 \om^2 = v_{n\ell}^2 k^2 + m_{n\ell}^2 + \eps^2 b_{n\ell} + \OO(\eps^4)
 \ee
with
 \bea \label{Kew}
 v_{n\ell}^2 &=& 1-a_{n\ell}\, \eps^2 \ ,\nonumber \\
  a_{n\ell} &=& 4  \langle n,\ell |\cos^4 z | n,\ell\rangle\ , \nonumber\\
  b_{n\ell} &=& \langle n,\ell | h(z) | n,\ell\rangle\ ,
 \eea
where $|n,\ell\rangle$ are the eigenfunctions of the Hamiltonian
with the unperturbed potential $V^0$ of (\ref{sbn}), with wave functions
\be\label{UnperturbedWaveFn}
\psi_{n\ell}^0(z) = \Gamma\left(\ell+\frac{3}{2}\right) 2^{1+\ell}
\sqrt{\frac{ n\! \left(n+\ell+\frac{3}{2}\right)}
{\pi \Gamma\left(n+2\ell+3\right)}}
\left(\sin z \right)^{\frac{3}{2}+\ell} C_n^{(\ell+\frac{3}{2})}(\cos z)\ .
\ee
Using the recursion relations for the generalized
Gegenbauer polynomials $C_n^{(\alpha)}$~\cite{AbramowitzStegun},
$a_{n\ell}$ can be evaluated analytically, yielding
\be\label{anlresult}
a_{n\ell}=
 2-{(n+2l+1)(n+2l+2)\ov 4(n+l+1/2)(n+l+3/2)}
                    -{(n+1)(n+2)\ov 4(n+l+3/2)(n+l+5/2)} \ .
\ee
So, for the ground state with $n=\ell=0$, $a_{00}=18/15$.
$b_{n\ell}$ can be computed numerically, but we will not do so here.
The dispersion relation (\ref{doss}) is valid for $\eps^2\ll1$ and $\eps^2k^2\ll1$,
meaning that at small $\eps$ it is valid for $k\ll1/\eps$.
No matter how small $\eps$ is, the perturbation
theory breaks down for $ k \sim {1 \ov \eps} $ and (\ref{doss})
does not apply. In other words, the low temperature $\eps \to 0$
limit and the high meson momentum $k \to \infty$ limits do not commute. 
Even though (\ref{doss}) cannot be used to determine the meson velocity 
at large $k$, it is suggestive.
We shall see below that in the large-$k$ limit, the meson velocity
is indeed $1- \OO(\eps^2)$, but the coefficient of $\eps^2$ is not
given by (\ref{anlresult}).  

\subsubsection{Low temperature at fixed, large, $\eps k$}

\FIGURE{
\centerline{\hbox{\psfig{figure=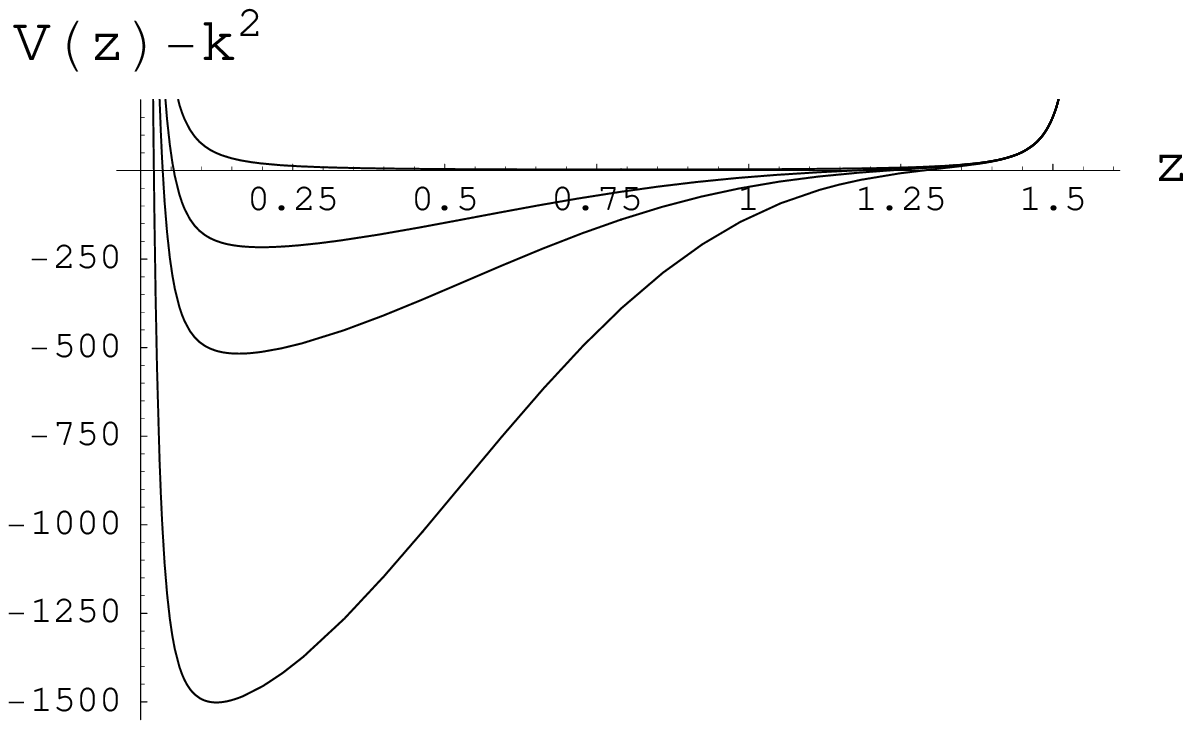,width=11cm}}}
\caption{The potential (\ref{pes}) with $\eps=0.756$ and $k=5$, 20 and 100.
We see that as $\Lambda=\eps^2 k^2$ increases, the minimum
of the potential moves towards $z=0$, the potential deepens,
and the curvature around the minimum increases. 
 \label{fig:pot-Lambda}
 }}

To explore the behavior of the dispersion relations in the
large-$k$ limit, we now consider the following scaling limit
 \be \label{adsL}
 \eps \to 0, \qquad k \to \infty, \qquad {\rm with } \qquad \Lam^2 = k^2 \eps^2 = {\rm
 finite.}
 \ee
In this limit, the potential (\ref{pot}) again greatly simplifies and, consistent
with (\ref{nepo}), becomes
 \be \label{pes}
 V (z) = k^2 + {4 \al_\ell \ov\sin^2 2z}-1 - 4 \Lam^2 \cos^4 z \ .
 \ee
This potential is valid in the limit (\ref{adsL}) for 
any value of $\Lambda$, small or large. If $\Lambda$ is small,
the dispersion relation can be determined using perturbation
theory as before, yielding (\ref{doss}) without the $\eps^2b_{n\ell}$ term.
In order to analyze the large-$k$ regime, we now consider $\Lambda \gg 1$,
and seek to evaluate the dispersion relation as an expansion in $1/\Lambda$.
For this purpose, we notice that as $\Lam \to \infty$
the potential (\ref{pes}) develops a minimum at
 \be
 z_0 = \le({\al_\ell \ov 8 \Lam^2}\ri)^{1 \ov 4}
  \to 0 \quad {\rm for}\ \ \Lam \to \infty\ ,
 \ee
as depicted
in Fig.~\ref{fig:pot-Lambda}. 
The curvature about the minimum is $V'' (z_0) \propto \Lam^2$.
Thus, if we imagine watching how
the wave function changes as we take the 
large-$\Lam$ limit, we will see the wave function getting
more and more tightly localized around the point $z=z_0$
which gets closer and closer to $z=0$.  That is, 
the wave function will be localized around the tip of the
brane $z=0$. This motivates us to expand the potential around
$z=0$, getting
 \be
 V (z) - k^2 + 1 = \al_l \le({1\ov z^2} + {4 \ov 3} + {16 z^2\ov 15} + \ldots \ri)
 - 4 \Lam^2 \le(1 - 2z^2 + {5 z^4 \ov 3} + \ldots \ri) \ .
 \ee
If we now introduce a new variable $\xi = \Lam^\ha z$, the
Schr\"odinger equation (\ref{eow}) becomes
 \be \label{dke}
 \le(-\p_\xi^2 + {\al_\ell \ov \xi^2} + {1 \ov 4} \Om^2 \xi^2 \ri)  \psi + \tilde V \psi = E \psi
 \ee
where
 \be\label{Omega32}
 \Om^2 = 32, \qquad  E = {1 \ov \Lam} (\om^2 - k^2 + 4 \Lam^2 )\ , 
  \ee
 and
$\tilde V$ contains only  terms that are  higher order in $1/\Lam$:
 \be
 \tilde V = {1 \ov \Lam} \le({4 \al_\ell \ov 3}- 1 - {20 \ov 3} \xi^4 \ri) +
 \OO(1/\Lam^2) \ .
 \ee
Thus to leading order in the large $\Lam$ limit, we can drop the
$\tilde V$ term in (\ref{dke}).  Upon so doing, and using the
expression (\ref{sbn}) for $\al_\ell$,
the equation (\ref{dke}) 
becomes that of a harmonic oscillator in 4 dimensions with mass
$\ha$ and frequency $\Om$. This quantum
mechanics problem can be solved exactly, with
wave functions given by
 \begin{equation}
\label{eq:eig} \psi_{nl} =  \xi^{3/2 + \ell} L^{(\ell+1)}_\nu\left(\frac{\Omega}{2}\xi^2\right)
e^{- \frac{\Omega}{4} \xi^2}\ ,
\end{equation}
up to a normalization constant,  and with eigenvalues given by
 \be
 E_{n} = \Om (n+2), \quad n=0,1,\ldots
 \ee
In (\ref{eq:eig}), $L^{(\al)}_\nu$ is the generalized Laguerre polynomial of
order 
\be\label{nuDefn}
\nu =\frac{ n - \ell}{2}\ .
\ee
The allowed values of $\ell$ are determined by
the requirement that $\nu$ 
must be a non-negative integer. The
degeneracy of $n$-th energy level is $\frac{(n+3)!}{3! n!}$. Higher
order corrections in $1/\Lam$  can then be obtained using
perturbation theory. For example, with the next order correction
included, the degeneracy among states with different $\ell$ and
the same $n$ is lifted and the eigenvalues are given by
 \be
 E_{n\ell} = \Om (n+2) + {c_{n\ell} \ov \Lam} + \OO(1/\Lam^2)
 \ee
with
 \be
 c_{n\ell} = -{5 \ov 4} (n+2)^2 +{7 \ov 4} \ell(\ell +2) \ .
 \ee
Thus, 
in the small-$\eps$ limit with $\Lambda$ fixed and large,
we find using (\ref{Omega32})  that  the
dispersion relation becomes
 \be \label{Osa}
 \om^2_{n\ell} = (1- 4 \eps^2) k^2 + 4 \sqrt{2} (n+2)  k \eps + c_{n\ell}
 + O(1/k)\ .
 \ee
Notice that $c_{n\ell}$ is negative for the ground state, and indeed
for any $n$ at sufficiently small $\ell$.  We learn from this calculation
that in the large-$k$ limit, at low temperatures mesons move with a velocity given
to order $\eps^2$ 
by $v=\sqrt{1-4\eps^2}=1-2\eps^2$.   Recalling that to the order
we are working $\ep_\infty=\eps$, this result  can be expressed
in terms of $T$, $m_q$ and $\lambda$ using (\ref{epsinftydefn}).
In the next subsection,
we shall obtain the meson velocity at large $k$ for any $\eps$ .


\subsection{Large-$k$ dispersion relation at generic temperature}

The technique of the previous subsection can be 
generalized to 
analyze the dispersion relation in
the large-$k$ limit at a generic temperature below
the dissociation temperature. For general $\eps < 1$, one again
observes that the potential has a sharper and sharper minimum near
the tip of the brane $z=0$ as $k$ becomes larger and larger. Thus,
in the large $k$ limit, we only need to solve the Schr\"odinger
equation near $z=0$.

To find the potential $V(z)$ as a power series in $z$ near $z=0$,
we need to know the solution $y(\rho)$
of (\ref{eq:em}) near the tip of the
brane at $\rho=0$:
 \begin{align}
y=1+\frac{\rho^2}{\eps^{-4}-1}+\frac{\eps^4(5+5\eps^4-3\eps^8)}{3(\eps^4-1)^3}\rho^4+
\mathcal{O}(\rho^4)\ .
 \label{eq:yexpan}
\end{align}
At small $\rho$, using the expansion of $y$ in (\ref{eq:yexpan}),
we find the tortoise coordinate $z$ has the expansion
   \begin{align}
  z  ={\sqrt{1+\eps^{2}}\ov 1-\eps^2}\rho +
         \mathcal{O}(\rho^3).
         \label{eq:tort}
            \end{align}
Using (\ref{eq:yexpan}) and (\ref{eq:tort}) in (\ref{pot}), after
some algebra we find
 \begin{equation}
\label{eq:oned}
 V_s(z) = k^2 \le( v_0^2  + {1 \ov 4} \Omega^2 \eps^2 z^2 +
\beta_\ell  z^4 + \ldots \ri) + {\al_\ell \ov z^2} +  \gamma_{s\ell} + \OO(z^2) \ ,
\end{equation}
where
 \begin{eqnarray}
v_0 &=& \frac{1-\eps^2}{1+\eps^2} \,,\label{v0result}\\
\Omega^2 &=& \frac{ 32
 (1-\eps^2)^2 (1+\eps^4)}{(1+\eps^2)^5}\ ,\label{OmegaResult}\\
\beta_\ell &=  &  -  \Omega^2 \eps^2 \frac{  5 - 36 \eps^2 + 28 \eps^4 - 36
\eps^6 + 5 \eps^8}{24 (1+\eps^2)^3}\ , \\
\gamma_{1\ell} &=& \frac{ \ell(\ell+2) \left( \frac{4}{3} + 4 \eps^2 + \frac{4}{3}
\eps^4 + 4 \eps^6 + \frac{4}{3} \eps^8 \right) -56 \eps^4}{(1+\eps^2)^3}\ ,\\
\gamma_{2\ell} &=& \gamma_{1\ell}  + \frac{80 \eps^4}{(1+\eps^2)^3}\ ,
\end{eqnarray}
and where $\al_\ell$ is given by (\ref{sbn}).  
We can  understand why the leading difference between the 
potentials $V_1$ and $V_2$ for the mesons $\psi_1$ and $\psi_2$
arises in this approximation in the constant terms $\gamma_{1\ell}$
and  $\gamma_{2\ell}$ as follows.  We see
from (\ref{pot}) that the difference between $V_1$ and $V_2$ comes
only from $m_1^2$ and $m_2^2$, which do not enter multiplied by $k^2$
and so cannot affect $v_0$, $\Omega^2$ or $\beta_\ell$. Furthermore,
$m_1^2$ and $m_2^2$
are curvature invariants,
see (\ref{effM}), and must therefore be smooth as 
$\rho\rightarrow 0$ because for Minkowski embeddings the
D7 brane is smooth at $\rho=0$.  This means that $m_1^2$ and $m_2^2$
cannot affect the coefficient of $1/z^2$ in (\ref{eq:oned}).

We can now obtain the dispersion relations from the Schr\"odinger
equations with potentials (\ref{eq:oned}) as we did in the
previous subsection.
After making the rescaling $z = k^{-1/2} \xi$, the Schr\"odinger
equation (\ref{eow}) takes exactly the form (\ref{dke}), 
with
\be
E = {1 \ov k} (\om^2 - v_0^2 k^2)\ ,
\ee
where $\Om$ and $v_0$ are given by (\ref{v0result}) and (\ref{OmegaResult}) respectively,
and where $\tilde V_s(z)$ contains only terms that are subleading in
the $1/k$ expansion, and is given by
 \be
 \tilde V_s(z) = {1 \ov k} \le(\gamma_{s\ell} + \beta_\ell \xi^4 \ri) + \OO(k^{-2})\ .
 \ee
Thus, we find the large-$k$ dispersion relation
\begin{equation}
\label{eq:disp} \omega_s^2 = k^2 v_0^2 + k \Omega \eps (n+2) + d_{sn\ell} +
\OO(1/k)
\end{equation}
with
 \begin{align}
 d_{1n\ell} & = {1 \ov (1+
\eps^2)^3} \le[\frac{4}{3} \ell(\ell+2) \left( 1 + 3 \eps^2 +
\eps^4 + 3 \eps^6 +  \eps^8 \right)  \right. \nonumber \\
 & \qquad \left. - \le(
{5 \ov 4} -
 9 \eps^2 + 7 \eps^4 - 9 \eps^6 +{5 \ov 4} \eps^8 \ri) (n+2)^2
 -56 \eps^4 \ri]
  \label{BA}
 \end{align}
 and
  \be
   d_{2n\ell} = d_{1n\ell} + \frac{80 \eps^4}{(1+\eps^2)^3}\ .
  \label{BA2}
 \ee
Restoring dimensionful quantities in 
the dispersion relation (\ref{eq:disp}), i.e. undoing (\ref{eq:rce}), means
multiplying the $k$ and constant terms by $L_0/R^2$
and $L_0^2/R^4$, respectively.

We can easily obtain an explicit expression for
the wave functions themselves if we neglect the $\beta_\ell$,
$\gamma_{s\ell}$ and higher order terms, as the potential
(\ref{eq:oned}) is then that in the radial wave equation for
a four-dimensional harmonic oscillator.
To this order, the wave functions are given up to a normalization constant by
 \begin{equation}
\label{eq:eig2} \psi = z^{3/2 + \ell} L^{(\ell+1)}_\nu \le(\ha \Om \eps k z^2
\ri) \exp\lt( -\frac{1}{4} \Omega \eps k z^2 \rt) \ ,
\end{equation}
where, as before, $\nu=(n-\ell)/2$ is the order of
the generalized Laguerre polynomial $L_\nu^{(\ell+1)}$.

The dispersion relations (\ref{eq:disp}) are the central 
result of Section 5.  We shall analyze (\ref{eq:disp}) and
discuss its consequences at length in Sections 5.4
and 6.  First, however, we close this
more technical discussion with a few remarks related
to the approximation that we have used to obtain
the large-$k$ dispersion relations:

\begin{enumerate}

\item 
The wave function is localized at the
tip of the brane, near $\rho = 0$ which is the 
fixed point of the $SO(4)$ symmetry at which the $S^3$ shrinks
to zero size and the fluctuations are fluctuations in $R^4$.
This is the reason why we find a four-dimensional harmonic oscillator.

\item 
Our approximation is valid for wave functions that are tightly
localized near $z=0$.  Evidently, this approximation must break
down for mesons with high enough $n$, whose wave functions explore more of
the potential.  More precisely,
if we increase $n$ and $\ell$ while keeping $\nu$ fixed and small,
the wave functions are peaked at 
$z_0 \sim \le({n \ov k \Om\eps } \ri)^\ha$ with 
a width ${1 \ov (k \Om\eps)^\ha}$.
Or, if we increase $n$ and $\nu$ while keeping $\ell$ fixed and small,
the wave functions become wider, with $\nu$ oscillations over
a range of $z$ from near zero to 
near $z_0\sim \le({n \ov k \Om\eps } \ri)^\ha$ and hence a wavelength
$\sim {1 \ov (n k \Om\eps)^\ha}$.
In either case, 
our approximation must break down for $n \sim k$,
since for $n$ this large $z_0$ is no longer small and
the wave function is no longer localized
near $z=0$.

\item 
We must ask at what $k$  (or, at what $\omega$)
stringy effects that we have neglected throughout may become important
in the dispersion relations for the mesons that we have analyzed.
We can answer this question by comparing
the length scale over which the meson wave functions that we have computed
varies to the string length scale $\alpha'^{\frac{1}{2}}$.
Considering first the case where $\nu$ is small,
we see from (\ref{eq:mtort}) that
the proper distance between the maximum of the wave function
at $z=z_0$ and the tip of the brane at $z=0$ is
 \be \label{Epw}
 l_0 \sim \sqrt{f(0)}\, R\, z_0 \sim \frac{1-\eps^2}{\sqrt{1+\eps^2}} \,R \le({4n \ov k
\Om\eps} \ri)^\ha
 \ee
and the width of the wave function is
\be
\delta l \sim \frac{1-\eps^2}{\sqrt{1+\eps^2}}\, R
\le({1 \ov k \Om\eps} \ri)^\ha\ .
\ee
Stringy effects can be neglected as long as 
$\delta l \gg  \alpha'^{\frac{1}{2}}$, meaning 
 \be\label{Stringyk}
  k  <  \OO(\lam^{1 \ov 4} M)\ , 
 \ee
 where in the last expression we have restored the dimensions of
 $k$ using (\ref{msma}) and (\ref{eq:rce}).  (Since $\omega = v_0 k$ at
 at large $k$, this parametric criterion is the same for $\omega$ as for $k$.)
 If $\nu$ is large, the wavelength of the wave function should be compared
 to $\alpha'^{\frac{1}{2}}$ meaning that $\delta l$ is reduced by a factor
 $\sim 1/\sqrt{\nu}$ and stringy effects can be neglected only as long as
   \be\label{Stringyk2}
  k  <  \OO(\lam^{1 \ov 4} M/\nu)\ . 
 \ee 
 We can conclude from either (\ref{Stringyk}) 
 or (\ref{Stringyk2}) that we are justified
 in using the dispersion relation that we have derived in the $k\rightarrow\infty$
 limit, as long as we take the $\lambda\rightarrow\infty$ limit 
 first.\footnote{Recall that although the mesons that we have focussed on
 have masses
 of order $M\sim m_q/\sqrt{\lambda}$, there
are also higher-lying stringy mesonic excitations with masses
of order $M\lambda^{\frac{1}{4}}\sim m_q/\lambda^{\frac{1}{4}}$.    
Requiring $\lambda^{1/4}$ to be large is what justifies
our neglect of these stringy mesonic excitations, just as it justifies our neglect
of stringy corrections to the dispersion relations of the low-lying
mesons.  Note also that the latter
becomes important at an $\omega$ of order the mass of the
former.}

\item Notice that as $\eps \to 1$ (i.e. approaching the critical embedding),
both $v_0$ and $\Om$ vanish. Our approximation will therefore break
down at the critical
embedding. (One way to see this is to note that in 
the leading terms in (\ref{eq:oned}) we will
then have zero times infinity, meaning that it is no longer obvious
that these {\it are} the leading terms.)
However, the first order phase transition occurs at $\eps = 0.756$, long before this happens.

\end{enumerate}

\subsection{Numerical results}\label{sect:NumericalResults}

\begin{figure}[h!]
$\begin{array}{cc} 
\hspace{-2cm} \psfig{figure=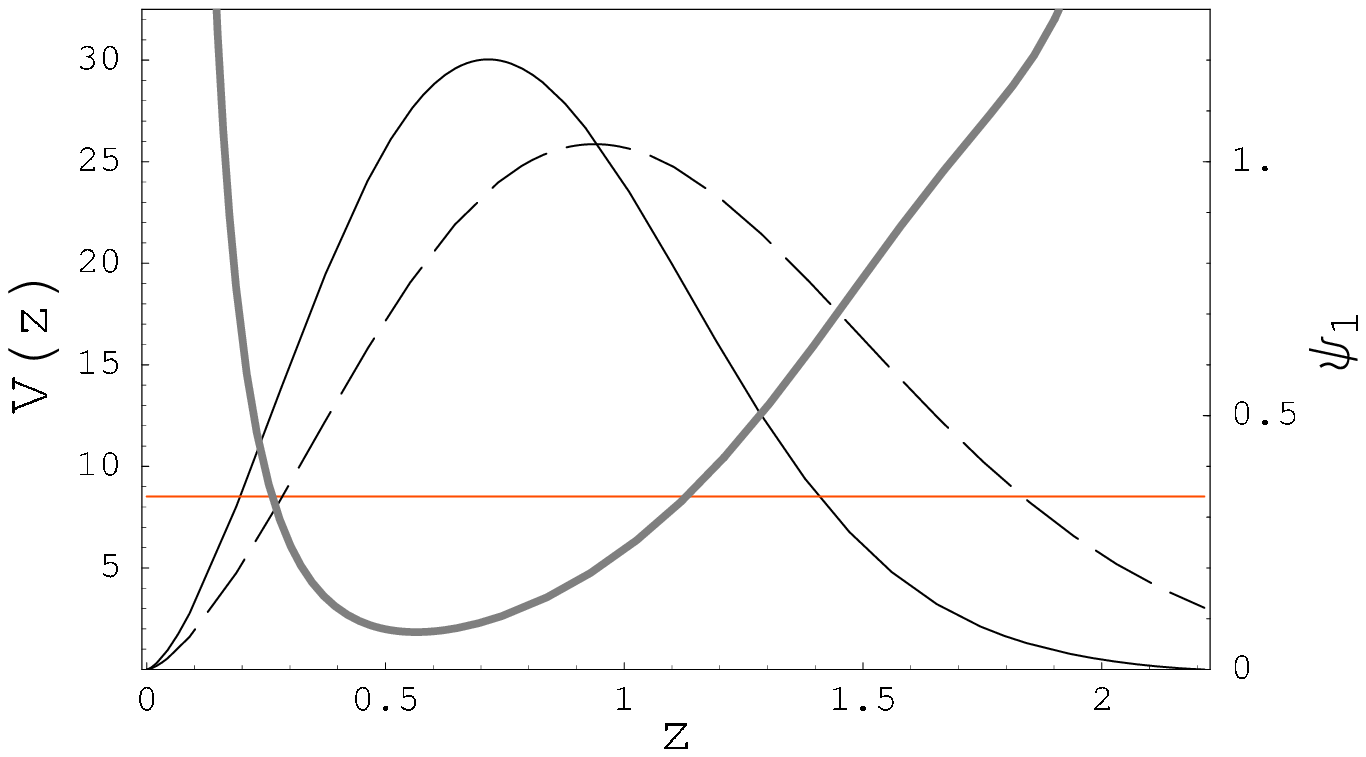,width=10.5cm,height=5.7cm} &
\hspace{-1.5cm}\psfig{figure=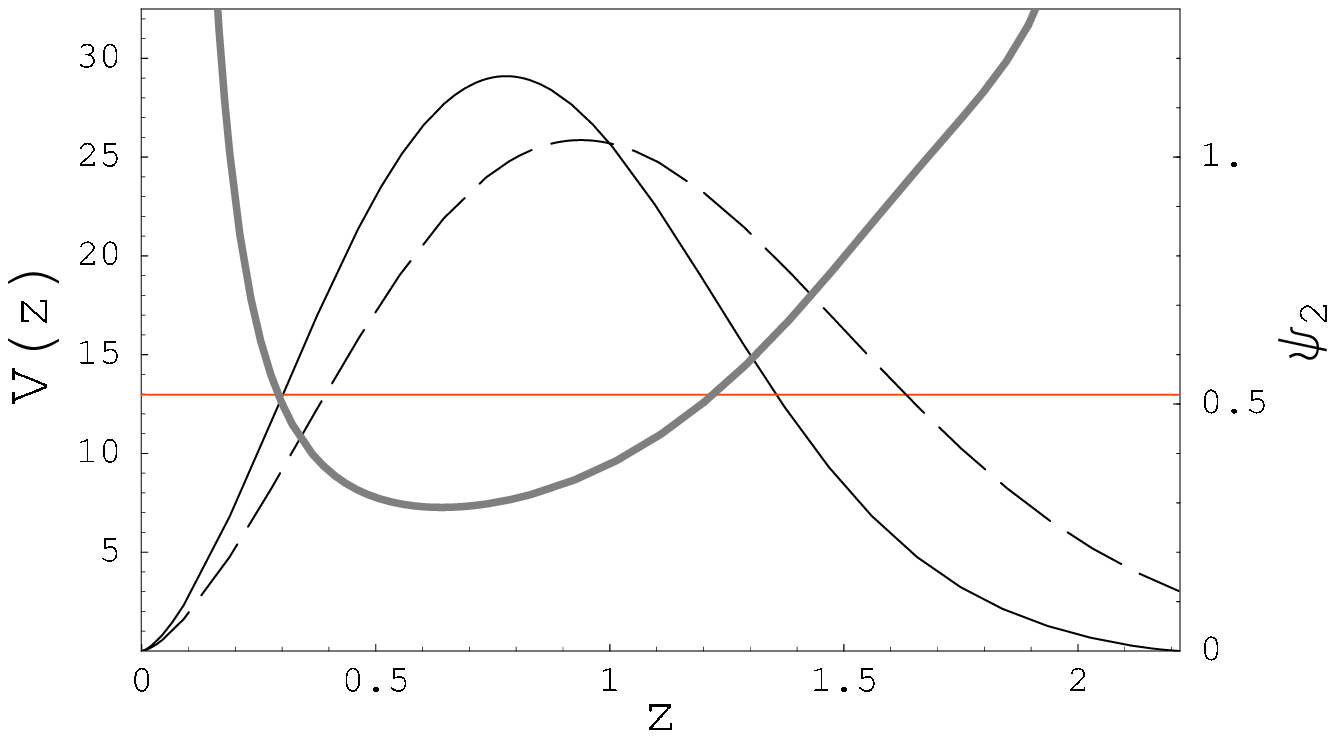,width=10.5cm,height=5.7cm} \\
 \hspace{-2cm}\psfig{figure=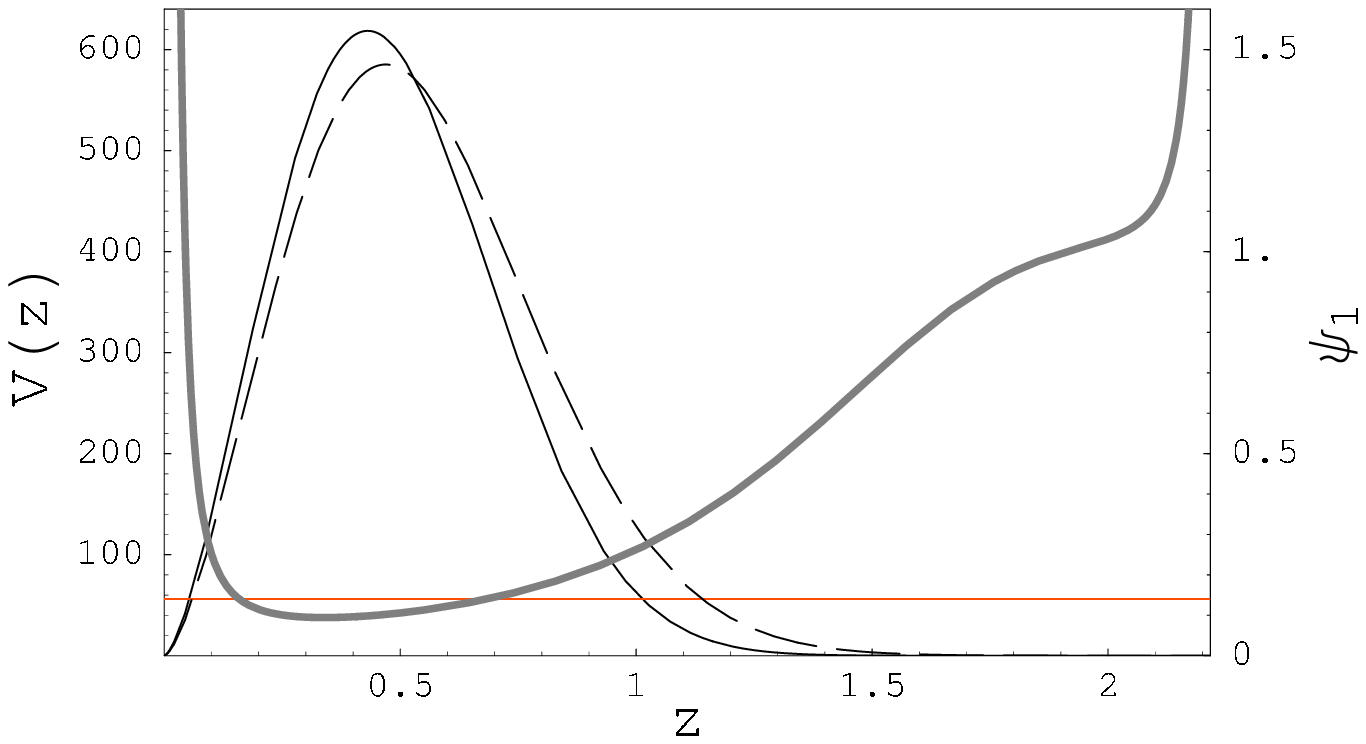,width=10.5cm,height=5.7cm} &
\hspace{-1.5cm}\psfig{figure=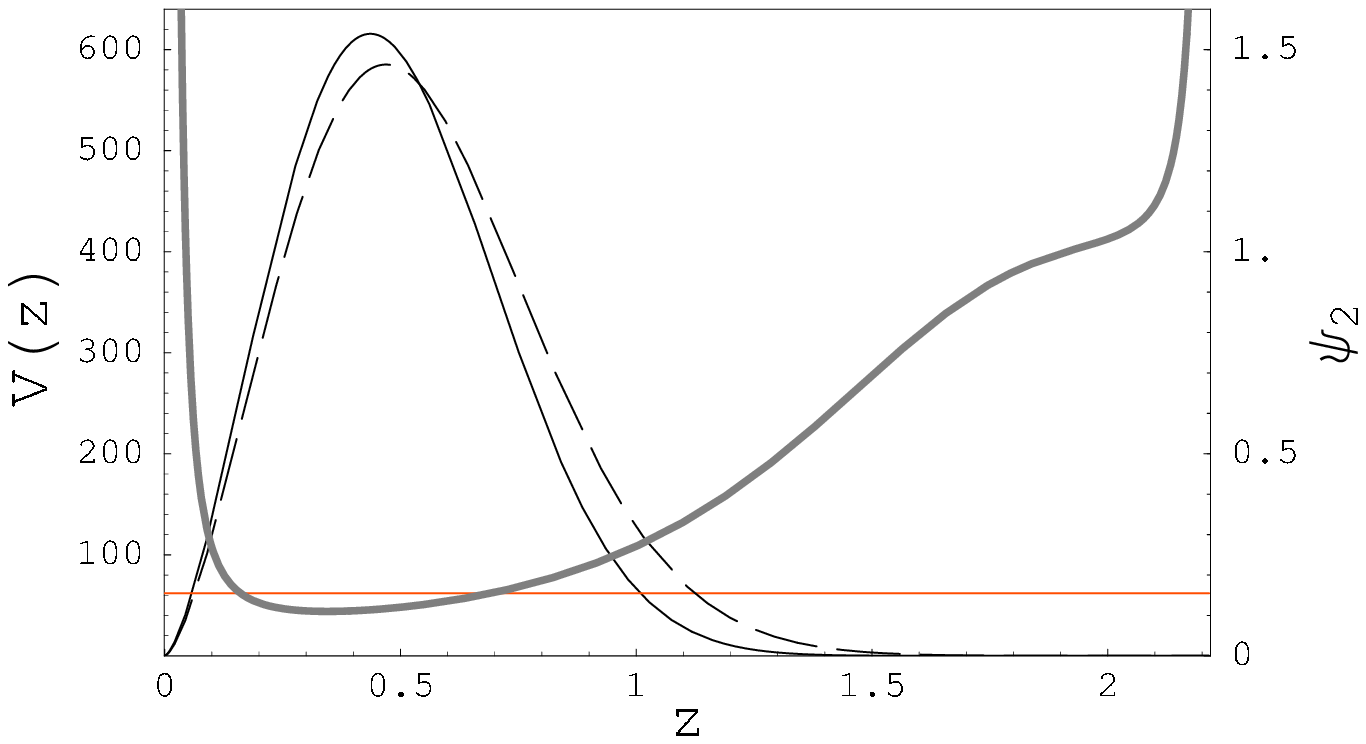,width=10.5cm,height=5.7cm} \\
 \hspace{-2cm}\psfig{figure=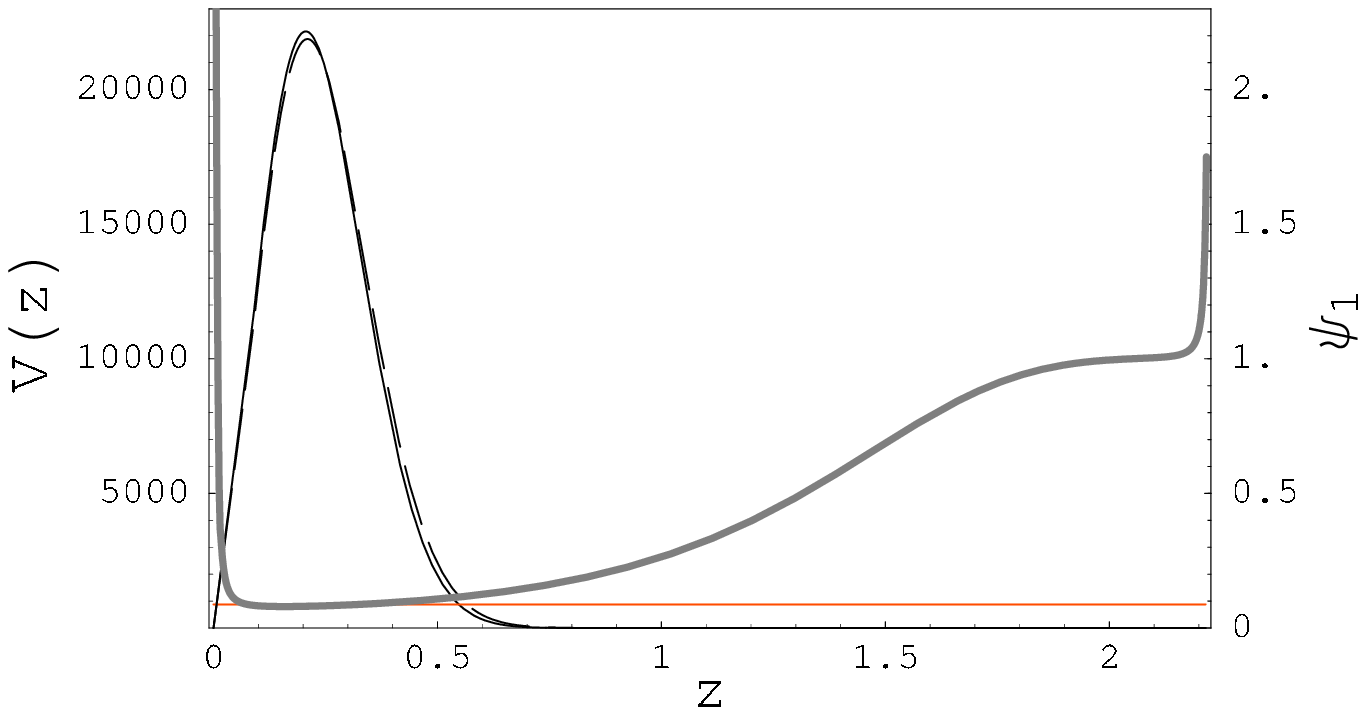,width=10.5cm,height=5.7cm} &
\hspace{-1.5cm}\psfig{figure=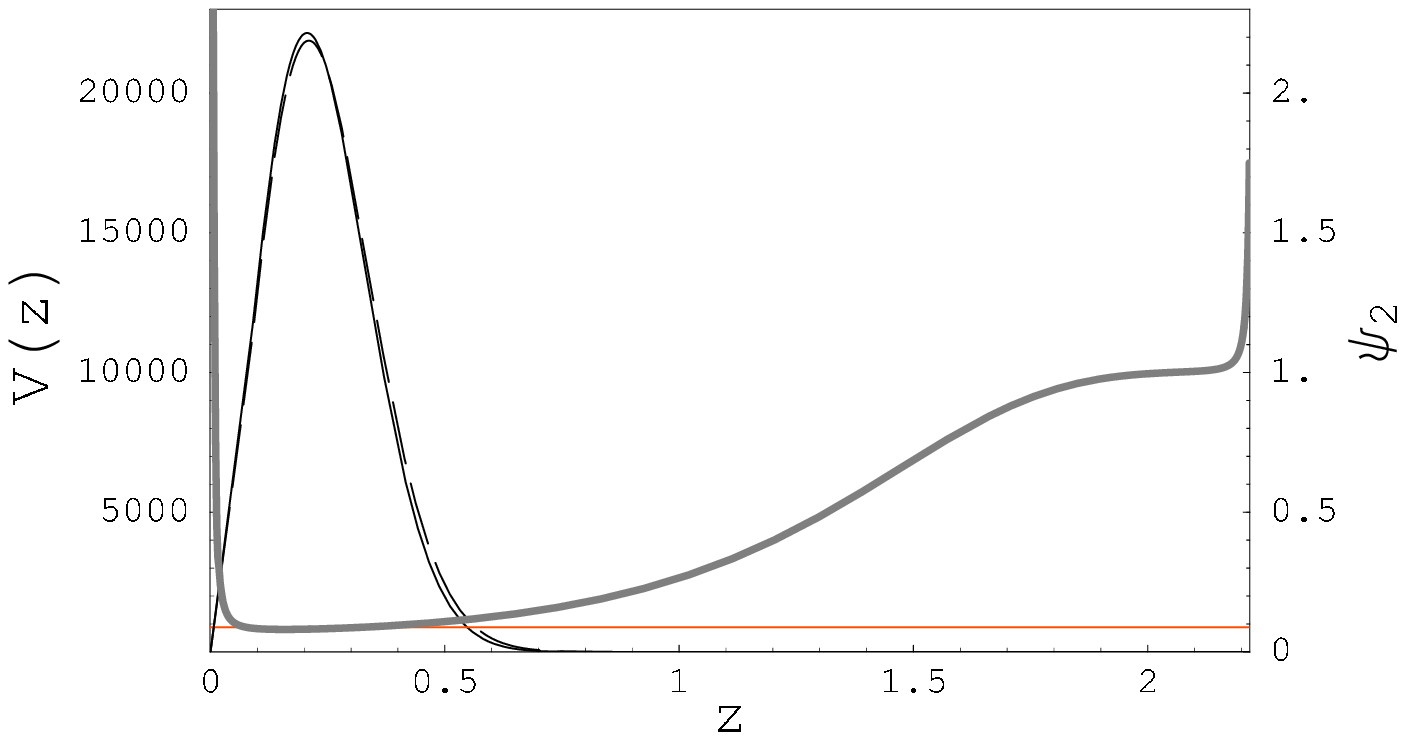,width=10.5cm,height=5.7cm}
\end{array}$
\caption{Potential and ground state wave function for 
$\psi_1$ (left three panels)  and $\psi_2$ 
(right three panels) for $k$ given by 5, 20 and 100 (top to bottom).
All plots have $\eps=0.756$, corresponding to the
Minkowski embedding at the dissociation transition.  $V(z)$ 
and the ground state ($n=\ell=0$) solutions
to the Schr\"odinger equation in the potentials $V$ 
are both shown as solid lines, and the ground 
state energies are indicated by the horizontal (red)
lines.  The dashed lines
show the approximation 
(\protect\ref{eq:eig2}) to the wave functions. 
\label{fig:num}}
\end{figure}

We can also obtain the meson wave functions and dispersion
relations numerically, without making either a small $\eps$
or a large-$k$ approximation.  In this subsection we plot a few
examples of such results, and compare them to
the analytic expressions that we have derived above upon making
the large-$k$ approximation.

In Fig.~\ref{fig:num} we plot the potentials (\ref{pot}) and 
ground state wave functions for those potentials that
we have obtained numerically for three values of $k$.   
Note the changing vertical
scale in the plots of $V$; as $k$ increases, $V$ 
deepens.  We 
see that as $k$ increases and the potential deepens, 
the wave function gets
more and more localized near $z=0$ 
and, correspondingly,
the expression (\ref{eq:eig2}) for the
wave function that we have derived in the large-$k$ limit
using the fact that the wave function becomes localized becomes
a better and better approximation to the exact wave function.

\FIGURE[t]{
\centerline{\hbox{\psfig{figure=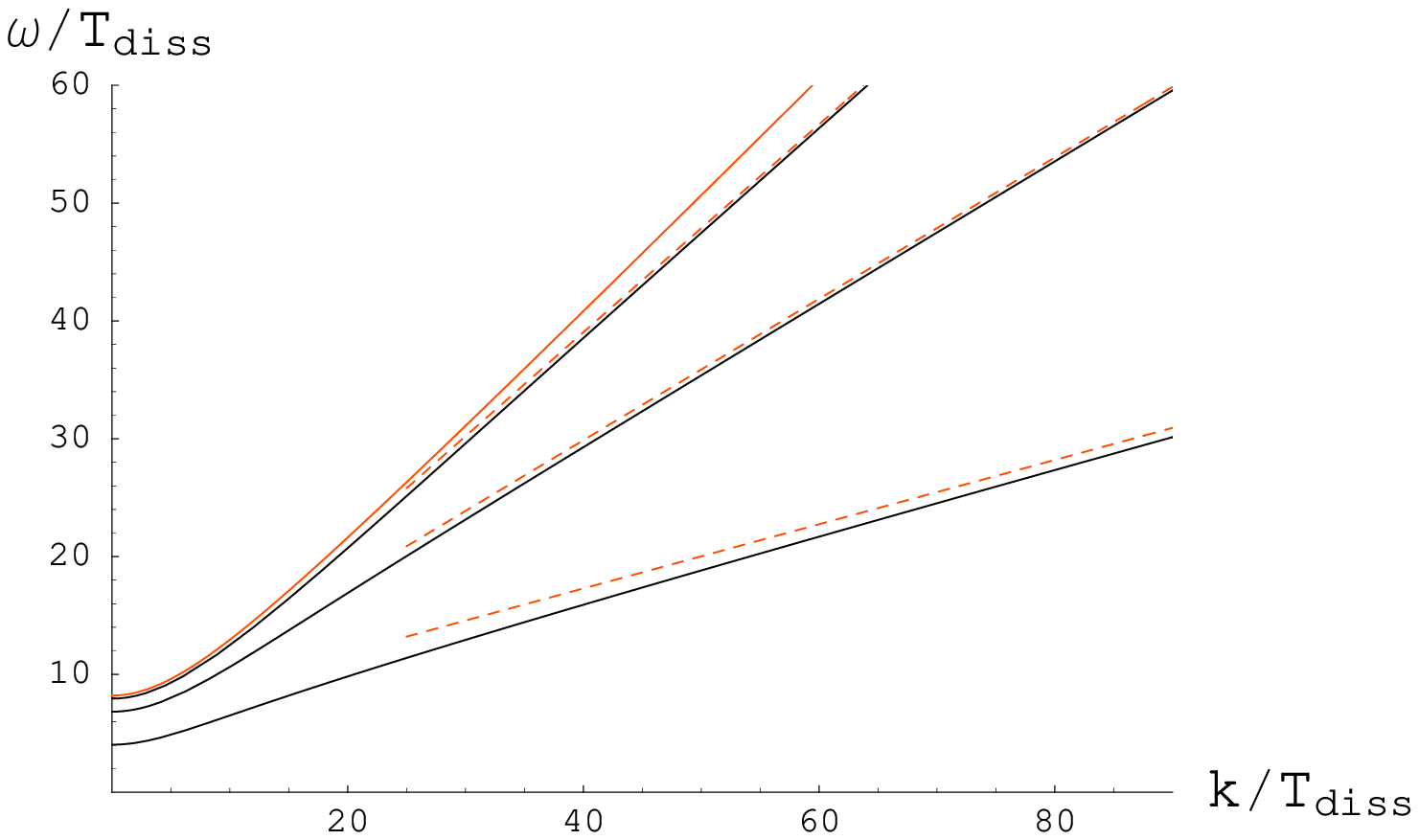,width=15cm}}}
\caption{Dispersion relations for the 
ground state $\psi_1$ meson with $n=\ell=0$  at
various values of $\eps$ (i.e. at various temperatures).  
The top (red) curve is the zero temperature dispersion
relation $\omega=\sqrt{k^2+m^2}$ with $m$ given by (\protect{\ref{mesp}})
and with a group velocity that approaches 1 at large $k$, as required in vacuum
by Lorentz invariance.
The next three solid (black) curves are
the dispersion relations for $\eps=0.25$, 0.5 and 0.756, top to bottom, the latter corresponding
to the Minkowski embedding at the temperature $T_{\rm diss}$ at which the first
order phase transition occurs.  The dashed (red) lines are the large-$k$
approximation discussed in
Section~\protect{\ref{sect:DispRelSummary}},
given by $\omega(k)=v_0 k + \Omega\eps L_0 / (v_0 R^2)$ with $\Omega$ specified by 
(\protect{\ref{eq:const}}).   We see that the dispersion relations
approach their large-$k$ linear behavior from below. The limiting velocity $v_0$ decreases
with increasing temperature.  Had we plotted dispersion relations for $0.756 < \eps < 1$
corresponding to metastable Minkowski embeddings with $T>T_{\rm diss}$, we would have
seen $v_0\rightarrow 0$ as  $\eps \rightarrow 1$, approaching the critical embedding.
\label{fig:NumericalResults}}
}

\FIGURE[t]{
\centerline{\hbox{\psfig{figure=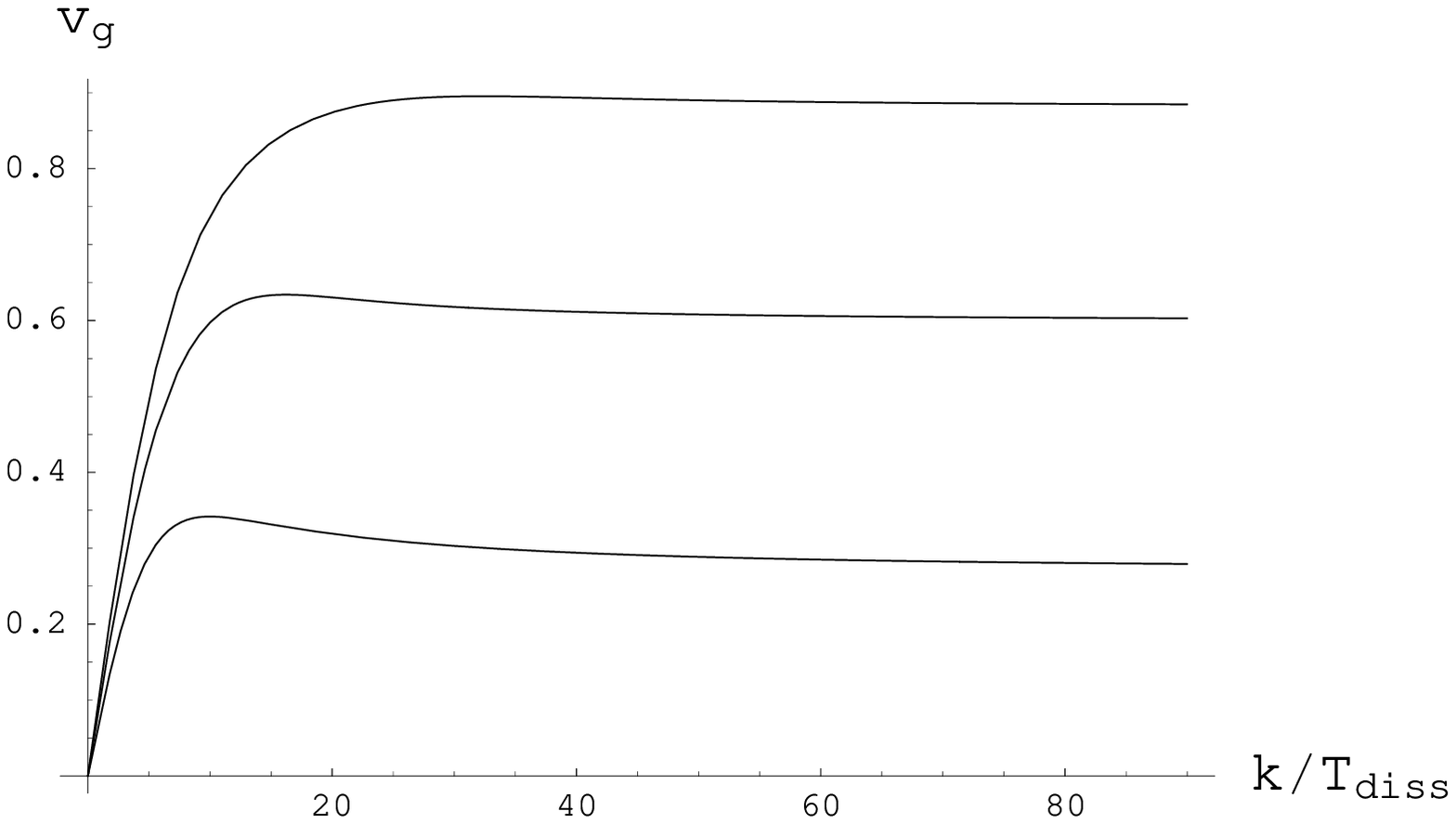,width=13cm}}}
\caption{Group velocities $v_g=d\omega/dk$ for the 
dispersion relations 
from
Fig.~\protect{\ref{fig:NumericalResults}},
with $\eps=0.25$, 0.5 and 0.756 (top to bottom).
We see that the
group velocity approaches its large-$k$ value $v_0$ from above.
And, we see $v_0$ decreasing with increasing temperature. (Again,
$v_0$ would approach zero if we included the metastable Minkowski
embeddings with $T>T_{\rm diss}$.) 
\label{fig:GroupVelocity}}
}

In Fig.~\ref{fig:NumericalResults} we show dispersion relations 
obtained numerically for the ground state
$\psi_1$ meson at several values of the temperature. At each $k$,
we solve the Schr\"odinger equation to find the ground state (using
the shooting method) and from the eigenvalue we obtain $\omega^2$ 
and hence a point on the dispersion relation.  By doing this at many $k$'s,
we obtain the curves plotted.  
We also overlay the linear approximation to the large-$k$
dispersion relations that we shall discuss in Sect.~\ref{sect:DispRelSummary}.
In Fig.~\ref{fig:GroupVelocity}, we plot the corresponding group velocities.

\subsection{Summary, limiting velocity 
and dissociation temperature}\label{sect:DispRelSummary}

In this Section we restate our central result for the dispersion relation
and then discuss its implications vis \`a vis a limiting velocity for
mesons at a given temperature as well as a limiting temperature
below which mesons with a given velocity are found, and above which they
are not.

In Section 5.2, we have derived the large-$k$ approximation to the meson
dispersion relations at any temperature below the dissociation transition.  We have
checked this result against numerical solutions valid at any $k$ in Section 5.3.
We begin by restating the analytic result (\ref{eq:disp}):
\be
 \label{sisp} 
 \omega^2 = v_0^2 k^2 +  \Omega \eps(n+2) \,\frac{L_0}{R^2} \,k 
+ d_{sn\ell} \, \frac{L_0^2}{R^4}\ ,
+ \OO(1/k)
\end{equation}
where
 \begin{equation}
 \label{varsD}
v_0 = \frac{1-\eps^2}{1+\eps^2} \,,\quad \Omega^2 =\frac{ 32
 (1-\eps^2)^2 (1+\eps^4)}{(1+\eps^2)^5}\ .
\end{equation}
The constant term $d_{sn\ell}$ (which depends on whether
we are discussing the $\psi_1$ or $\psi_2$ mesons --- $s=1$ or $s=2$ --- and
on the quantum numbers $n$ and $\ell$) was given 
in (\ref{BA}) and (\ref{BA2}).  
In writing the dispersion relation (\ref{sisp}) we have restored dimensions
by undoing the rescaling (\ref{eq:rce}).  The dimensionful quantity that we had
scaled out and have now restored can be written as
\be
\frac{L_0}{R^2} = \left(\frac{ 2 \pi m_q}{\sqrt{\lambda}}\right) \sqrt{\frac{\ep_\infty}{\eps}}\ ,
\label{L0overR2}
\ee
where we have used (\ref{3.6}), (\ref{epsinftydefn})  and (\ref{epsilondefn}). 
The first factor in (\ref{L0overR2}) is a (dimensionful) constant. The quantity
$\ep_\infty/\eps$
appearing in the second, dimensionless,
factor is weakly temperature dependent: it can be
read from Fig.~\ref{fig:epsepsinf}, and is not constant to the degree that
the curve in this plot is not a straight line (in the relevant regime 
$0<\eps<0.756$, as $\eps=0.756$ corresponds to $T=T_{\rm diss}$.)
Although using dimensionless variables obtained via
scaling by the temperature-dependent $L_0/R^2$ was very convenient 
in deriving all our results, in plotting the dispersion relation and 
group velocity in Figs.~\ref{fig:NumericalResults} and \ref{fig:GroupVelocity}
we have instead plotted $\omega$ and $k$ in units of 
$T_{\rm diss}=2.166\,m_q/\sqrt{\lambda}$, which is a relevant, constant,
physical, quantity comparable in magnitude to $L_0/R^2$.
In the remainder of this section, we shall analyze (\ref{sisp}).

In the large-$k$ limit, the asymptotic value of the group
velocity $d\omega/dk$ is given by $v_0$.  This velocity
decreases with increasing temperature, and vanishes
as $\eps\rightarrow 1$  on the critical embedding that
separates Minkowski and black hole embeddings in Figs.~\ref{fig:1}
and \ref{fig:epsepsinf}.  At the temperature at which
the first order dissociation transition occurs, $\eps=0.756$ and
$v_0=0.273$.   

There is a natural explanation within the dual gravity theory
for how the asymptotic velocity $v_0$ can arise.  
Using (\ref{vqu}), it is easy to show that $v_0$ in (\ref{varsD}) 
can also be written as
 \be
 v_0^2 = {f(\rho=0) \ov r^2(\rho=0)}\ ,
 \ee
which  we see from (\ref{eyr}) 
is precisely the local speed of light at the tip of the
D7-brane.    (The local speed of light is $1$ at $u=\infty$,
and decreases with decreasing  $u$, decreasing to $v_0$
at the tip of the D7-brane where $\rho=0$ and $u=y=1$.)
Since we have seen that in the large-$k$ limit
the wave function of the meson fluctuations becomes more 
and more localized closer and closer to the tip of the D7-brane,
this makes it natural that $v_0$ emerges as the asymptotic
velocity for mesons with large $k$.

In the low temperature (equivalently, heavy quark) limit, we find 
(either directly from (\ref{varsD}) or, initially, in (\ref{Osa}) in Section 5.1)
that
\be\label{v0smalleps}
 v_0^2 \approx 1 - 4 \eps^2\ .
 \ee
Since $\ep_\infty \approx \eps$ at small $\eps$, using (\ref{epsinftydefn}) we have
\be\label{v0smalleps2}
v_0^2 \approx  1 - {\lambda^2 T^4 \ov 16 m_q^4}\ ,
 \ee
which is precisely the critical velocity (\ref{oep}) obtained 
in~\cite{Liu:2006he} from the screening
calculation  as the velocity above which
the potential between two moving quarks of mass $m_q$ cannot be 
defined.
This is the first of two quantitative comparisons that we will be able 
to make between our present results for meson propagation and
results obtained previously via the screening calculation.    We see 
from Fig.~\ref{fig:crit} that (\ref{v0smalleps}) works very well
where $T \ll m_q/ \sqrt{\lambda}$, which is
where it was derived (both here
and in~\cite{Liu:2006nn}).

 \FIGURE[t]{
\centerline{\hbox{\psfig{figure=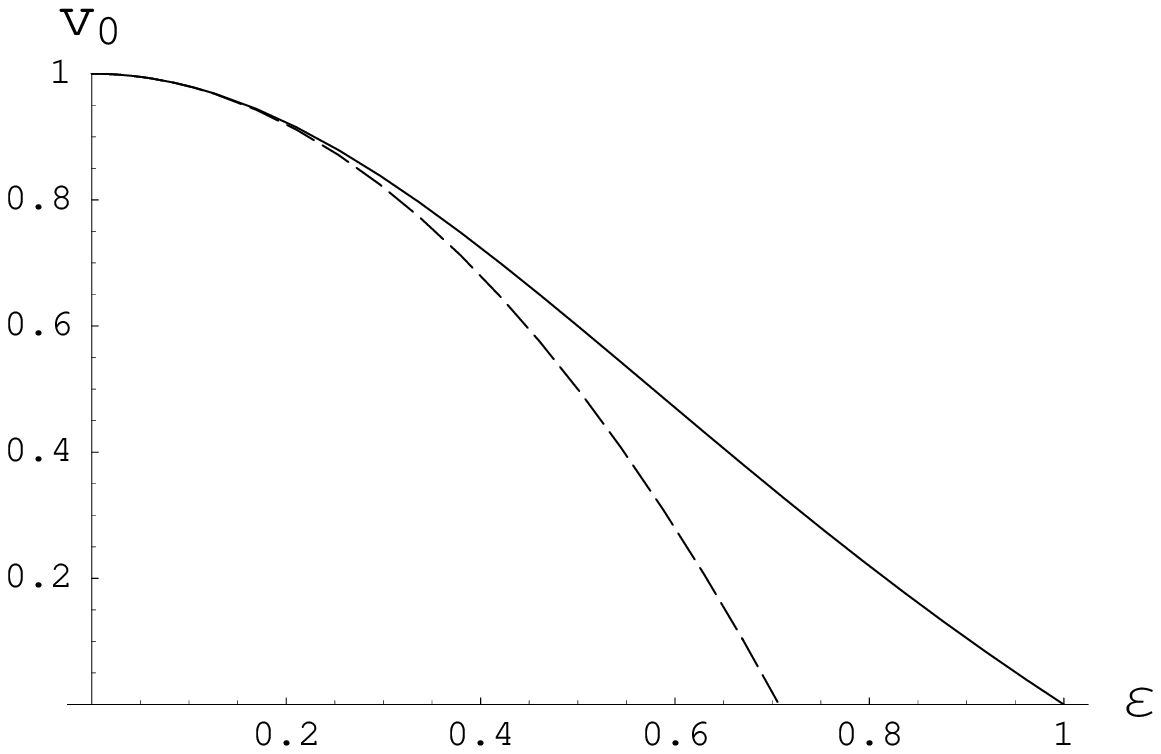,width=12cm}}}
\caption{The asymptotic velocity $v_0$ from (\ref{varsD})
as a function of $\eps$.  The low 
temperature approximation (\ref{v0smalleps}) is plotted as a dashed line.
Recall that the dissociation transition occurs at $\eps=0.756$.\label{fig:crit}
 }}

In order to analyze (\ref{sisp}) beyond the $k^2$ term, it is instructive
to rewrite it as a large-$k$ approximation to the dispersion
relation $\omega$ itself rather than to $\omega^2$, yielding
\be\label{eq:disprelitself}
\omega(k)  = v_0 k + \frac{\Omega \eps (n+2) L_0}{2 v_0 R^2}  + 
\frac{ 4 d_{sn\ell} v_0^2 - \Omega^2 \eps^2 (n+2)^2}{8 v_0^3}\, \frac{L_0^2}{R^4} \, \frac{1}{k}
+\OO(1/k^2)\ ,
\ee
in the form we discussed in Section 1.
We see that the term linear in $k$ in (\ref{sisp}) yields a constant
shift in the meson energies in (\ref{eq:disprelitself}).   Whereas $v_0$ is independent
of $s$, $n$ and $\ell$, the constant term in (\ref{eq:disprelitself})  results 
in evenly spaced dispersion relations for mesons with differing
$n$ quantum number, separated by 
\be\label{eq:const}
{\Om\eps L_0 \ov 2v_0 R^2}
       =\left(\frac{2 \pi m_q}{\sqrt{\lambda}}\right) \sqrt{ {8 \ep_\infty \eps (1+\eps^4)\ov (1+\eps^2)^3}} \ ,
\ee
which we plot in Fig.~\ref{fig:levspace}.

\FIGURE[t]{
\centerline{\hbox{\psfig{figure=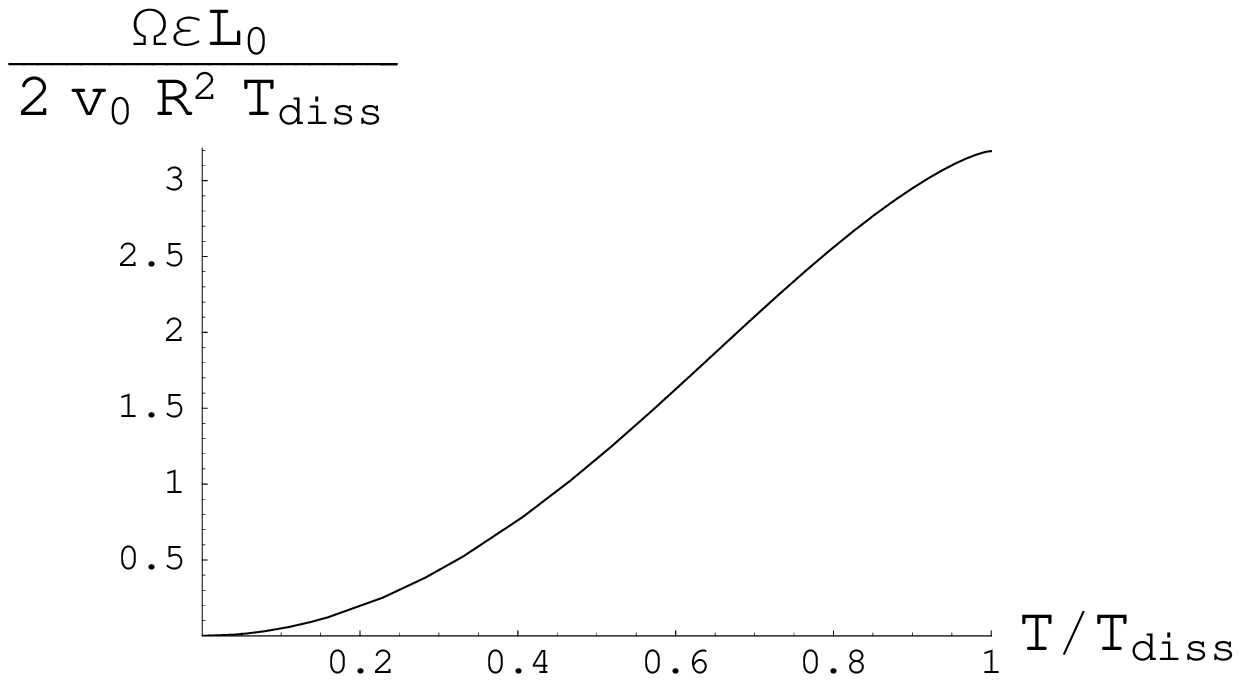,width=13cm}}}
\caption{The $k$-independent spacing $\Omega\eps L_0 /2 v_0 R^2$
between the dispersion relations
for any two mesons whose $n$ quantum numbers differ by 1,  in
units of $T_{\rm diss}$.  See ({\protect\ref{eq:const}}).  
\label{fig:levspace}}
 }

If we neglect the $\OO(1/k)$ and higher order terms in (\ref{eq:disprelitself}),
the dispersion relations are
the same for
mesons $\psi_1$ and $\psi_2$
and are
 independent of $\ell$.  These degeneracies are broken
at order $\OO(1/k)$, where $d_{sn\ell}$ first appears.    
We find that the coefficient of $1/k$ in $\omega(k)$ of (\ref{eq:disprelitself}) is 
typically negative: it is negative at all $\eps<1$ if $\ell=0$ for any $n$;
it can become positive only if $\eps$, $n$ and $\ell$ are all large enough.
When this coefficient is negative, it means that $\omega(k)$ approaches
its large-$k$ asymptotic behavior (which is a straight line with slope $v_0$
offset by the constant term in (\ref{eq:disprelitself})) from below. 
This means that $d^2\omega/dk^2<0$ at large $k$ and means that
the group velocity $v=d\omega/dk$ approaches $v_0$ from above at large $k$,
as shown in Fig.~\ref{fig:GroupVelocity}.
However, at $k=0$ the group velocity vanishes and $d^2\omega/dk^2>0$.
(We have shown this analytically at small $\eps$ in Section 5.1,
see (\ref{doss}), and our numerical results as in Section 5.3 indicate
that this is so at all $\eps$.)  So, as a function of increasing $k$,
the group velocity begins at zero, increases to some maximum
value that is greater than $v_0$, and then decreases to $v_0$
as $k\rightarrow\infty$ as depicted in
Fig.~\ref{fig:GroupVelocity}.\footnote{This behavior is not inconsistent 
with our identification of $v_0$
with the local speed of light at the tip of the brane: it is only for $k\rightarrow\infty$
that the meson wave function is squeezed down to the tip of the brane; at
finite $k$,  the wave function 
is peaked where the local speed of light exceeds $v_0$.}
Although $v_0$ is not the maximum possible group velocity,
it appears that the maximal velocity exceeds $v_0$ only by
a small margin.  For example, for the ground state $\psi_1$ meson
whose dispersion relations are given in Figs.~\ref{fig:NumericalResults} and
\ref{fig:GroupVelocity},
we find that $v_0=0.882$, 0.6, and 0.273 for $\eps=0.25$, 0.5, and 0.756 whereas the maximal
velocities  are 0.896, 0.634 and 0.342, respectively.   We shall 
therefore simplify the following discussion
by taking the maximal possible meson velocity
at a given temperature to be the limiting velocity $v_0$, neglecting
the slight imprecision that this introduces.

\begin{figure}[t]
\begin{center}
$\begin{array}{cc}
\psfig{figure=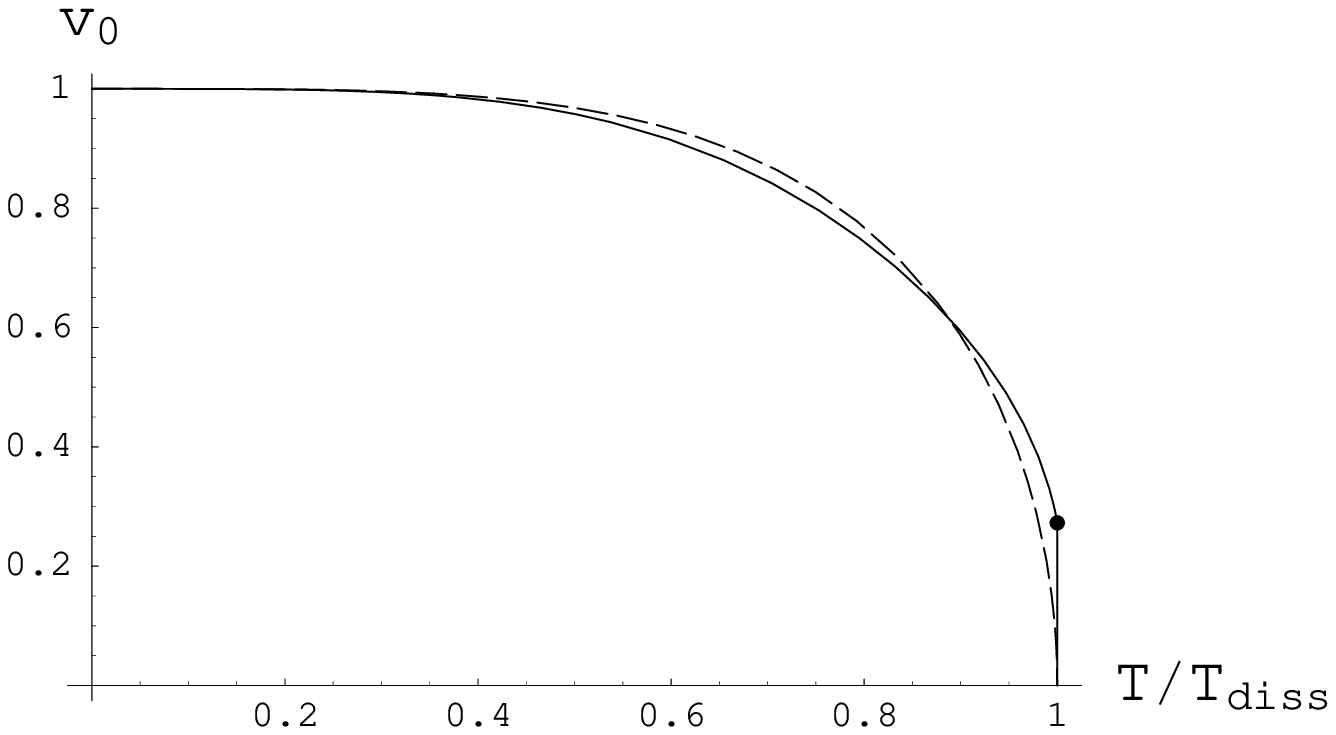,width=10cm} &
\psfig{figure=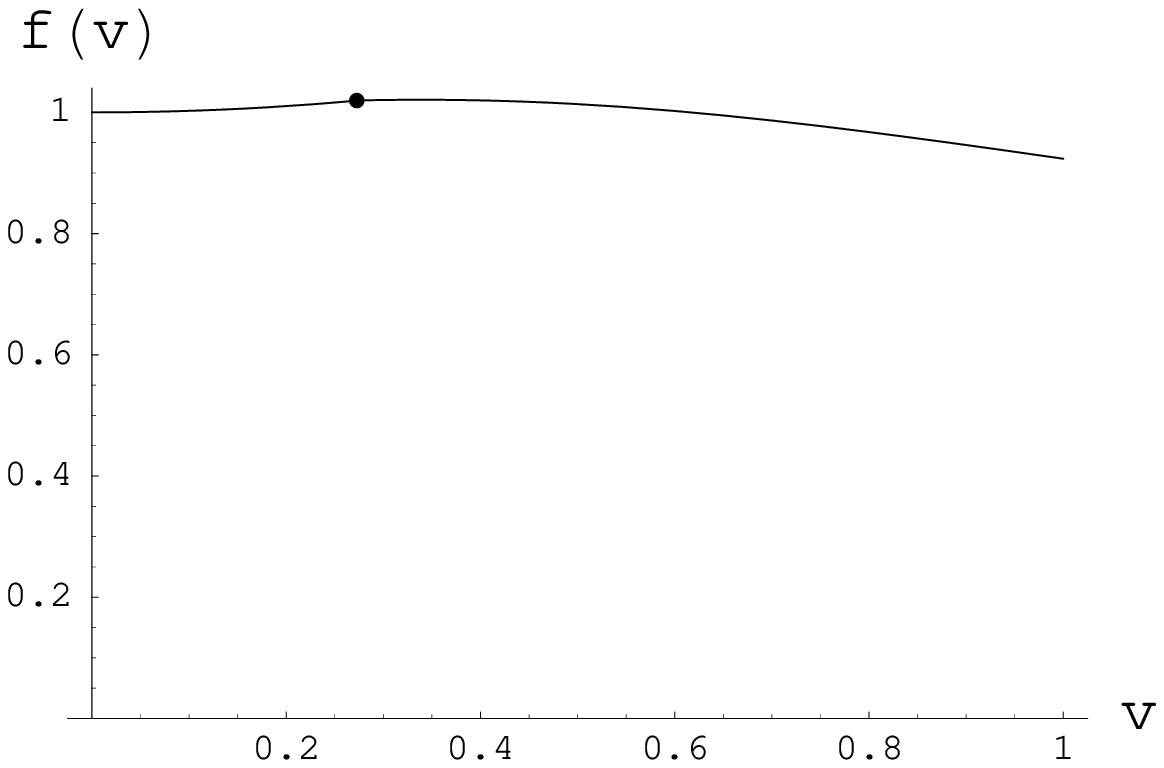,width=8cm}
\end{array}$
\caption{Left panel: The solid curve
is the  limiting velocity $v_0$ 
as a function
of $T/T_{\rm diss}$, where $T_{\rm diss}$ is the temperature
of the dissociation transition at zero velocity.  
The dissociation
transition occurs 
at the dot, 
where $v_0\approx 0.273$.   The dashed curve is
the approximation obtained by setting $f(v)=1$ in (\protect\ref{eq:fv}).
Right panel: $f(v)$, the ratio of the solid and dashed curves
in the left panel at a given $v$. We see that $f(v)$ is within a few percent of 1
at all velocities.
\label{fig:fv}}
\end{center}
\end{figure}

We now wish to compare our results for the limiting meson velocity
$v_0$ at a given temperature to the result (\ref{rro}) inferred 
(qualitatively) from the analysis of screening in a hot wind in~\cite{Liu:2006he}.
We must first convert $v_0(\eps)$ into $v_0(T)$, meaning that we must
convert from $\eps$ to $\ep_\infty$ as discussed in and around 
Fig.~\ref{fig:epsepsinf}.   The result is the solid curve in the
left panel of Fig.~\ref{fig:fv}, where we have plotted $v_0$ versus
$T/T_{\rm diss}$.  We have derived this curve as a limiting 
meson velocity at a given temperature. However, it can just
as well be read (by asking where it cuts horizontal lines rather
than vertical ones)
as giving $T_{\rm diss}(v)$, the temperature below which mesons
with a given velocity $v$ are found and above which no mesons with
that velocity exist.   We see that $T_{\rm diss}(v)\rightarrow 0$
for $v\rightarrow 1$, the regime where $v_0$
is given
by (\ref{v0smalleps2})
and $T_{\rm diss}(v)$ is therefore given by (\ref{PWw}).
In order to compare our result for $T_{\rm diss}(v)$ at all velocities
to (\ref{rro}), we parametrize our result as
 \begin{equation}
T_{\rm diss}(v) = f(v) (1-v^2)^{1/4} T_{\rm diss}(0)\ .
\label{eq:fv}
\end{equation}
In the left panel of Fig.~\ref{fig:fv} we compare our result (the solid curve)
to (\ref{eq:fv}) with $f(v)$ set to 1, which is of course (\ref{rro}). In
the right panel, we plot $f(v)$.  We see that this function is
close to 1 at all velocities, varying between 1.021 at its maximum  and 0.924 at $v=1$.
The weakness of the dependence of $f(v)$ on $v$ is a measure
of the robustness with which the simple scaling (\ref{rro}) describes
our result for the meson dissociation temperature at all velocities.

\section{Discussion and Open Questions}

We have used the AdS/CFT correspondence to compute the dispersion relation
$\omega(k)$ for the heavy ``quarkonium'' mesons that exist as 
stable bound states in the 
strongly coupled plasma of ${\cal N}=4$ SYM to which heavy fundamental
quarks with mass $m_q$ have been added.  In Section 4 we have introduced
a new, and more geometrical, method of analyzing these mesons that has allowed
us, in Section 5, to obtain the dispersion relations at large-$k$  analytically in the 
form (\ref{eq:disprelitself}), which we can summarize as in Section 1 by
writing
\be
\omega(k)=v_0 k + a + \frac{b}{k} + \ldots\ .
\label{SimpleForm}
\ee
We have computed $a$ and $b$ explicitly and analytically in Section 5, but at
present we have no argument that the behavior of these coefficients, which
depend on the meson quantum numbers, could teach us lessons that generalize
beyond the particular theory in which we have computed them.  
On the other hand, the limiting large-$k$ meson velocity $v_0$
seems to encode much physics
that may generalize to meson bound states in other strongly coupled gauge theory
plasmas.  
\begin{itemize}
\item
Our explicit result is
\be
v_0=\frac{1-\eps^2}{1+\eps^2}\ ,
\ee
where $\eps$ is related to $\ep_\infty = \lambda T^2/(8 m_q^2)$ as in Fig.~\ref{fig:epsepsinf}.
We see that $v_0$ depends on the temperature (in the
combination $\sqrt{\lambda} T / m_q$) but not on the meson quantum numbers.
We see in Figs.~\ref{fig:NumericalResults} and \ref{fig:GroupVelocity} 
that $v_0$ decreases with increasing temperature, becoming much less than 1
as the temperature approaches $T_{\rm diss}$, the temperature at which
mesons at rest dissociate.  We see in these figures that the coefficient $b$ 
in (\ref{SimpleForm}) can be negative, meaning that the group velocity approaches
its large-$k$ value $v_0$ from above.  Thus, $v_0$ is the limiting meson velocity
at large $k$, but the maximal velocity occurs at finite $k$ and is slightly larger
than $v_0$. We describe this quantitatively in 
Section 5, but it is a small effect and in this discussion we shall ignore the
distinction between $v_0$ and the maximal velocity.
\item
We find that $v_0$, which in the
gauge theory is the limiting velocity of the mesons that they attain
at large $k$,
also has a nice interpretation in the dual gravity theory.  It
is precisely the local velocity of light 
at the ``tip'' of 
the D7-brane, namely where the D7-brane reaches closest to the black hole.
This is physically sensible because we have shown
that the D7-brane fluctuations --- i.e. the mesons in the dual gravity theory ---
are localized at the D7-brane tip in the large-$k$ limit.  
\item
At low temperatures or, equivalently, for heavy quarks we find
\be
v_0 \approx 1- \frac{\lambda^2 T^4}{32 m_q^4}\ .
\label{LowTempResult}
\ee
This is precisely, i.e. even including the
numerical factor, the criterion for meson dissociation inferred from a completely
different starting point in \cite{Liu:2006he}.   The logic there was that
the screening length that characterizes the potential between 
a quark and antiquark moving with $v>v_0$ is shorter than
the quark Compton wavelength, meaning that if a quark and antiquark
moving with $v>v_0$ are separated by more than a Compton
wavelength, to leading order in $\sqrt{\lambda}$ they feel no attractive force.
By inference, no mesons should exist with $v>v_0$.  We now see this result
emerging by direct calculation of meson dispersion relations, rather than by inference.  
\item
We have a result for $v_0(T)$, the limiting velocity beyond which there
are no meson bound states, at all $T<T_{\rm diss}$ not just at low temperatures,
see Fig.~\ref{fig:fv}.  We can just as well read this as determining a temperature
$T_{\rm diss}(v)$ above which no meson bound states with velocity $v$ exist.
We find that up to few percent corrections, see Fig.~\ref{fig:fv}, this is given
by 
\be
T_{\rm diss}(v) = (1-v^2)^{1/4} T_{\rm diss}\ .
\label{ThirdTime}
\ee
Once again, this is a result that was previously inferred from analysis
of the velocity dependence of the screening length characterizing
the potential between a quark and antiquark moving through the
plasma~\cite{Liu:2006nn}. We have now derived this result and
the (few percent) corrections to it for the mesons whose
dispersion relations we have explicitly
constructed. We should also note that it is
a slight abuse of terminology to call $T_{\rm diss}(v)$ at $v>0$
a ``dissociation'' temperature: although
it {\it is} a temperature above which no mesons with velocity $v$ exist, 
if we imagine heating the plasma through $T_{\rm diss}(v)$ we have not
shown that any mesons present therein dissociate --- they may simply slow down.
The question of what happens in this hypothetical context is a dynamical one
that cannot be answered just from the dispersion relations we have analyzed.

\item
As we discussed in Section 2, the 
result (\ref{ThirdTime}) can be read as saying
that no mesons with velocity $v$ exist when the energy density of
the strongly coupled plasma exceeds $\rho_{\rm diss}(v)$ where,
up to small corrections,
\be
\rho_{\rm diss}(v)=(1-v^2) \rho_{\rm diss}\ ,
\label{ThirdTimeEnergy}
\ee
with $\rho_{\rm diss}$ the energy density 
at which mesons at rest dissociate.  Correspondingly, 
the low temperature result (\ref{LowTempResult})
can be written as 
\be
1-v_0 = {\rm constant}\frac{\rho}{\rho_{\rm diss}}\ ,
\label{v0andrho}
\ee
valid when $\rho\ll \rho_{\rm diss}$ and $v_0$ is close to 1.
Thinking as in \cite{Caceres:2006ta},
we can ask whether the same result holds in other theories.
It will be interesting to address this question in $(3+1)$-dimensional
gauge theories that are in various senses more QCD-like than ${\cal N}=4$
SYM.  At present, however, 
we have only investigated the $(p+1)$-dimensional
gauge theories 
described by $N$ D$p$-branes~\cite{Itzhaki:1998dd}
into which fundamental
quarks, and hence mesons, have been introduced by embedding
a D$q$-brane~\cite{Arean:2006pk,Myers:2006qr,Mateos:2006nu,Mateos:2007vn}.  
The D$p$-branes fill coordinates $0,1,\ldots,p$.  The
D$q$-brane fills the first $d+1$ of these coordinates $0,1,\ldots,d$, 
where $d$ may be less than or equal to $p$, as well as $q-d$
of the remaining $9-p$ coordinates. In Appendix B, we sketch
an investigation of those theories for which
$p-d+q-d=4$.  (The case that we have analyzed throughout
the rest of this paper is $p=d=3$, $q=7$.)
These theories are not conformal for $p\neq 3$, as their
coupling constant $\lambda$ has dimension $p-3$. It is convenient
to introduce a dimensionless
$\lambda_{\rm eff} \equiv \lambda T^{p-3}$.
We have not repeated our
entire construction for the D$p$-D$q$-brane theories.  However, we expect that
the wave functions for large-$k$ mesons will
again be localized at the tip of the D$q$-brane,
and therefore expect that in these theories $v_0$ will again be
given by the local velocity of light at this location. We compute
this velocity in Appendix B. Assuming that this is indeed
the limiting meson velocity, we find
\be
v_0 = \left( \frac{ 1 - \eps^{(7-p)/2} }{ 1+ \eps^{(7-p)/2}}\right)\ ,
\label{DpDqLimitingVelocity}
\ee
where $\eps$ is given at small $T/m_q$ by
\be
\eps \approx \ep_\infty \propto \left(\frac{T}{m_q}\right)^2
\lambda_{\rm eff}^{2/(5-p)}
= \frac{\lambda^{2/(5-p)} T^{4/(5-p)}  }{m_q^2}\ .
\ee
(Relating $\eps$ to $\ep_\infty$ beyond the small $T/m_q$ limit
requires solving the embedding equation given in Appendix B.)
In these theories, the energy density of the plasma 
depends on parameters according to~\cite{Itzhaki:1998dd}
\be
\rho \propto N^2 T^{p+1} \lambda_{\rm eff}^{(p-3)/(5-p)}
= N^2 \lambda^{(p-3)/(5-p)} T^{(14-2p)/(5-p)}\ ,
\ee
and zero-velocity mesons dissociate at some energy
density $\rho_{\rm diss}$ corresponding to $\eps=\eps_{\rm diss}$
where $\eps_{\rm diss}={\cal O}(1)$.  
From these results we
notice that at small $\eps$
\be
\eps^{(7-p)/2} \propto \frac{\lambda^{(7-p)/(5-p)}T^{(14-2p)/(5-p)}}
{m_q^{7-p}}\propto \frac{\rho}{\rho_{\rm diss}}\ ,
\ee
meaning that the velocity $v_0$ of (\ref{DpDqLimitingVelocity}) 
can be written in the form (\ref{v0andrho}) for all values of $p$!
In Appendix B, we describe the verification that (\ref{ThirdTime})
also holds, but only when phrased as in (\ref{ThirdTimeEnergy}) in terms of energy density
rather than temperature.
\end{itemize}

Emboldened by these successes, we advocate investigating the consequences
that follow from hypothesizing that 
$\Upsilon$ and $J/\Psi$ mesons in the strongly coupled 
quark-gluon plasma of QCD propagate with a dispersion relation
(\ref{SimpleForm}) with $v_0$ dropping dramatically as the temperature
approaches $T_{\rm diss}$ from below, and with no bound states with
velocity $v$ possible if $T>T_{\rm diss}(v)$ given by (\ref{ThirdTime}).
In applying (\ref{ThirdTime}) to QCD, it is important to
scale $T_{\rm diss}(v)$ 
relative to the $T_{\rm diss}$ for $\Upsilon$ and $J/\Psi$ mesons in QCD
itself.   The 
result $T_{\rm diss}=2.166 m_q/\sqrt{\lambda}$ for the mesons
that we have analyzed is surely affected by the fact that they are
deeply bound and so should not be used as a guide to quarkonia
in QCD. For example, it seems to overestimate
$T_{\rm diss}$ for $J/\Psi$ mesons by a factor of 2 or 3.  However, as argued
in \cite{Liu:2006nn,Liu:2006he} and as
we have discussed above, the velocity scaling (\ref{ThirdTime}) 
may transcend the detailed meson physics in any one theory and apply
to mesonic bound states in any strongly coupled plasma.  The successful
comparison of our detailed results to this simple scaling form supports
this conjecture.

As we have explained at length in Section 1, meson
propagation is only one piece of the physics that must be treated in order
to understand quarkonium suppression in heavy ion collisions.  
Introducing the dispersion relation and limiting velocity that we have found
into such a treatment is something we leave to the future, instead making
only a few qualitative remarks.

First, from the dispersion relations alone we {\it cannot} conclude that
if a quark-antiquark pair is produced (from an initial hard scattering)
with a velocity $v>v_0(T)$, with $v_0(T)$ the limiting meson velocity
in the plasma of temperature $T$ in which the quark-antiquark pair finds
itself, then the quark-antiquark pair do not bind into a meson.
The reason that we cannot make this inference is that the dispersion
relations describe stable mesons with arbitrarily large momentum $k$, making
it a logical possibility that  a high velocity quark-antiquark pair 
with arbitrarily high momentum interacts 
with the medium in some way such as to slow down and
lose energy while
conserving its momentum, and thus in some way
dresses itself into a meson with arbitrarily
high momentum $k$, and velocity $v_0$.   That is, since the dispersion
relations describe the propagation of mesons with arbitrarily large momentum,
by themselves they do not require that quarkonium production is suppressed
when the precursor quark-antiquark pair has velocity $v>v_0(T)$.   Excluding
this possibility, allowed by the kinematics, requires some consideration
of the dynamics.  The heuristic
argument of \cite{Liu:2006he} provides guidance:
the precursor quark-antiquark pair with $v>v_0(T)$
do not attract each other and so even though it is kinematically allowed
by the meson dispersion relations for them to slow down and form a meson,
instead they will propagate independently through the medium.
Thus, the $p_T$-dependent
quarkonium suppression pattern
proposed in \cite{Liu:2006nn},
with the production of 
quarkonium states with $T_{\rm diss}$ higher than the
temperature reached in  a given heavy ion collision experiment
nevertheless becoming suppressed above a threshold
transverse momentum at which a quark-antiquark pair
with that transverse momentum has velocity $v_0(T)$,
rests upon the dynamical argument of \cite{Liu:2006he}.
It is natural that 
analyzing quarkonium suppression requires consideration of both
the precursor quark-antiquark pair and the putative meson, and only
the latter is described by the meson dispersion relation.
It is then nice to discover that
the limiting meson velocity $v_0(T)$
agrees precisely with the velocity at which quark-antiquark pairs
can no longer feel a force at order $\sqrt{\lambda}$.

We have just argued that the very large-$k$ region 
of the meson dispersion
relation is
unlikely to be populated in heavy ion collisions. 
But, whether or not
such large-$k$ modes are excited, it 
is clear from Fig.~\ref{fig:NumericalResults} that 
at temperatures near to $T_{\rm diss}$
mesons at any $k$ move much more slowly than they would if they 
propagated with
their vacuum dispersion relation.  There are several in-principle-observable
signatures of the slow velocity of quarkonium mesons.  First, it increases the
separation in space long after the collision
between those mesons that are produced at the surface
of the fireball moving outwards, and hence escape into vacuum promptly, and those
which are produced in the center of the plasma and hence move more slowly
than if they had their vacuum dispersion relation.  An increase in the typical
separation of identical mesons because of this slow velocity
effect will shift the onset of Bose-Einstein enhancement in the two particle
momentum correlation to a lower relative momentum.
This simple idea underlies a technique widely used in heavy ion physics and often referred 
to as Hanbury-Brown Twiss (HBT) two-particle interferometry, in which identical two-particle 
momentum correlations are used to determine spatio-temporal characteristics of the collision 
region. For a review, see Ref.~\cite{Lisa:2005dd}. Quarkonium HBT interferometry would thus
in principle be able to find signatures of a depressed meson velocity. 
Second, non-identical two-particle correlation functions are sensitive to 
whether one particle species $A$ is emitted from the medium on average before 
or after another particle species $B$. Such a difference in average emission times
could result, for instance, if the maximal velocities in the dispersion relations for $A$ 
and $B$ differ because of their different mass. The analysis of the effect 
of a difference in average emission times on 
non-identical two-particle correlation functions can be found in~\cite{Lednicky:1995vk}.
In principle, this provides a second way of finding signatures of a depressed
velocity 
for those mesons for which the plasma reaches temperatures close
to their dissociation temperature.

Quarkonium mesons in the quark-gluon plasma of  QCD have nonzero width.
In contrast, the mesons we have analyzed at $T<T_{\rm diss}$ are
stable, with zero width.  The dispersion relations that we have found have
no imaginary part.  This is certainly an artifact of the large number
of colors $N$ and large coupling $\lambda$ limits that we have taken
throughout.  Processes in which one meson decays into two mesons
are suppressed by $1/N$.  And, thermal fluctuations which unbind
a meson whose binding energy is $2 m_q$ are suppressed by the
Boltzmann factor 
\be
\exp(-2 m_q/T) = \exp( -0.92  \sqrt{\lambda} T_{\rm diss}/T )\ , 
\ee
which at some fixed $T/T_{\rm diss}$ is nonperturbative in an
expansion about infinite $\lambda$.  
A calculation of the imaginary
part of the meson dispersion relations at finite $\lambda$ remains
for the future, but this simple consideration is enough to be sure
that it is nonzero, as is the case in QCD at weak coupling~\cite{Laine:2006ns}.
As soon as the mesons have nonzero width, their slow velocity
has a further consequence in the context
of heavy ion collisions: because they move more slowly, they 
spend a longer time in the medium giving the absorptive imaginary
part more time to effect the dissociation of the meson than would
otherwise be the case.

Our discussion in this Section has highlighted three different
avenues of further investigation opened up by our analysis of
meson dispersion relations in a strongly coupled gauge theory
plasma.  The first is the investigation of the phenomenological consequences
for $J/\Psi$ and $\Upsilon$ suppression in heavy ion collisions
of a dispersion relation of the form (\ref{SimpleForm}) with (\ref{ThirdTime}).
Second, it appears to us that the most interesting open question about
the mesons whose dispersion relations we have analyzed is extending
the calculation to finite $\lambda$ and analyzing the width of the mesons.
And, third, we could gain significant confidence in the application of
the lessons we have learned to QCD by repeating our analysis for
heavy quark mesons in the 
plasma of other strongly coupled gauge theories, in particular those
with a controlled degree of nonconformality.

\medskip
\section*{Acknowledgments}
\medskip

HL is supported in part by the A.~P.~Sloan Foundation and the U.S.
Department of Energy (DOE) OJI program. HL is also supported 
in part by the Project of Knowledge Innovation
Program (PKIP) of Chinese Academy of Sciences. HL would like to thank KITPC
(Beijing) for hospitality during the last stage of this project.
This research was supported in part by the DOE Offices of Nuclear 
and High Energy Physics under grants
\#DE-FG02-94ER40818 and \#DE-FG02-05ER41360.

\medskip
\section*{Appendices}

\appendix

\section{General discussion of brane embedding and fluctuations}

In this appendix we present a general discussion of brane
embedding in a curved spacetime (in the absence of fluxes) and its
small fluctuations. We then specialize to the case of D7-branes embedded in the
$AdS_5 \times S_5$ black hole geometry.

\subsection{General discussion}

Consider a $p+1$-dimensional brane in a $D$-dimensional target
space whose action is 
  \be
 S_{Dp} = -\mu_p \int d^{p+1} \xi \, \sqrt{-\det \tilde {h}_{ij}}\ ,
\label{daction1}
 \ee
where $\xi^i, i=0,1,\ldots, p$ denote the worldvolume coordinates
and ${\tilde h}_{ij}$ is the induced metric in the worldvolume
 \be \label{eie1}
  {\tilde h}_{ij} = G_{\mu \nu} (X) {\p X^\mu
  \ov \p \xi^i} {\p X^\nu \ov \p\xi^j}, \qquad \mu =0,1, \ldots, D-1
   \ .
  \ee
Suppose that 
$X_0^\mu(\xi^i)$ solves the equations of motion following
from (\ref{daction1}), thus describing an embedding of the brane
in the target spacetime. We are interested in understanding the
behavior of small fluctuations around $X_0$. For this purpose, let
\begin{equation}
X^\mu(\xi) = X_0^\mu(\xi^i) + \delta X^{\mu}(\xi^i) \, .
\label{eq:pert-2}
\end{equation}
The action for $\delta X^\mu$ can then be obtained
straightforwardly from (\ref{daction1}). The resulting action and
equations of motion for $\delta X^\mu$ are, however, not
geometrically transparent. This is due to the fact that $\delta
X^{\mu}(\xi^i)$ is the difference between coordinates and thus
does not have good properties under coordinate transformations.
 A more convenient way to
parameterize $\delta X^\mu (\xi)$ is to use the exponential map to
express it in terms of a vector in the tangent space at $X_0^\mu$,
as we now describe.
(Such techniques have also been used in the calculation of string
worldsheet beta functions~\cite{Callan:1989nz}.)  
Given a
vector $\eta^\mu$, we shoot out geodesics of unit affine parameter
from $X_0$ with tangent $\eta^\mu$. The end point of such a
geodesic is identified with $X_0^\mu + \delta X^\mu$. Such a map
should be one-to-one within a small neighborhood of $X_0$.
 To second order in $\eta$ one
may solve the geodesic differential equation, finding
\begin{equation}
\label{eq:eta1} \delta X^{\mu} = \eta^\mu - \frac{1}{2}
\Gamma^\mu_{\alpha\beta} (X_0) \eta^\alpha \eta^\beta + \ldots\ .
\end{equation}
Note that the appearance of $\Gamma$ is consistent with the
coordinate dependence of $\delta X$; they can  both be shown to
have the same variation under a coordinate transformation.

Using the parametrization (\ref{eq:eta1}), we find that
 \begin{align}
\label{eq:ind-2} \tilde{h}_{ij} & = G_{\mu\nu} ( X_0 + \delta X)
\partial_i ( X_0^\nu + \delta X^\nu)
\partial_j ( X_0^\mu + \delta X^\mu) \\
 & =  h_{ij} + 2 G_{\mu\nu}
\lambda^\mu_{(i}\nabla_{j)} \eta^\mu + G_{\mu\nu} \nabla_i
\eta^\mu \nabla_j \eta^\nu \nonumber  + \eta^\alpha \eta^\beta
\lambda^\mu_{(i} \lambda^\nu_{j)} R_{\nu\beta\alpha\mu}
\end{align}
with
 \bea
 h_{ij} & = & G_{\mu \nu} (X_0) \p_i X^\mu_0 \p_j X^\nu_0 = \lam_i^\mu \lam_{j \mu} , \qquad
 \nabla_i = \lam_i^\mu \nabla_\mu, \qquad \lam_i^\mu = \p_i
 X_0^\mu \ .
 \eea
The simplest way to find (\ref{eq:ind-2}) is to use the Riemann
Normal coordinates at $X_0$ in which the Christoffel symbols
vanish. $h_{ij}$ is the induced metric in the worldvolume theory
and below indices $i,j$ will be raised and lowered by $h$. 
To quadratic order in $\eta$ we have
 \begin{align}
\label{eq:determ} \sqrt{-\textrm{det}\tilde{h}_{ij}} =
\sqrt{-\textrm{det} h_{ij}} &\left( 1 + \lambda^i_\nu
\nabla_j\eta^\nu + \frac{1}{2} \nabla^i\eta^\mu \nabla_i \eta_\mu
- (\lambda_{i\mu} \nabla_j \eta^\mu) ( \lambda^{(i}_\nu
\nabla^{j)} \eta^\nu) \right. \nonumber \\
& \left. + \frac{1}{2} ( \lambda^i_\nu \nabla_i\eta^\nu)^2 +
 \ha \eta^\alpha \eta^\beta h^{ij} \lam^\mu_i \lam^\nu_j R_{\alpha\mu \nu\beta}  \right)
  \ .
\end{align}
We now take $\eta^\mu$ to be orthogonal to the brane
worldvolume (which corresponds  to choosing the static gauge),
i.e.
 \begin{equation} \label{PK}
\eta^\mu = \chi_s n_s^\mu (X_0), \qquad s = 1, \ldots , D-p-1\ ,
\end{equation}
where $n_s (X_0)$ are unit vectors orthogonal to the worldvolume
direction. Note that $\lam_i^\mu$ and $n_s^\mu$ together span the
full tangent space at $X_0$. i.e.
 \be \label{NAa}
 \lam_i^\mu n_{s \mu} = 0, \qquad n_{s \mu} n_{t}^{ \mu} =
 \delta_{st}, \qquad \lam_i^\mu \lam_{\mu}^j = \delta_{i}^j \ ,
 \ee
and
 \be\label{DDa}
 \delta^\mu_\nu = \lam_i^\mu \lam^i_{\nu} +  n_{s}^\mu
 n_{s\nu} \ .
 \ee
We now introduce
 \be
 K_{sij} = \lam_i^\mu \lam_j^\nu \nabla_\mu n_{s \nu}, \qquad K_s
 = K_{sij} h^{ij} , \qquad
 U^i_{st} = n_s^\nu  \nabla_i n_{t \nu}  = n_s^\nu \lam_i^\mu \nabla_\mu n_{t \nu}\ .
 \ee
$K_{sij}$ is the extrinsic curvature of the brane in the $s$-direction, and is symmetric
in $i,j$. 
(This follows from the fact that a
surface orthogonal to $n_s^\mu$ satisfies $\nabla_{[\mu} n^{t}_{\nu]}
= \sum_{s} v^s_{[\mu} n^s_{\nu]}$ for some one-form $v^s_\mu$.
Note also that $K_{sij}$ can be written as $K_{sij} = \ha L_{n_s} h_{ij}$, 
where $L_n$ is the Lie derivative along $n$-direction.)
$U^i_{st}$, which is antisymmetric in $s,t$, is an
$SO(D-1-p)$ connection for the transverse directions. Note that
the choice of $n_s^\mu$ (and thus $\chi_s$) is not unique. One can
choose a different set of basis vectors by making an arbitrary local
$SO(D-1-p)$ transformation. Thus $\chi_s$ transforms as a vector
under the $SO(D-1-p)$ ``gauge'' symmetry and $U^i_{st}$ transform
as a connection. Note that this gauge symmetry is  not dynamical.
 With these definitions we can now write
 \be \label{PA}
 \nabla_i \eta_{\mu} = (D_i \chi_s) n_{s\mu} +  K_{sij} \chi_s \lam_\mu^j\, ,
 \ee
where
 \be
D_i \chi_s = \p_i \chi_s + U_{i st} \chi_t
 \ee
  is an $SO(D-p-1)$ covariant derivative.
Using (\ref{PA}) in (\ref{eq:determ}), we now find that
 \begin{align} \label{cPNA}
S_{Dp} = - \mu_p \int d^{p+1} \xi \, &\sqrt{-\textrm{det}h_{ij}} \left(
1+ \chi_s  K_s + \frac{1}{2} D_i \chi_s D^i \chi^s + \frac{1}{2}
\chi_s \chi_t \left(  - K_{s ij} K_t^{ij} + R_{sijt} h^{ij}  + K_s
K_t  \right) \right)
\end{align}
with $R_{sijt} = n_s^\al n_t^\beta \lam_i^\mu \lam_j^\nu
 R_{\al \mu \nu \beta}$. For $X_0$ to satisfy the equations of motion, the 
 terms in (\ref{cPNA}) that are linear in the $\chi$'s have to vanish.
 This implies that
  \be\label{embeddingequationsA}
   K_s = K_{sij} h^{ij} = 0, \qquad s =1, \ldots, D-p-1 \ .
   \ee
These are the embedding equations for the background. 
Thus, the action (\ref{cPNA}) for the small
fluctuations to quadratic order becomes
\begin{align} \label{PNA}
S_{Dp} =  \mu_p \int d^{p+1} \xi \, &\sqrt{-\textrm{det}h_{ij}} \left(
 - \frac{1}{2} D_i \chi_s D^i \chi^s - \frac{1}{2}
\chi_s \chi_t \left(  - K_{s ij} K_t^{ij} + R_{sijt} h^{ij}
\right) \right)\ .
\end{align}
We have used both the embedding equations (\ref{embeddingequationsA})
and the action for the small fluctuations (\ref{PNA}) in Section 4. 

The action (\ref{PNA}) can be further simplified if $n_s^\mu$
satisfies additional constraints. For example, if $n_s^\mu$ is
proportional to a Killing vector, then
  \be
 K_{sij} = 0\ .
 \ee
This follows from the fact that $n_s^\mu$ satisfies $\nabla_{(\mu}
n^s_{\nu)} = v_{(\mu} n^s_{\nu)}$ for some $v_\mu$. If in addition
$n_s^\mu$ is a hypersurface orthogonal, i.e. if it satisfies
$\nabla_{[\mu} n^s_{\nu]} = w_{[\mu} n^s_{\nu]}$ for some one form
$w_\mu$, then
 \be
 U_{i st} =0, \quad {\rm for \;\; all} \;\; t \ .
 \ee
We have used this simplification in Section 4.

Finally, note that equation (\ref{PNA}) was written using the
coordinate split (\ref{NAa}). One can write it and other equations
in a more covariant way by introducing
\be
 h_{\mu \nu} = h^{ij} \lam_{i \mu} \lam_{j \nu} , \qquad
 h_{\mu}{}^{\nu} =  h^{ij} \lam_{i \mu} \lam_{j}^{\nu}, \qquad
 h^{\mu \nu} =  h^{ij} \lam_{i}^{ \mu} \lam_{j}^{\nu}\ ,
\ee
and using these objects in place of $h_{ij}$ and $\lam_i^\mu$ in
various places. 
$h_{\mu \nu} = g_{\mu \nu} - n_{s \mu} n_{s \nu}$
is the covariant induced metric on the brane and $h_{\mu}{}^{\nu}$
is the projector onto the worldvolume directions.

\subsection{D7-branes in $AdS_5 \times S_5$ black hole}

We now specialize to the case of D7-branes considered in the main
text, where we have two transverse directions with
 \begin{equation}
n_1^\nu = \frac{1}{N_1} \left( \dbyd{y}^\nu - y_0'(\rho)
\dbyd{\rho}^\nu \right)\,,  \quad n_2^\nu = \frac{1}{N_2}
\dbyd{\phi}^\nu\ ,
\end{equation}
where $N_{1,2}$ are normalization factors. In this case $U_{st}^i$
is proportional to the 
two-dimensional antisymmetric tensor $\ep_{st}$.
It is easy to see that $n_2^\nu$ is both hypersurface orthogonal
and proportional to a Killing vector (since nothing depends on
$\phi$). We thus have $K_{2ij} =0$ and $U^i_{12} =0$. The action
(\ref{PNA}) now reduces to the form we have used in Section 4, namely
 \begin{equation}
 \label{epep}
S_{D7} = \mu_7 \int d^8 \xi \sqrt{-h} \left( 1+ \frac{1}{2} (\partial
\phi_1)^2 + \frac{1}{2} (\partial \phi_2)^2 + \frac{1}{2} m_1^2
\phi_1^2 + \frac{1}{2} m_2^2 \phi_2^2 \right)\ ,
\end{equation}
where the ``masses'' are given by
\begin{eqnarray}\label{BB}
m_1^2 &=& - R_{11} - R_{2112} - K_{1ij}K_{1}^{ij}\ ,  \\
m_2^2 &=& - R_{22} - R_{2112} \ ,\label{AA}
\end{eqnarray}
with $R_{2112}$, $R_{11}$ and $R_{22}$ as defined in (\ref{R2112defn}).
In writing  (\ref{epep})--(\ref{AA}) we have used the identities
 \be
 R_{sijt} h^{ij} = n_s^\al n_t^\beta h^{ij} \lam_i^\mu \lam_j^\nu
 R_{\al \mu \nu \beta} = -  R_{s t} - R_{s11t} - R_{s22t}, \qquad
 s,t =1,2
 \ee
and the fact that $R_{12} =0$ for the $AdS_5 \times S_5$ black hole
spacetime. We can also use the generalization of the Gauss-Codazzi relation
for a codimension two surface which we derive
in~Section~\ref{sec:UF}, see Eq.~(\ref{NNS}), to write
 \begin{equation}
 K_{1ij}K_{1}^{ij} = - \left.^{(8)}R\right. +  R - 2 R_{11} - 2 R_{22} - 2 R_{2112}\ .
 \label{eq:gc}
\end{equation}
Therefore, $m_1^2$ in (\ref{BB}) can equivalently be written as
 \be
 m_1^2 = R_{11} + R_{2112} + 2 R_{22} +  \left.^{(8)}R\right. - R
 \ ,
 \ee
which is the form that we used in Section 4.

\subsection{Gauss-Codazzi relations for co-dimension 2 \label{sec:UF}}

Define the covariant derivative on the D7 brane as
\begin{equation}
D_\alpha s^\beta \equiv h^\mu_\alpha h^\beta_\nu \nabla_\mu s^\nu\ .
\end{equation}
This is equivalent to the covariant derivative defined with
respect to $h_{ij}$. We can now use $D_\alpha$ to define the curvature
of the D7-brane and then relate it to the curvature of the full
space. Calculations similar to those in \cite{WaldCitation}
reveal that
\begin{equation}
 \label{NNS}
\left.^{(8)}R_{ijk}^{\hphantom{ijk}l}\right. =
P(R)_{ijk}^{\hphantom{ijk}l} + (K^s)_{ik} (K^s)_j^l - (K^s)_{jk}
(K^s)_i^l\ ,
\end{equation}
where $s$ labels the two directions perpendicular to the brane and is summed
over. $P(R)$ is the projection of the full Riemann tensor onto the D7-brane,
\begin{equation}
P(R)_{ijk}^{\hphantom{ijk}l} = \lambda_i^\mu  \lambda_j^\nu \lambda_k^\alpha
\lambda^l_\beta R_{\mu\nu\alpha}^{\hphantom{\mu\nu\alpha}\beta}\ .
\end{equation}
Taking further contractions of Eq.~(\ref{NNS}) with $\delta_j^l$
and $h^{ik}$ and using Eq.(\ref{DDa}) gives
\begin{equation}
\left.^{(8)}R\right. = R - 2 R_{ss} - R_{tsst} + K_s K_s - (K_{sij}
K_{s}^{ij})\, ,
\end{equation} 
where $s,t$ are both summed.  In the case of interest, where $K_{2ij}=0$
because $n_2^\mu$ is proportional to a Killing vector and where
$K_s=0$ is the embedding equation, we obtain (\ref{eq:gc}).

\section{D$p$-D$q$-Brane Theories}

It will be of interest in future  to study the degree to
which the meson
dispersion relations that we have derived, together
with their consequences like (\ref{LowTempResult}) 
and (\ref{ThirdTime}), change as one modifies the gauge
theory to make it more QCD-like.  In this appendix,
we report on a check that we have mentioned in Section 6
in which the gauge theory is modified,
albeit not in the direction of QCD.  We consider
the $(p+1)$-dimensional
gauge theories 
described by $N$ D$p$-branes~\cite{Itzhaki:1998dd} into which fundamental
quarks, and hence mesons, have been introduced by embedding
$N_f$ D$q$-branes~\cite{Arean:2006pk,Myers:2006qr,Mateos:2006nu,Mateos:2007vn}.  
The D$p$-branes fill coordinates $0,1,\ldots,p$.  The
D$q$-branes fill coordinates $0,1,\ldots,d$, 
where $d \leq p$, as well as $q-d$ of the remaining $9-p$ coordinates.
In the large-$N$ limit, the near horizon geometry of the D$p$-branes
is dual to a $(p+1)$-dimensional supersymmetric
Yang-Mills theory with 16 supercharges that is nonconformal
for $p\neq 3$.  We will restrict to $p<5$. 
In the $N_f/N \rightarrow 0$ approximation,
the D$q$-branes live in the background D$p$-brane geometry, and
their back-reaction on the geometry can be neglected.   
Strings which stretch between the D$q$- and the D$p$-branes
are dual to $N_f$ fundamental quarks in the gauge theory.
We shall set $N_f=1$.  And, scalar mesons in the gauge theory
are represented by fluctuations of the position of the Dq-brane.
The specific case that we have analyzed throughout most of this paper
is $p=d=3$, $q=7$.  In this more general setting, as in the
specific case, there is a dissociation transition at some $T_{\rm diss}$
at which the
spectrum of meson fluctuations changes from discrete to continuous.

The background D$p$-brane geometry is described by
the metric~\cite{Itzhaki:1998dd}
\begin{equation}
ds^2 = R^2 
\left(\frac{R}{L_0}\right)^{(3-p)/2}
\left( -f dt^2 + r^{(7-p)/2} dx_p^2 + \frac{r^{(p-3)/2}}{ u^{2}} 
\left( d\rho^2 + dy^2
+ \rho^2 d\Omega_{q-d-1}^2 + y^2 d\Omega_{8-p-q+d}^2 \right) \right) 
\label{DpDqMetric}
\ee
and the dilaton
\be
e^{\phi} = \left(\frac{L_0}{R}\right)^{(p-3)(7-p)/4}
 g_s r^{(p-3)(7-p)/4}\ ,
\label{DpDqDilaton}
\ee
where
\begin{eqnarray}
f &=& u^{-(7-p)/2} \frac{\left( u^{7-p} - \eps^{(7-p)/2}\right)^2  }{ u^{7-p} 
+ \eps^{(7-p)/2}}\ ,  \\
r^{(7-p)/2} &=& u^{-(7-p)/2} \left( u^{7-p} + \eps^{(7-p)/2} \right) \ ,\\
u^2 &=& y^2 + \rho^2\ ,
\label{eq:YMdpdq}
\end{eqnarray}
and where we are using dimensionless coordinates as in (\ref{eyr}).
The black hole horizon is located at $u=u_0\equiv \sqrt{\eps}$.
$L_0$ specifies the position where the D$q$-brane that we introduce
will sit, as follows.  We shall embed a D$q$-brane described, in
the absence of fluctuations, by a curve $y(\rho)$ with the D$q$-brane
placed such
that its tip is located at $\rho=0$ and $y=L_0$, and then use
$L_0$ to rescale metric coordinates such that the tip of the D$q$-brane
is at $y(0)=1$.  After this rescaling, the metric and dilaton are
given by (\ref{DpDqMetric}) and (\ref{DpDqDilaton}).
The holographic dictionary determines the coupling, number of colors,
and temperature in the gauge theory via
\begin{eqnarray}
\lambda &=& \frac{(16 \pi^3)^{(p-3)/2}}{\Gamma\left(\frac{7-p}{2}\right)}  \,R^{7-p} \alpha'^{p-5}\ ,
\label{DpDqlambda}\\ 
\frac{\lambda}{N} &=& 2^{p-1}\pi^{p-2}g_s\alpha'^{(p-3)/2}\ ,\label{DpDqgstring}\\
T &=& \frac{(7-p) 2^{(5-p)/(7-p)} }{4\pi} \, u_0^{(5-p)/2} R^{-1} \left( \frac{L_0}{R} \right)^{(5-p)/2}\ .
\label{DpDqT}
\end{eqnarray}
Note that $\lambda$ has dimension $p-3$, making it useful to
define the dimensionless coupling
\begin{equation}
\lambda_{\rm eff} \equiv \lambda T^{p-3}\ .
\end{equation}

The differential equation that specifies the shape of the
embedding curve $y(\rho)$ can be derived as we did
in obtaining (\ref{eq:yrom}). For the special case in
which $p-d+q-d=4$, the embedding equation simplifies, becoming
\begin{equation}
\frac{y''}{1+y'^2} + \frac{(q-d-1) y'}{\rho} = 
\frac{ 2 \eps^{(7-p)/2} (y - y' \rho)}{u^2} 
\left( \frac{(3-d) u^{7-p} + (q-d) \eps^{(7-p)/2}}{u^{2(7-p)} - \eps^{7-p}} 
\right)\ .
\end{equation}
We have scaled our variables so that the tip of the D$q$-brane  
is at $y(0)=1$; in order to have a smooth embedding we
require $y'(0)=0$; using these boundary conditions, we can
then solve the embedding equation and obtain 
$y(\infty)$, which defines $\ep_\infty$ via $y(\infty)=\sqrt{ \eps/\ep_\infty}$, .  Finally, we can determine
what the mass $m_q$ of the quarks that we are analyzing is
via
\begin{equation}
\label{eq:mqdpdq}
m_q^2 =  \frac{\eps L_0^2}{4\pi^2 \ep_\infty \alpha'^2 }\ .
\end{equation}
From (\ref{DpDqlambda}), (\ref{DpDqT}) and (\ref{eq:mqdpdq}) we find that
\begin{equation}
\ep_\infty = a_p \left( \frac{T}{m_q}\right)^2 \lambda_{\rm eff}^{2/(5-p)}
= a_p \frac{\lambda^{2/(5-p)} T^{4/(5-p)}  }{m_q^2}\ ,
\end{equation}
where the constant $a_p$ is given by
\be
a_p = \frac{2^{(10-2p)/(7-p)}\, \pi^{(3-p)/(5-p)}}{(7-p)^{4/(5-p)}}\,
 \left( \Gamma\left(\frac{7-p}{2}\right) \right)^{2/(5-p)}\ .
\ee
We also note that the energy density of the  plasma is given by~\cite{Itzhaki:1998dd}
\begin{equation}
\rho = b_p N^2 T^{p+1} \lambda_{\rm eff}^{(p-3)/(5-p)}
=b_p  N^2 \lambda^{(p-3)/(5-p)} T^{(14-2p)/(5-p)}\  ,
\label{DpDqEnergy}
\end{equation}
where the constant $b_p$ is given by
\be
b_p=\frac{(9-p) \,  2^6 \,  \pi^{(13-3p)/(5-p)}    }{(7-p)^{(19-3p)/(5-p)} } 
\left( \Gamma\left(\frac{7-p}{2} \right)\right)^{2/(5-p)}\ .
\ee
This means that
\begin{equation}
\label{eq:ratiodpdq}
\left(\frac{\ep_\infty}{\ep_\infty^{\rm diss}} \right)^{(7-p)/2} =
\frac{\rho (T)}{\rho_{\rm diss}} \ ,
\end{equation}
where the zero-velocity mesons dissociate at a temperature $T_{\rm diss}$ corresponding
to $\rho=\rho_{\rm diss}$  and $\ep_\infty=\ep_\infty^{\rm diss}$, with
$\ep_\infty^{\rm diss}$ a constant of order unity.

We shall not repeat our construction of the meson wave functions and dispersion
relations for the D$p$-D$q$ system here.  Instead, we shall assume that
in the large-$k$ limit the meson wave functions become localized at the
tip of the D$q$ brane at $\rho=0$ and $y=1$, as we found for the D3-D7 system.
As a consequence, the limiting meson velocity will be given by the local
speed of light at the tip of the D$q$-brane.  This velocity 
can be read from the metric (\ref{DpDqMetric}),
and is given by
\begin{equation}
v_0 = \left(\frac{ 1 - \eps^{(7-p)/2}}{1+ \eps^{(7-p)/2}} \right)\ .
\label{v0DpDqAppendix}
\end{equation}
In Section 6 we have analyzed this result in the small $\eps$ limit, showing
that in this limit it takes on the form (\ref{v0andrho}) for any $p$.  This illustrates
the generality of the result (\ref{LowTempResult}) when it is phrased in terms
of the energy density.  Here, we shall analyze (\ref{v0DpDqAppendix}) at
arbitrary $\eps<1$, seeking to compare it to (\ref{ThirdTime}).   
From (\ref{v0DpDqAppendix}) and (\ref{eq:ratiodpdq}) 
we see that the critical velocity satisfies
\be
\frac{1-v_0}{1+v_0}=\frac{1-v_0^2}{(1+v_0)^2} = \eps^{(7-p)/2} 
= \frac{\rho}{\rho_{\rm diss}}
\left(  \ep_\infty^{\rm diss} \, \frac{\eps}{\ep_\infty} \right)^{(7-p)/2} \ .
\label{v0vsrho}
\ee
Recall that $\ep_\infty^{\rm diss}$ is a constant
of order unity and that
$\eps/\ep_\infty$ is a weak function of temperature and hence
of $\rho$, obtained
by solving the embedding equation and making a plot
of $\ep_\infty$ vs. $\eps$ as in Fig.~\ref{fig:epsepsinf}, and
reading off the ratio.  

Much as we did in Section 6, we can see (\ref{v0vsrho}) either as giving 
the limiting velocity $v_0$ as a function of $\rho$, or as giving 
$\rho_{\rm diss}(v)$, the energy density above which no mesons
with velocity $v$ exist, via
\be
\rho_{\rm diss}(v) = (1-v^2) \rho_{\rm diss} \left[ \frac{1}{(1+v)^2} 
\left( \ep_\infty^{\rm diss}\, \frac{\eps}{\ep_\infty} \right)^{(p-7)/2} \right] \  .
\label{rhodissv}
\ee
This is the generalization of (\ref{ThirdTime}) to the D$p$-D$q$ system.
It is written somewhat implicitly, since $\eps/\ep_\infty$ which occurs
within the square brackets is a weak function of $\rho_{\rm diss}(v)$.
It is nevertheless manifest that 
the entire expression in the square brackets is a weak
function of $v$, varying from one constant of order one at $v=0$ to some different
constant of order one at $v=1$.   As in (\ref{eq:fv}), we can then define a function $f(v)$
by rewriting (\ref{rhodissv})  as
\be
\rho_{\rm diss}(v) = \left[ f(v)\right]^{(14-2p)/(5-p)} 
\frac{\rho_{\rm diss}(0)}{\gamma^2}, 
\label{rhodissScaling}
\ee
where $\gamma=1/\sqrt{1-v^2}$ is the Lorentz boost factor. Equivalently,
using (\ref{DpDqEnergy}) we can write
\be
T_{\rm diss}(v) = f(v) \frac{T_{\rm diss}(0)}{\gamma^{(5-p)/(7-p)}}\ .
\label{TdissScaling}
\ee
We have seen in Fig.~\ref{fig:fv} that for the D3-D7 brane system, $f(v)$
is everywhere close to $1$, with $f(1)=0.924$ being the farthest it 
gets from 1.  We have also done the exercise of solving the embedding
equations for $p=4$, the D4-D6 brane system with $d=3$, and find
in that case that the farthest that $f(v)$ gets from $f(v)=1$ is $f(1)=1.048$.

Given its derivation via (\ref{v0vsrho}), it would have been reasonable to
try writing
\be
\rho_{\rm diss}(v) = \left[ \tilde f(v)\right]^{(14-2p)/(5-p)} 
\rho_{\rm diss}(0)\frac{1-v}{1+v} 
\label{WrongrhodissScaling}
\ee
instead of (\ref{rhodissScaling}). This does not work as well, yielding
a $\tilde f(v)$ that reaches 1.306 for the D3-D7 system and 1.261
for the D4-D6 system.  So although there is no important parametric difference
between (\ref{WrongrhodissScaling}) and (\ref{rhodissScaling}), we have
focussed on the form (\ref{rhodissScaling}), and hence (\ref{TdissScaling}),
throughout this paper.  

The most important conclusion from our D$p$-D$q$ investigation in
this Appendix comes by comparing (\ref{rhodissScaling}) and (\ref{TdissScaling}).
We see that in all the D$p$-D$q$ systems we 
analyze, the leading velocity dependence of 
$\rho_{\rm diss}(v)$ is that it is proportional to $1/\gamma^2$,
as if the mesons see a boosted energy density as we discussed
in Section 2.    In contrast, $T_{\rm diss}(v)$ scales with a power of $\gamma$
that depends on $p$.


\end{document}